\documentclass[12pt]{article}
\usepackage{amsmath}
\usepackage{graphicx}
\usepackage[round]{natbib}
\usepackage{url} % not crucial - just used below for the UR
\usepackage[latin1]{inputenc} % for PCs
\newtheorem{theorem}{Theorem}

\newtheorem{lemma}{Lemma}

\newtheorem{remark}{Remark}
\newtheorem{proposition}{Proposition}

\usepackage{amssymb}
\usepackage{bm}
\usepackage{algorithmic}
\usepackage{booktabs}
\usepackage{amsmath}
\usepackage{algorithm}
\usepackage{algorithmic}
\renewcommand{\algorithmicrequire}{\textbf{Input:}}    \renewcommand{\algorithmicensure}{\textbf{Output:}}
\usepackage{color}
\usepackage{algorithm}
\usepackage{verbatim}%????
\usepackage{algorithmic}
\usepackage{makecell}
\usepackage{enumerate}
\usepackage{graphicx}
\usepackage{threeparttable}
\usepackage{multirow}
\usepackage{mathrsfs}
\usepackage{hyperref}
\usepackage{subfigure}
\usepackage{setspace}
 \hypersetup{
     colorlinks=true,
     linkcolor=magenta,
     filecolor=blue,
     citecolor = blue,
     urlcolor=cyan,
     }

\newcommand*{\QEDA}{\hfill\ensuremath{\blacksquare}}

\newcommand{\blind}{0}

\definecolor{darkgreen}{rgb}{0,0.6,0.2}

\begin{document}

\def\spacingset#1{\renewcommand{\baselinestretch}%
{#1}\small\normalsize} \spacingset{1}

%%%%%%%%%%%%%%%%%%%%%%%%%%%%%%%%%%%%%%%%%%%%%%%%%%%%%%%%%%%%%%%%%%%%%%%%%%%%%%

\if0\blind
{
  \title{\bf Randomized Spectral Clustering in Large-Scale Stochastic Block Models}
  \author{ Hai Zhang$^{\dagger}$, Xiao Guo$^\dagger$
\thanks{Xiao Guo is the corresponding author}, Xiangyu Chang$^\ddag$\hspace{.2cm}  \\
    $^\dagger$ Center for Modern Statistics,\\ School of Mathematics, Northwest University, China\\
    $^\ddag$
     Center for Intelligent Decision-Making and Machine Learning, \\School of Management, Xi'an Jiaotong University, China}
  \maketitle
} \fi     

\if1\blind
{
  \bigskip
  \bigskip
  \bigskip
  \begin{center}
    {\LARGE\bf Randomized Spectral Clustering in Large-Scale Stochastic Block Models}
\end{center}
  \medskip
} \fi

\bigskip
\begin{abstract}
Spectral clustering has been one of the widely used methods for community detection in networks. However, large-scale networks bring computational challenges to the {eigenvalue decomposition} therein. In this paper, we study the spectral clustering using randomized sketching algorithms from a statistical perspective, where we typically assume the network data are generated from a stochastic block model that is \emph{not} necessarily of full rank. To do this, we first use the recently developed sketching algorithms to obtain two randomized spectral clustering algorithms, namely, the random projection-based and the random sampling-based spectral clustering. Then we study the theoretical bounds of the resulting algorithms in terms of the approximation error for the population adjacency matrix, the misclassification error, and the estimation error for the link probability matrix. It turns out that, under mild conditions, the randomized spectral clustering algorithms lead to the same theoretical bounds as those of the original spectral clustering algorithm. We also extend the results to degree-corrected stochastic block models. Numerical experiments support our theoretical findings and show the efficiency of randomized methods. A new R package called \textsf{Rclust} is developed and made available to the public.

\end{abstract}

\vfill

\newpage
\spacingset{1.5} % DON'T change the spacing!
\section{Introduction}
\label{sec:intro}
Extraordinary amounts of data are being collected in the form of arrays across many scientific domains, including sociology, physics, and biology, among others. In particular, network data and network data analysis have received a lot of attention because of their wide-ranging applications in these areas \citep{newman2018networks,goldenberg2010survey,kolaczyk2009statistical}. Community detection is one of the fundamental problems in network analysis, where the goal is to find groups of nodes that are, in some sense, more similar to each other than to the other nodes. Past decades have seen various procedures on community detection including modularity maximization, spectral clustering, likelihood methods, semidefinite programming, among others; see \citet{abbe2017community} for a recent survey.
However, large networks, say, networks with millions of nodes, bring great challenges to these community detection procedures despite the increasing computational power. {Taking the spectral clustering that we will focus on in this paper as an example, the full eigenvalue decomposition therein is time demanding when the dimension becomes large.
}

Randomization has become one popular method for modern large-scale data analysis; see \citet{mahoney2011randomized}, \citet{drineas2016randnla}, and references therein. The general idea is that depending on the problem of interest, {one uses} a degree of randomness to construct a small ``sketch'' of the full data set, and {then uses} the resulting sketched data instead to reduce the computational burden. Random projection and random sampling are the two general approaches to obtain such a sketch matrix. Roughly speaking, random projection reduces the computational cost by projecting the data matrix to a smaller dimensional space in order to approximate the data. While random sampling algorithms lighten the computational burden by sampling and rescaling the data in some manner. The randomization techniques have been applied to the least squares regression \citep{drineas2006sampling,drineas2011faster,drineas2012fast}, and the low-rank matrix approximation \citep{halko2011finding,martinsson2016randomized,witten2015randomized,mahoney2009cur}, among many others. Most works in this area were analyzed from an algorithmic perspective, where the randomized algorithm could lead to approximately as good performance as the full data at hand does for some problems of interest. However, from a statistical perspective, the aim is not only to obtain randomized algorithms which perform well on a particular data set but also to understand how well they perform under some underlying mechanisms. In the context of regression, there have been a few works that study the randomized algorithms under underlying regression models--for example, the ordinary linear regression \citep{ma2015statistical,raskutti2016statistical,wang2019information}, the logistic regression \citep{wang2018optimal,wang2019more}, the ridge regression \citep{wang2017sketched}, the constrained regressions \citep{pilanci2016iterative,pilanci2017newton}, and the spatial autoregressive (SAR) models \citep{zhou2017estimating,li2019randomized}, among others.

Just like they have studied how well the randomized algorithms can estimate the underlying regression model, it is natural and important to study how well we can use the randomization techniques to detect the communities in a ``true'' network model. The stochastic block model (SBM) \citep{holland1983stochastic} is a simple but expressive network model that captures the community structure of networks observed in the real world. In an SBM, nodes are partitioned into several distinct communities and conditioned on the underlying community assignments, the edges are generated independently according to the community membership of their end nodes. Nodes within the same community are generally more likely to be connected than the other nodes. The SBM is popular among statisticians because it can be rigorously studied coupling with various network community detection procedures; see \citet{abbe2017community} for an excellent review.

In this work, we focus on studying how randomization can be used to reduce the computational cost of spectral clustering, and understanding how well the resulting randomized spectral clustering algorithms perform under the SBMs. Spectral clustering is a popular and simple algorithm for clustering which consists of the following two steps. One first conducts the eigenvalue decomposition of the adjacency matrix or the Laplacian matrix and then runs the $k$-means on several leading eigenvectors to obtain the nodes clusters or communities \citep{von2007tutorial}. It is well known that the full eigenvalue decomposition in the first step generally requires $O(n^3)$ time where $n$ denotes the number of nodes, which is time demanding when $n$ becomes huge. Regardless of the computational issues, it has been shown to enjoy good theoretical properties within the SBM framework; see, \citet{rohe2011spectral,choi2012stochastic,qin2013regularized,lei2015consistency,sarkar2015role,joseph2016impact,su2019strong,
yang2020simultaneous,tang2021asymptotically,deng2021strong,levin2021limit}, among many others. {Facing large networks, it is thus desirable to study whether these properties would retain under certain randomization of the algorithms.} In this paper, we utilize the idea of randomization to obtain two kinds of randomized spectral clustering algorithms; namely, the random projection-based and the random sampling-based spectral clustering, and in particular, we study their theoretical properties under the SBMs.

We focus on the adjacency matrix $A$ of the network. The random projection-based method is motivated as follows. Note that the adjacency matrix inherits a low-rank structure {approximately} {since it is assumed to be sampled from a SBM~\citep{lei2015consistency,rohe2011spectral}}. Therefore, if one can make use of such low-rank structure to derive a matrix with a lower dimension which captures the essential information of $A$, then the eigenvalue decomposition of this matrix can help to derive that of $A$, which in turn reduces the computational cost. Indeed, the recently developed randomized low-rank matrix approximation algorithms provide a powerful tool for performing such low-rank matrix approximation \citep{halko2011finding,witten2015randomized,martinsson2016randomized}. Specifically, these techniques utilize some amount of randomness to compress the columns and rows of $A$ to $l$ ($l\ll n$) linear combinations of the columns and rows of $A$. The eigenvalue decomposition on the resulting $l$-dimensional matrix can be largely reduced since $l$ is far smaller than $n$. The random projection-based spectral clustering refers to the original spectral clustering with its first step replaced by the randomized eigenvalue decomposition. On the other hand, the computational cost of the original spectral clustering can be reduced via the random sampling. Note that we only need to find a few leading eigenvectors of $A$, which can be obtained using many fast iterative methods, such as the orthogonal iteration and Lanczos iteration; see \citet{baglama2005augmented,calvetti1994implicitly}, among others. And it is well known that the time complexity of iterative algorithms {is in direct proportion to }the number of non-zero elements of $A$ multiplied by the number of iterations. Therefore, if we sample the elements of $A$ in some way to obtain a sparser matrix, then the time for computing its leading eigenvectors will be largely reduced. There have been a few works on the randomized matrix sparsification; see \citet{gittens2009error,achlioptas2007fast,arora2006fast,li2020network}, among others. In particular, \citet{li2020network} apply the sampling technique to study the network cross-validation problem. In this work, we use a simple sampling strategy to obtain a sparsified matrix; that is, sample pair $(i,j)$'s of nodes with probability $p_{ij}$'s, then use the iteration method of \citet{calvetti1994implicitly} to find its leading vectors, and after that perform the $k$-means algorithm on these eigenvectors, which we refer to the random sampling-based spectral clustering.

We theoretically justify the randomized spectral clustering algorithms in terms of the approximation error that measures the deviation of the randomized matrix $\tilde{A}$ of the adjacency matrix $A$ from the population matrix $P$ and the misclassification error. In addition, although the spectral clustering is nonparametric in nature, we develop a simple method to estimate the link probability matrix $B$ based on the output clusters where $B_{kl}$ is the edge probability between any node pairs in communities $k$ and $l$, and provide its theoretical bound.
It is worth noting that our analysis does \emph{not} rely on the common assumption in most SBM literatures that $B$ is of full rank. In particular, we analyze the true eigen-structure of $P$ in the rank-deficient scheme and provide an explicit condition under which the nodes from different communities are separable. It turns out that the approximation error bound in terms of the spectral norm, namely, $\|\tilde{A}-P\|_2$, attains the minimax optimal rate in SBMs \citep{gao2015rate,gao2020discussion} under mild conditions, {indicating that the optimization error from randomization, namely, $\|\tilde{A}-A\|_2$, is dominated by the statistical error from SBMs, namely, $\|A-P\|_2$. The misclassification error bounds are identical to the original spectral clustering \citep{lei2015consistency} and are optimal provided that the community number $K$ is fixed \citep{ahn2018hypergraph}.} We also generalize the results to degree-corrected block models--an extension of SBMs incorporating the degree heterogeneity \citep{karrer2011stochastic}.

The contributions of this paper are as follows. First, we utilize randomization tools to obtain two kinds of randomized spectral clustering algorithms and theoretically study the resulting algorithm under the SBMs. The results provide statistical insights of randomization on spectral clustering. From the statistical perspective, the randomization does not deteriorate the error bound of $\|\tilde{A}-P\|_2$, because the latter already attains the minimax optimal rate in SBMs. Second, extending the full-rank assumption in most works on SBMs, we also study the rank-deficient SBMs, analyze the true eigen-structure of these models, and provide sufficient conditions under which the spectral clustering may succeed, which is rarely mentioned in SBMs works and of independent interest. Third, we develop a new R package called \textsf{Rclust}\footnote{\url{https://github.com/XiaoGuo-stat/Rclust}} to facilitate the use of the randomized spectral clustering algorithms studied in this work.

The remainder of this paper is organized as follows. Section \ref{sec:pre} defines the notation, introduces and analyzes the SBM and spectral clustering in more detail. Section \ref{sec:rspec} includes the random projection-based and random sampling-based spectral clustering schemes that we consider. Section \ref{sec:theo} presents the theoretical results. Section \ref{sec:dcsbm} contains extensions to degree-corrected block models.
Section \ref{sec:related} reviews and discusses related works. Section \ref{sec:simu} and \ref{sec:real} display the simulation and real experiments that verify the theoretical results and show the effectiveness of the proposed methods. Section \ref{sec:dissc} concludes with discussion. Proofs are provided in the online supplemental material.

\section{Preliminaries }
\label{sec:pre}
In this section, we provide some notation and briefly introduce the SBMs and the spectral clustering algorithm. In particular, the rationality of spectral clustering under SBMs is analyzed.
\subsection{Notation}
Let $\mathbb M_{n,K}$ be the set of all $n\times K$ matrices that have exactly one 1 and $K-1$ 0's in each row. Any $\Theta\in \mathbb M_{n,K} $ is called a \emph{membership matrix} where each row represents the community membership of a node in a network with $K$ communities; for example, node $i$ belongs to community $g_i\in\{1,...,K\}$ if and only if $\Theta_{ig_i}=1$. For $1\leq k\leq K$, let $G_k=G_k(\Theta)=\{i\in[n]:g_i=k\}$, where $[n]:=\{1,2,\dots,n\}$. $G_k$ consists of nodes with their community membership being $k$, and denote $n_k=|G_k|$. For any matrix $A_{n\times n}$ and $I,J\subseteq [n]$, $A_{I\ast}$ and $A_{\ast J}$ denote the submatrix of $A$ consisting of the corresponding rows and columns, respectively.
$\|A\|_{\tiny {\rm F}}$ and $\|A\|_\infty$ denote the Frobenius norm and the element-wise maximum absolute value of $A$, respectively. We use $\|\cdot\|_2$ to denote the Euclidean norm of a vector and the spectral norm of a matrix.  In addition, ${\rm diag}(A)$ denotes the matrix with its diagonal elements being the same as those of $A$ and non-diagonal elements being 0's.
\subsection{Stochastic block model}
\label{sub:sbm}
The SBM introduced by \citet{holland1983stochastic} is a class of probabilistic model for networks with well-defined communities. For a potential network with $n$ nodes and $K$ communities, the model is parameterized by the membership matrix $\Theta\in\mathbb M_{n,K}$, and the link probability matrix $B\in[0,1]^{K\times K}$ where $B$ is symmetric, and the entry of $B$; for example, $B_{kl}$, represents the edge probability between the community $l$ and $k$. Here, $B$ is \emph{not} necessarily of full rank, and we assume ${\rm rank}(B)=K'\leq K$. Given $\Theta$ and $B$, the network adjacency matrix $A=(a_{ij})_{1\leq i,j\leq n}\in\{0,1\}^{n\times n}$ is generated as
\[a_{ij}=\begin{cases}
\label{2.1}
{\rm Bernoulli } (B_{g_ig_j}) & \mbox{if }\; i<j, \\
0,& \mbox{if } \;i=j,\\
a_{ji}, & \mbox{if } \;i>j.\tag{2.1}
\end{cases}\]
Define $P=\Theta B\Theta^{\intercal}$, then it is easy to see that $P$ is the population version of $A$ in the sense that $\mathbb E(A)=P-{\rm diag}(P)$. Under the SBMs, the goal of community detection is to use the adjacency matrix $A$ to recover the membership matrix $\Theta$ up to column permutations.

\subsection{Spectral clustering}
Spectral clustering is a popular and simple algorithm for community detection in networks \citep{von2007tutorial}. It generally consists of two steps. The first step is to perform the eigenvalue decomposition of a suitable matrix representing the network, where we consider the simple adjacency matrix $A$, and then put the eigenvectors of $A$ corresponding to the $K'$ largest eigenvalues into a $n\times K'$ matrix $\hat{U}$. Here and throughout the following paper, we should keep in mind that the target rank is $K'$ while the target community number is $K$ in the SBMs defined in Subsection \ref{sub:sbm}. In the next step, we treat each row of $\hat{U}$ as a point in $\mathbb R^K$ and run $k$-means on $\hat{U}$ with $K$ clusters. In this paper, for simplicity, solving $k$-means is to use the standard and efficient heuristic Lloyd's algorithm. The resulting clustering labels are arranged as $\tilde{\Theta}\in \mathbb M_{n,K}$, and the $K$-dimensional centroid vectors are collected as $\tilde{X}\in \mathbb R^{K\times K}$, where the $i$th row of $\tilde{X}$ corresponds to the centroid of the $i$th cluster. We summarize the spectral clustering in Algorithm \ref{spectral}, where note that we use $\hat{U}$ to denote the eigenvectors of $A$ by contrast to those of the population $P$ denoted by $U$, and we use ${\tilde{U}}$ to denote the estimator of $\hat{U}$ obtained by $k$-means.

The spectral clustering is interpretable in SBMs because the population matrix $P$ has eigenvectors that reveal the true clusters as shown in the next lemma.

\begin{lemma}
\label{lem:eigen}
For an SBM with $K$ communities parameterized by $\Theta\in \mathbb M_{n,K}$ and $B\in [0,1]^{K\times K}$ with ${\rm rank}(B)=K'(K'\leq K)$, suppose the eigenvalue decomposition of $P=\Theta B\Theta^{\intercal}$ is $U_{n\times K'}\Sigma_{K'\times K'} U^{\intercal}_{K'\times n}$. Define $\Delta={\rm diag}(\sqrt{n_1},...,\sqrt{n_{K}})$ and denote the eigenvalue decomposition of $\Delta B\Delta$ by $L_{K\times K'}D_{K'\times K'}L^\intercal_{K'\times K}$. Then the following arguments hold.

(a) If $B$ is of full rank, i.e., $K'=K$, then for $\Theta_{i\ast}=\Theta_{j\ast}$, we have ${U}_{i\ast}={U}_{j\ast}$; while for $\Theta_{i\ast}\neq\Theta_{j\ast}$, we have $\|{U}_{i\ast}-{U}_{j\ast}\|_2=\sqrt{(n_{g_i})^{-1}+(n_{g_j})^{-1}}$.

(b) If $B$ is rank deficient, i.e., $K'<K$, then for $\Theta_{i\ast}=\Theta_{j\ast}$, we have ${U}_{i\ast}={U}_{j\ast}$; while for $\Theta_{i\ast}\neq\Theta_{j\ast}$, if $\Delta^{-1}L$'s rows are mutually distinct such that there exists a deterministic sequence $\{\xi_n\}_{n\geq 1}$ satisfying
\begin{align}
\label{A1}
{\rm min}_{\textcolor{red}{k\neq l}}\|\frac{L_{k\ast}}{\sqrt{n_{k}}}-\frac{L_{l\ast}}{\sqrt{n_{l}}}\|_2\geq \xi_n >0,\tag{A1}
\end{align}
then $\|{U}_{i\ast}-{U}_{j\ast}\|_2=\|\frac{L_{g_{i}\ast}}{\sqrt{n_{g_i}}}-\frac{L_{g_{j}\ast}}{\sqrt{n_{g_j}}}\|_2\geq \xi_n >0 $.
\end{lemma}

Lemma \ref{lem:eigen} says that when $B$ is of full rank, two rows of $U$ are identical if and only if the corresponding nodes are in the same community. These results have already been obtained in \cite{lei2015consistency,rohe2011spectral}, among others. While when $B$ is rank deficient, we additionally assume (\ref{A1}) holds in order to make sure that two rows of $U$ are separable when the corresponding nodes are in distinct communities. The next lemma provides an explicit condition on $B$ that suffices for (\ref{A1}). In particular, the within-community probabilities should dominate the between-community probabilities in the sense of (\ref{2.2}).
\begin{lemma}
\label{lem:eigencondition}
For an SBM with $K$ communities parameterized by $\Theta\in \mathbb M_{n,K}$ and $B\in [0,1]^{K\times K}$ with ${\rm rank}(B)=K'<K$, suppose the eigenvalue decomposition of $P=\Theta B\Theta^{\intercal}$ is $U\Sigma U^{\intercal}$. If there exists two deterministic sequences $\{\eta_n\}_{n\geq 1}$ and $\{\iota_n\}_{n\geq 1}$ such that
\begin{align}
\label{2.2}
{\rm min}_{1\leq k<l\leq K}B_{kk}+B_{ll}-2B_{kl}\geq \eta_n >0,\nonumber
\tag{2.2}
\end{align}
and for any $1\leq i\leq K'$, $0<\Sigma_{ii}\leq \iota_n$, then (\ref{A1}) holds with $\xi_n=\sqrt{\eta_n/\iota_n}$.
\end{lemma}

Lemma \ref{lem:eigen} and \ref{lem:eigencondition} indicate that the spectral clustering could work well if the $K'$ leading eigenvectors of $A$ are close to those of the population $P$. While when $n$ is large, the full eigenvalue decomposition is time consuming. In the following sections, we will make use of the recently developed randomization techniques--namely, the random projection and the random sampling, to accelerate the spectral clustering. In the meanwhile, we will theoretically study how the randomized spectral clustering methods interact with the assumptions of SBMs.

\begin{algorithm}[htb]
\footnotesize

\renewcommand{\algorithmicrequire}{\textbf{Input:}}

\renewcommand\algorithmicensure {\textbf{Output:} }

\caption{Spectral clustering for $K$ clusters}~\label{alg:spectral_cluster} %Ëã·¨µÄ±êÌâ

\label{spectral} %¸øËã·¨Ò»¸ö±êÇ©£¬ÕâÑù·½±ãÔÚÎÄÖÐ¶ÔËã·¨µÄÒýÓÃ

\begin{algorithmic}[1] %Õâ¸ö1 ±íÊ¾Ã¿Ò»ÐÐ¶¼ÏÔÊ¾Êý×Ö

\REQUIRE ~\\ %Ëã·¨µÄÊäÈë²ÎÊý£ºInput

Cluster number $K$, {target rank $K'$}, adjacency matrix $A\in \mathbb{R}^{n\times n}$;\\

\ENSURE ~\\ %Ëã·¨µÄÊä³ö£ºOutput
Estimated membership matrix $\tilde{{\Theta}}\in\mathbb M_{n,K}$ and centriods $\tilde{{X}}\in \mathbb{R}^{K\times K'}$ ;\\
Estimated eigenvectors ${\tilde{U}}=\tilde{{\Theta}}\tilde{{X}}$;\\
~\\
\STATE Find the $K'$ leading eigenvectors $\hat{U}$ of $A$ corresponding to the $K'$ largest eigenvalues of $A$. \\
\STATE Treat each row of $\hat{U}$ as a point in $\mathbb{R}^{K'}$ and run the Lloyd's algorithm on these points with $K$ clusters. Let
$(\tilde{\Theta},\tilde{X})$ be the solution.
 \\
\end{algorithmic}
\end{algorithm}

\section{Randomized spectral clustering }
\label{sec:rspec}
In this section, we use the randomization techniques to derive two kinds of randomized spectral clustering--namely, random projection-based spectral clustering and random sampling-based spectral clustering.
\subsection{Randomized spectral clustering via random projection}
Recall that $A$ is generated from a low-rank matrix $P=\Theta B\Theta^{\intercal}$, hence $A$ inherits a low-rank structure naturally. Therefore, if one can make use of such low-rank structure to derive a smaller matrix that captures the essential information of $A$, then the eigenvalue decomposition of the smaller matrix can help to derive that of $A$, which in turn reduces the computational cost. Fortunately, randomization is a powerful tool for performing such low-rank matrix approximation \citep{halko2011finding,witten2015randomized,martinsson2016randomized}. These techniques utilize some amounts of randomness to compress the input matrix to obtain a low-rank factorization efficiently, which is called \emph{random projection}. In this section, we introduce the random projection strategy in the context of eigenvalue decomposition.

Let us see how the random projection can help reduce the time for the eigenvalue decomposition of adjacency matrix $A$. For a symmetric matrix $A\in\mathbb{R}^{n\times n}$ with target rank $K'$, we aim to find an orthonormal basis $Q\in \mathbb{R}^{n\times K'}(K'\leq n)$ such that
$$A\approx QQ^{\intercal}AQQ^{\intercal}:= \tilde{A}^{\rm rp},$$
where $\tilde{A}^{\rm rp}$ is essentially a low-rank approximation of $A$. Before constructing $Q$, we here provide some insights. $Q\in \mathbb{R}^{n\times K'}$ can be thought as a low-rank approximation of the column (row) space of matrix $A$. To see this, suppose the eigendecomposition of $A$ is $A=\hat{U}_{n\times m}\hat{\Sigma}_{m\times m}\hat{U}^{\intercal}_{m\times n}$, where $m$ is the rank of $A$ and $\hat{U}$ represents the column (row) space of $A$. Then, when $Q=\hat{U}$ and $m=K'$, it is straightforward to see $A=QQ^{\intercal} AQQ^{\intercal}$. In addition, $QQ^{\intercal}$ is a projection operator which projects any vector $x\in\mathbb{R}^n$ to the column space of $Q$, i.e., $\|x-QQ^{\intercal}x\|_2^2=\mbox{min}_{y\in \mathbb{R}^K}\;\|x-Qy\|_2^2$. $Q$ can be obtained using the following steps \citep{halko2011finding}:\\
\textbf{\emph{Step 1:}} Form a random test matrix $\Omega=(\omega_1,...,\omega_{K'})\in \mathbb R^{n\times K'}$, where $\{\omega_i\}_{i=1}^{K'}$ are $n$-dimensional random vectors independently drawn from a distribution.    \\
\textbf{\emph{Step 2:}} Form the sketch matrix $Y=(y_1,...,y_K)=A\Omega\in\mathbb{R}^{n\times K'}$.\\
\textbf{\emph{Step 3:}} Obtain $Q$ via the QR decomposition $Y=:QR$.\\
Once $Q$ is obtained, we can perform the eigenvalue decomposition on the smaller matrix $C:=Q^{\intercal}AQ\in \mathbb{R}^{K'\times K'}$, and then post process it to obtain the approximate eigenvectors of $A$. In this way, {the computational cost of the original spectral clustering could be largely reduced when we incorporate the aforementioned steps into Algorithm \ref{alg:spectral_cluster} to provide the approximate eigenvectors of the adjacency matrix. We call this procedure random projection-based spectral clustering.
}

The random test matrix $\Omega$ can be generated in various ways, specifically, the entries of $\omega_i$ can be i.i.d. standard Gaussian, uniform, and Rademacher distributions, among many others. The oversampling strategy is often used to improve the empirical performance of the randomized low-rank approximation \citep{halko2011finding,witten2015randomized,martinsson2016randomized}. As most data matrices do not have exact rank $K'$, it is desirable to use $l:=K'+r$ random projections instead of exact $K'$ projections to form the random sketch of $A$. In practice, $r=\{5,10\}$ often suffices to make sure that the obtained basis $Q$ is close to the best possible basis, namely, the $K'$ leading eigenvectors of $A$, with high probability  \citep{martinsson2016randomized}. Besides the oversampling scheme, the power iteration is another way to improve the quality of low-rank approximation. For some data matrices, the eigenvalues decay slowly that may lead to information loss. Thus instead of forming the sketch $Y$ on the basis of $A$, several authors incorporate $q$ steps of a power iteration before constructing the sketch matrix $Y$. Formally, it is defined as
$$Y:=(AA^{\intercal})^qA\Omega=A^{2q+1}\Omega.$$ In practice, $q=1$ or $q=2$ often suffices to make the spectrum decay fast \citep{halko2011finding}. We summarize the random projection-based spectral clustering procedure with such power iteration and the aforementioned oversampling strategies in Algorithm \ref{randsp}.

\begin{remark}
The time complexity of Algorithm \ref{randsp} is dominated by the matrix multiplications when forming $Y$ and $C$ in Step 2 and Step 4, which take $O((2q+1)n^2(K'+r))$ and $O(n^2(K'+r))$ time, respectively. In particular, the time complexity of Step 2 can be improved to $O((2q+1)n^2{\rm log}(K'+r))$ by using {\rm structured} random test matrices, for example, the {\rm subsampled random Fourier transform} \citep{halko2011finding,erichson2019randomized}. Moreover, the matrix-vector multiplications in Step 2 can be paralleled to further reduce the computation cost.
\end{remark}

\begin{algorithm}[htb]
\footnotesize

\renewcommand{\algorithmicrequire}{\textbf{Input:}}

\renewcommand\algorithmicensure {\textbf{Output:} }

\caption{Randomized spectral clustering via random projection } %Ëã·¨µÄ±êÌâ

\label{randsp} %¸øËã·¨Ò»¸ö±êÇ©£¬ÕâÑù·½±ãÔÚÎÄÖÐ¶ÔËã·¨µÄÒýÓÃ

\begin{algorithmic}[1] %Õâ¸ö1 ±íÊ¾Ã¿Ò»ÐÐ¶¼ÏÔÊ¾Êý×Ö

\REQUIRE ~\\ %Ëã·¨µÄÊäÈë²ÎÊý£ºInput

Cluster number $K$, target rank $K'$, adjacency matrix $A\in \mathbb{R}^{n\times n}$, oversampling parameter $r$, and exponent $q$;\\

\ENSURE ~\\ %Ëã·¨µÄÊä³ö£ºOutput
Membership matrix ${\tilde{\Theta}}^{\rm rp}\in\mathbb M_{n,K}$ and centriods ${\tilde{X}}^{\rm rp}\in \mathbb{R}^{K\times K'}$ ;\\
${\tilde{U}}^{\rm rp}={\tilde{\Theta}}^{\rm rp}{\tilde{X}}^{\rm rp}$;\\
~\\
\STATE Draw a $n\times( K+r)$ random test matrix $\Omega$.\\
\STATE Form the matrix $Y=A^{2q+1}\Omega$.\\
\STATE Construct $Q$ via orthonomalizing the columns of $Y$, i.e., $Y=:QR$.\\
\STATE Form $C=Q^{\intercal}AQ$ and denote $ \tilde{A}^{\rm rp}\equiv QCQ^{\intercal}$.\\
\STATE Compute the eigenvalue decomposition of the small matrix: $C={U_s}\Sigma_s {U}_s^{\intercal}$.\\
\STATE Set ${U}^{\rm rp}$ to be the column subset of $Q{U}_s$ corresponding to the $K'$ largest values of $\Sigma_s$.
\STATE Treat each row of ${U}^{\rm rp}$ as a point in $\mathbb{R}^{K'}$ and run the Lloyd's algorithm on these points with $K$ clusters. Let $({\tilde{\Theta}}^{\rm rp},{\tilde{X}}^{\rm rp})$ be the solution.
\end{algorithmic}
\end{algorithm}

\subsection{Randomized spectral clustering via random sampling}
The random sampling strategy is to first do element-wise sampling from the adjacency matrix $A$, and then use fast iterative methods, say orthogonal iteration or Lanczos iteration, to find a nearly-optimal best rank $K'$ approximation of $A$. The motivation is that in spectral clustering, we aim to find the first $K'$ eigenvectors of $A$, or the best rank $K'$ approximation of $A$. And there exist many fast iterative methods for computing such low-rank matrix approximation; see \citet{calvetti1994implicitly,baglama2005augmented,allen2016lazysvd,lehoucq1995analysis}, among many others. The time complexity of iterative methods is generally proportional to the number of non-zero elements of $A$ multiplied by the number of iterations. Hence, if we sample the elements of $A$ in some way to obtain a sparser matrix, then the time for computing its rank $K'$ approximation will be largely reduced. In the meantime, we hope that the sampling scheme does not deteriorate the accuracy too much. In the sequel, we introduce the random sampling procedure and the corresponding randomized spectral clustering.

We adopt a simple sampling strategy to obtain a sparsified version of $A$. That is, randomly select pairs $(i,j)$'s of the adjacency matrix $A$ independently with probability $p_{ij}$'s, and the randomized sparsified matrix
$\tilde{A}^{\rm s}$ is defined as
\[\tilde{A}_{ij}^{\rm s}=\begin{cases}
\label{3.1}
\frac{A_{ij}}{p_{ij}}, & \mbox{if }\; (i,j) { \mbox{ is selected},} \\
0,& \mbox{if } \;(i,j) {\mbox{ is not selected}},\tag{3.1}
\end{cases}\]
for each $i<j$, and $\tilde{A}_{ji}^{\rm s}=\tilde{A}_{ij}^{\rm s}$ for each $i>j$. Once $\tilde{A}^{\rm s}$ is obtained, we can apply an {iterative algorithm} for the eigenvalue decomposition of $\tilde{A}^{\rm s}$ to attain the nearly-optimal rank $K'$ approximation of $\tilde{A}^{\rm s}$ such that
$$\tilde{A}^s\approx{U}^{\rm rs}_{n\times K'}{\Sigma}^{\rm rs}_{K'\times K'}({U}^{\rm rs})^\intercal_{K'\times n}:=\tilde{A}^{\rm rs}.$$
Then the Lloyd's algorithm can be applied on the rows of ${U}^{\rm rs}$ to find the clusters. Let $({\tilde{\Theta}}^{\rm rs},{\tilde{X}}^{\rm rs})$ be the solution. For reference, we summarize these steps in Algorithm \ref{randss}.

%Now we illustrate the steps in Algorithm \ref{randss} in more detail. There has been a few works on the randomized matrix sparsification; see \cite{achlioptas2007fast}; \cite{gittens2009error}; \cite{spielman2011graph}, and references therein. In particular, \cite{gittens2009error} provide theoretical bounds on the approximation error of $\|\tilde{A}^{\rm s}-A\|_2$ with respect to several norms, where $A$ is assumed to be fixed. In our context, we are interested in how the sparsified and then low-rank approximated matrix $\tilde{A}^{\rm rs}$ performs under the SBMs, namely, the derivation of $\tilde{A}^{\rm rs}$ from $P$. We will actually make use of the low-rank nature of $\tilde{A}^{\rm rs}$ to derive the approximation bound $\|\tilde{A}^{\rm rs}-P\|_2$. In addition, the selected elements of $A$ in {\emph{Step 1}} of Algorithm \ref{randss} are divided by $p_s$ to ensure that $\mathbb E(A^{\rm s})=A$ conditioned on $A$.

\begin{remark}
The sampling strategy is element-specific. The simplest choice is that $p_{ij}=p$ for all pairs of $(i,j)$. Note that it is equivalent to sampling $1$'s with probability $p$ and sampling $0$'s with probability $p'\;(p'<p)$. Another choice is to set $p_{ij}$ proportional to $\|A_{i\ast}\|_2$
which {enables that the edges from high-degree nodes would remain with higher probability}, but computing $\|A_{i\ast}\|_2$ brings additional time cost. In addition, for real applications where certain edges or all edges of certain nodes are forced to remain in $\tilde{A}^s$, one can use the element-subject sampling strategy.
\end{remark}

\begin{remark}
It should be noted that the iteration algorithms in \emph{Step 2} of Algorithm \ref{alg:sampling_spectral_cluster} yields the nearly-optimal solution instead of the exactly-optimal rank $K'$ approximation and it is acceptable to work with a nearly-optimal low-rank approximation. In the theoretical analysis, we treat \emph{Step 2} as a black box and suppose the best rank $K'$ approximation is obtained. We mainly deal with approximation error induced by \emph{Step 1}.
\end{remark}

\begin{algorithm}[htb] %Ëã·¨µÄ¿ªÊ¼
\footnotesize

\renewcommand{\algorithmicrequire}{\textbf{Input:}}

\renewcommand\algorithmicensure {\textbf{Output:} }

\caption{Randomized spectral clustering via random sampling }\label{alg:sampling_spectral_cluster} %Ëã·¨µÄ±êÌâ

\label{randss} %¸øËã·¨Ò»¸ö±êÇ©£¬ÕâÑù·½±ãÔÚÎÄÖÐ¶ÔËã·¨µÄÒýÓÃ

\begin{algorithmic}[1] %Õâ¸ö1 ±íÊ¾Ã¿Ò»ÐÐ¶¼ÏÔÊ¾Êý×Ö

\REQUIRE ~\\ %Ëã·¨µÄÊäÈë²ÎÊý£ºInput

Cluster number $K$, target rank $K'$, adjacency matrix $A\in \mathbb{R}^{n\times n}$, sampling probability matrix $\bar{P}=(p_{ij})$;\\

\ENSURE ~\\ %Ëã·¨µÄÊä³ö£ºOutput
Membership matrix ${\tilde{\Theta}}^{\rm rs}\in\mathbb M_{n,K}$ and centriods ${\tilde{X}}^{\rm rs}\in \mathbb{R}^{K\times K'}$ ;\\
${\tilde{U}}^{\rm rs}={\tilde{\Theta}}^{\rm rs}{\tilde{X}}^{\rm rs}$;\\
~\\
\STATE For each pair $(i,j)(i<j)$, randomly select pair $(i,j)$ of $A$ with probability $p_{ij}$. Form the sparsified matrix $\tilde{A}^{\rm s}$ according to (\ref{3.1}). \\
\STATE  Apply an {iterative algorithm} to obtain the nearly-optimal rank $K'$ approximation of $\tilde{A}^{\rm s}$ such that $$\tilde{A}^{\rm s}\approx{U}^{\rm rs}_{n\times K'}{\Sigma}^{\rm rs}_{K'\times K'}({U}^{\rm rs})^\intercal_{K'\times n}:=\tilde{A}^{\rm rs}.$$\\
\STATE  Treat each row of ${U}^{\rm rs}$ as a point in $\mathbb{R}^{K'}$ and run the Lloyd's algorithm on these points with $K$ clusters. Let $({\tilde{\Theta}}^{\rm rs},{\tilde{X}}^{\rm rs})$ be the solution.
\end{algorithmic}
\end{algorithm}

\section{Theoretical analysis}
\label{sec:theo}
In this section, we theoretically justify the performance of two randomization schemes on spectral clustering under the model set-up of SBMs. Specifically, for each method, we evaluate its performance from the following three aspects. First, we derive an upper bound on how the randomized matrix $\tilde{A}^{\rm rp}$ (or $\tilde{A}^{\rm rs}$) deviates from the population adjacency matrix of SBMs. Then, we use these results to bound the misclassification error rate of the randomized spectral clustering algorithms. At last, we use the estimated clusters to obtain an estimate of $B$, and provide its theoretical bounds.

\subsection{Random projection}
\label{subsec:rp}
The following notes and notation would be used throughout this subsection. Let $A$ be a $n\times n$ adjacency matrix generated from a SBM with $K$ communities parameterized by $\Theta\in \mathbb M_{n,K}$ and $B\in [0,1]^{K\times K}$ with ${\rm rank}(B)=K'(K'\leq K)$. Denote the eigenvalue decomposition of $P=\Theta B\Theta^{\intercal}$ by $U_{n\times K'}\Sigma_{K'\times K'} U^{\intercal}_{K'\times n}$. Let $\sigma_n$ and $\gamma_n$ be the largest and smallest nonzero eigenvalue of $P$. Let ${\tilde{\Theta}}^{\rm rp}$ be the output of Algorithm \ref{randsp} with the target rank being $K'$, the oversampling and the power parameter being respectively $r$ and $q$, and the test matrix $\Omega$ generating  i.i.d. standard Gaussian entries. The following theorem provides the deviation of $\tilde{A}^{\rm rp}$ from $P$.
\begin{theorem}\label{rproappro}
If
\begin{equation}
\label{A2}{\rm max}_{kl}B_{kl}\leq \alpha_n \;{\rm for\; some}\; \alpha_n\geq c_0\,{\rm log}n/n,\tag{A2}
\end{equation}
and
\begin{equation}
\label{A3} r\geq 4,\, r{\rm log}r\leq n,\, K'+r\leq n,\, q=c_1\cdot n^{1/\tau},\tag{A3}
\end{equation}
for some constant $c_0,c_1>0$ and any $\tau>0$,
then for any $s>0$, there exists a constant $c_2=c_2(s,c_0,c_1)$ such that
\begin{equation}
\label{4.1}\|\tilde{A}^{\rm rp}-P\|_2\leq c_2\sqrt{n\alpha_n}, \tag{4.1}
\end{equation}
with probability at least $1-6r^{-r}-2n^{-s}$.
\end{theorem}

The deviation of $\tilde{A}^{\rm rp}$ from $P$ arises from two sources, one is the deviation of $\tilde{A}^{\rm rp}$ from $A$ (optimization error), and the other is the deviation of $A$ from $P$ (statistical error). To bound the statistical error $\|A-P\|_2$, we pose condition (\ref{A2}), a weak condition on the population network sparsity, which has been used to obtain a sharp bound of $\|A-P\|_2$ \citep{lei2015consistency,gao2017achieving,chin2015stochastic}. To bound the optimization error $\|\tilde{A}^{\rm rp}-A\|_2$, we utilize the result in \citet{halko2011finding} and pose condition (\ref{A3}) on the order of the oversampling parameter $r$ and the power parameter $q$. It essentially indicates that the optimization error caused by random projection is dominated by the statistical error caused by sampling $A$ from $P$. Note that $q=c_1\cdot n^{1/\tau}$ is mild because $\tau$ can be sufficiently large. Under (\ref{A2}) and (\ref{A3}), the bound in (\ref{4.1}) attains the minimax optimal rate under the SBMs \citep{gao2015rate,gao2020discussion}. Thus in the sense of the spectral norm, the randomized matrix $\tilde{A}^{\rm rp}$ and the non-randomized matrix $A$ behave the same provided that $A$ is generated from an SBM, and thus the randomization pays no price theoretically ignoring the conditions that we imposed. Moreover, (\ref{A2}) could be removed if one consider regularized population adjacency matrix \citep{qin2013regularized} or using other trimming steps \citep{le2015sparse}. (\ref{A3}) could be relaxed if one use more advanced methods, say \citet{clarkson2017low,hu2021sparse,martinsson2020randomized}.

With the derivation of $\tilde{A}^{\rm rp}$ from $P$ at hand, we are ready to justify the clustering performance of Algorithm \ref{randsp}. We consider the following metric that measures the sum of the fractions of the misclustered nodes within each community,
\begin{align}
\label{4.2}
L_1({\tilde{\Theta}}, \Theta)=\underset{J\in E_K}{\rm min}\,\underset{1\leq k\leq K}{\sum}\;(2n_k)^{-1}\|({\tilde{\Theta}}J)_{G_{k}\ast}-\Theta_{G_{k}\ast}\|_0,
\tag{4.2}
\end{align}
where ${\tilde{\Theta}}$ is an estimate of $\Theta$, and $E_K$ is the set of all $K\times K$ permutation matrices. The following theorem provides an upper bound on $L_1$.

\begin{theorem}\label{rpromis}
Suppose that {(\ref{A1})}, {(\ref{A2})} and {(\ref{A3})} hold, and there exists an absolute constant $c_3>0$ such that,
\begin{equation}
\label{A4}\frac{K'n\alpha_n}{\gamma_n^2\delta_n^2 {\rm min}\, n_k}\leq c_3,\tag{A4}
\end{equation}
where $\delta_n:=\delta_{1n}$ when $K'=K$ and $\delta_n:=\delta_{2n}$ when $K'<K$ with
\begin{equation}
\label{4.3}\delta_{1n}:={\rm min}_{l\neq k}\;\sqrt{n_k^{-1}+n_l^{-1}},\tag{4.3}
\end{equation}
\begin{equation}
\label{4.4}\delta_{2n}:=\xi_n \;({\rm recall }\, (\ref{A1})),\tag{4.4}
\end{equation}
then with probability larger than $1-6r^{-r}-2n^{-s}$ for any $s>0$, there exist subsets $S_k\in G_k$ for $k=1,...,K$ such that
\begin{equation}
\label{4.5}L_1({\tilde{\Theta}}^{\rm rp}, \Theta)\leq\sum_{k=1}^K \frac{|S_k|}{n_k}\leq c_3^{-1}\frac{K'n\alpha_n}{\gamma_n^2\delta_n^2 {\rm min}\, n_k}\tag{4.5}
\end{equation}
Moreover, for $G=\cup _{k=1}^K(G_k\backslash S_k)$, there exists a $K\times K$ permutation matrix $J$ such that
\begin{equation}
\label{4.6}{\tilde{\Theta}}^{\rm rp}_{G\ast}J=\Theta_{G\ast}.\tag{4.6}
\end{equation}
\end{theorem}

The proof of Theorem \ref{rpromis} follows that in \citet{lei2015consistency}. (\ref{A1}) is required only when $K'<K$. (\ref{A2}) and (\ref{A3}) ensure the results of Theorem \ref{rproappro} hold. (\ref{A4}) is a technical condition which ensures the bound in (\ref{4.5}) vanishes and provides the range of parameters $(K,n,\alpha_n,\gamma_n, \delta_n)$ in which the result is appropriate. (\ref{A4}) is satisfied automatically if the bound in (\ref{4.5}) is $o(1)$. $S_k$ is actually the set of nodes in $G_k$ that are misclustered. $\delta_n$ measures the minimum distance of every pair of rows of true eigenvectors for nodes from different communities (recall Lemma \ref{lem:eigen}). As expected, larger $\delta_n$ and ${\rm min}\, n_k$ indicate better misclassification error rate. In particular, following Theorem \ref{rproappro}, the bound in (\ref{4.5}) is identical to that of the non-randomized spectral clustering when $K'=K$ \citep{lei2015consistency}.

The bound in (\ref{4.5}) is not explicit as $\gamma_n$ is related to $n$. To illustrate, we now consider a simple case. Suppose a SBM parameterized by $(\Theta,B)$ is generated with balanced communities size $n/K$, and
\begin{equation}
\label{4.7}P=\Theta B\Theta^{\intercal}=\Theta(\alpha_n\lambda I_K+\alpha_n(1-\lambda)1_K1_K^{\intercal})\Theta^{\intercal},\tag{4.7}
\end{equation}
where $1_K$ represents a $K$ dimensional vector of 1's and $\lambda$ is a constant. In the case, $\gamma_n=n\alpha_n\lambda/K$ \citep{rohe2011spectral}, and then the bound in (\ref{4.5}) reduces to $${\sum_{k=1}^K \frac{|S_k|}{n_k}= O({K^3}/{n\alpha_n}).}$$
Let us discuss some specific parameter settings now. For fixed $K$, $n\alpha_n$ needs to be of order $\omega(1)$, namely, $n\alpha_n\geq c$ for some constant $c$, to ensure a vanishing error bound. In such case, the bound ${O(1/{n\alpha_n})}$ is {optimal in the sense that there is no estimator which is weakly consistent when $n\alpha_n=O(1)$} (see \citet{ahn2018hypergraph} for example). On the other hand, when $\alpha_n=c_0{\rm log}n/n$, $K=o(({\rm log} n)^{1/3})$ is required to ensure a vanishing misclassification error rate. It should be noted that since the pure spectral clustering generally could not attain the optimal misclassification error rate under SBMs \citep{gao2017achieving} except some simple case ($K=2$, within-community and between-community probability being $\frac{b{\rm log} n}{n}$ and $\frac{{a\rm log} n}{n}$) considered in \citet{abbe2020entrywise}, our randomized version also has limitations in terms of misclassification rate. While the algorithms in \citet{gao2017achieving} that attain the statistical optimal error rate has {higher computational complexity} than the randomized spectral clustering we considered here do. The current error rate would be improved if one study more refined proof techniques of pure spectral clustering or develop variants of spectral clustering that has better error rates but without increasing the time complexity.

\begin{remark}
In the proof of Theorem \ref{rpromis}, we made an assumption that the $k$-means algorithm finds the optimal solution as in \citet{rohe2011spectral}. Alternatively, one can use more delicate $(1+\varepsilon)$-approximate $k$-means \citep{kumar2004simple,matouvsek2000approximate} to bridge the gap, where one can find a good approximate solution within a constant fraction of the optimal value.
\end{remark}

In the sequel, we discuss how we can utilize the estimated membership matrix ${\tilde{\Theta}}^{\rm rp}$ and $\tilde{A}^{\rm rp}$ to estimate the link probability matrix $B$. {Without loss of generality, we assume that the permutation matrix $J$ in (\ref{4.6}) is $I_{K\times K}$}. Noting that
\begin{equation*}
{B}_{ql}:=\frac{\sum_{1\leq i,j\leq n}P_{ij}\Theta_{iq}\Theta_{jl}}{\sum_{1\leq i,j\leq n}\Theta_{iq}\Theta_{jl}},\quad 1\leq q,l\leq K,
\end{equation*}
thus it is reasonable to estimate $B$ by the following ${\tilde{B}}^{\rm rp}=({\tilde{B}}^{\rm rp}_{ql})_{1 \leq q,l\leq K}$,
\begin{equation*}
{\tilde{B}}_{ql}^{\rm rp}:=\frac{\sum_{1\leq i,j\leq n}\tilde{A}^{\rm rp}_{ij}{\tilde{\Theta}}^{\rm rp}_{iq}{\tilde{\Theta}}^{\rm rp}_{jl}}{\sum_{1\leq i,j\leq n}{\tilde{\Theta}}^{\rm rp}_{iq}{\tilde{\Theta}}^{\rm rp}_{jl}},\quad 1\leq q,l\leq K.
\end{equation*}
The following theorem provides a theoretical bound for the estimator ${\hat{B}}^{\rm rp}$.
\begin{theorem}\label{rprolink}
Suppose that {(\ref{A2})}, {(\ref{A3})} and {(\ref{A4})} hold, then with probability larger than $1-6r^{-r}-2K^2n^{-s}$ for any $s>0$, there exists constant $c_4>0$ that,
\begin{equation}
\label{4.8}\|{\tilde{B}}^{\rm rp}-B\|_\infty \leq c_4\left( \frac{\sqrt{K'+r}\sqrt{n\alpha_n}}{{\rm min}\, n_k}+\frac{\sqrt{K'}\sigma_n}{{\rm min}\, n_k}\right)\left(1+(1-\Phi_n)^{-1}+\frac{2{\rm max}\, n_k}{{\rm min}\, n_k}(1-\Phi_n)^{-2}\right),
\tag{4.8}
\end{equation}
with $\Phi_n:=c_3^{-1}\frac{K'n\alpha_n}{\gamma_n^2\delta_n^2{\rm min}\, n_k}$ where $\delta_n=\delta_{1n}$ (see (\ref{4.3})) when $K'=K$ and $\delta_n=\delta_{2n}$ (see (\ref{4.4})) when $K'<K$.
\end{theorem}

Let us illustrate the bound in (\ref{4.8}) more explicitly. As a simple example, we consider the specific case in (\ref{4.7}). Suppose further that $\alpha_n=c_0{\rm log}n/n$, and then the bound in (\ref{4.8}) reduces to
\begin{equation}
\label{4.9}O\left(\frac{K^{3/2}{{\rm log} n}}{n}\Big(1+(1-\frac{K^3}{{\rm log} n})^{-1}\Big)\right).\tag{4.9}
\end{equation}
It turns out that $K=o(({\rm log} n)^{1/3})$ would lead to a vanishing bound. We note that \citet{tang2021asymptotically} established the asymptotic normality results for the estimation of $B$ in SBMs, where they assume $n\alpha_n=\omega(\sqrt{n})$. In particular,
when $B$ is of full-rank, $n\alpha_n=O(\sqrt{n})$ and the community size is balanced, the error rate for $B$ is $\frac{\sqrt{K}}{n^{3/4}}$ in \citet{tang2021asymptotically}, tighter than $\frac{K^{3/2}}{\sqrt{n}}$ in our work, which is partially because that we study the randomized spectral clustering while they considered the original spectral clustering. Note that the parameter range of $\alpha_n$ in this work is more friendly than theirs. In addition, it would be interesting to study the asymptotic properties of $\hat{B}$ under the setting of randomized spectral clustering.

\subsection{Random sampling}
Similar to the random projection-based spectral clustering, we will derive theoretical results on the random sampling method from three aspects--namely, the deviation of $\tilde{A}^{\rm rs}$ from $P$, the misclassification error rate, and the deviation of ${\tilde{B}}^{\rm rs}$ from $B$, where ${\tilde{B}}^{\rm rs}$ is an analog of ${\tilde{B}}^{\rm rp}$ with the estimators therein replaced by the counterparts under the random sampling scheme.

The SBM set-up is the same with that in Subsection \ref{subsec:rp}. We here recall some notation specific to the random sampling scheme. Let $\tilde{A}^{\rm rs}$ be the intermediate output in Algorithm \ref{randss} with the target rank being $K'$, i.e., the best rank-$K'$ approximation of the sparsified matrix $\tilde{A}^{\rm s}$ whose elements $(i,j)$'s are sampled from $A$ with probability $p_{ij}$'s. The next theorem provides an upper bound for the deviation of $\tilde{A^{\rm rs}}$ from $P$.
\begin{theorem}\label{rsamappro}
Suppose that (\ref{A2}) holds and assume
\begin{equation}
\label{A5}p_{ij}\geq p_{\rm min},\quad {\rm for\; all\;} 1\leq i<j\leq n.\tag{A5}
\end{equation}
Define $$I_1={\rm min}\Big\{\sqrt{\frac{n\alpha_n}{p_{\rm min}}},\sqrt{{\rm max}_i\sum_j\frac{1}{p_{ij}}}\Big\},$$
then there exist constants $c_5>0$ and $c_6>0$ such that
\begin{equation}
\small
\label{4.10}\|\tilde{A}^{\rm rs}-P\|_2\leq  c_{5}\, {\rm max}\Big\{I_1,\;\frac{\sqrt{{\rm log} n}}{p_{\rm min}},\, \sqrt{n\alpha_n^2(\frac{1}{p_{\rm min}}-1)},\;\sqrt{\alpha_n^2{\rm log}n{\rm max} \{1,\frac{1}{p_{\rm min}}-1\}^2}\Big\}:=\Psi_{n}, \tag{4.10}
\end{equation}
with probability larger than $1-c_6n^{-\nu}$, where constant $\nu>0$ depends on $c_5$.
\end{theorem}

It should be noted that the bound in (\ref{4.10}) is not obtained by simple combination of the $\|A-P\|_2$ and $\|\tilde{A}^{\rm rs}-A\|_2$. Instead, we make use of the low-rank nature of $P$, which would lead to the improved result. Theorem \ref{rsamappro} indicates that besides the minimum sampling probability ${ p_{\rm min}}$, the term $\sqrt{{\rm max}_i\sum_j\frac{1}{p_{ij}}}$ which measures the worst overall sampling probability of all edges of certain node may affect the bound. In particular, when $\alpha_n$ is fixed, and $p_{i1}, p_{i2},...,p_{in}$'s are highly heterogeneous for each fixed $i$, $I_1$ reduces to $\sqrt{{\rm max}_i\sum_j\frac{1}{p_{ij}}}$. It should be noted that when $p_{ij}$'s are uniform and fixed, the RHS of (\ref{4.10}) reduces to $\sqrt{n\alpha_n}$, being the same with the best concentration bound of the full adjacency matrix $A$ around its population $P$ \citep{lei2015consistency,gao2015rate}. In this sense, the sampled matrix $\tilde{A}^{\rm rs}$ can be regarded as a network sampled from the same SBM generating $A$, although the elements of $\tilde{A}^{\rm rs}$ are not binary.

The following theorem justifies the clustering performance of the randomized spectral clustering via the random sampling (Algorithm \ref{randss}).
\begin{theorem}\label{rsammmis}
Suppose that (\ref{A1}), (\ref{A2}) and {(\ref{A5})} hold, and assume there exists an absolute constant $c_7>0$ such that,
\begin{equation}
\label{A6}\frac{K'\Psi_n^2}{\gamma_n^2\delta_n^2 {\rm min}\, n_k}\leq c_7,\tag{A6}
\end{equation}
where $\delta_n:=\delta_{1n}$ (see (\ref{4.3})) when $K'=K$ and $\delta_n:=\delta_{2n}$ (see (\ref{4.4})) when $K'<K$.
Then with probability larger than $1-c_6n^{-\nu}$ for some $\nu>0$, there exist subsets $S_k\in G_k$ for $k=1,...,K$ such that
\begin{equation}
\label{4.11}
\small L_1({\tilde{\Theta}}^{\rm rs}, \Theta)\leq\sum_{k=1}^K \frac{|S_k|}{n_k}\leq c_7^{-1}\frac{K'\Psi_n^2}{\gamma_n^2\delta_n^2 {\rm min}\, n_k}\tag{4.11}
\end{equation}
where recall that $\Psi_n$ is defined in (\ref{4.10}).
Moreover, for $G=\cup _{k=1}^K(G_k\backslash S_k)$, there exists a $K\times K$ permutation matrix $J$ such that
\begin{equation}
\label{4.12}{\tilde{\Theta}}^{\rm rs}_{G\ast}J=\Theta_{G\ast}.\tag{4.12}
\end{equation}
\end{theorem}

The proof is similar to that of Theorem \ref{rpromis}, hence we omit it. Under the assumption of SBM in (\ref{4.7}) and let $p$ be fixed; then similar to the random projection scheme, the bound in (\ref{4.11}) reduces to ${O({K^3}/{n\alpha_n})},$ which is $o(1)$ under the parameter set-up that $\alpha_n=c_0{\rm log}n/n$ and $K=o(({\rm log} n)^{1/3})$. Also, the current bound could be improved potentially; see our discussion after Theorem \ref{rpromis}.

Next, we turn to the estimation of the link probability matrix $B$. Similar to the random projection setting, we define the following plug-in estimator ${\tilde{B}}^{\rm rs}=({\tilde{B}}^{\rm rs}_{ql})_{1 \leq q,l\leq K}$ for $B$,
\begin{equation}
 {\tilde{B}}_{ql}^{\rm rs}:=\frac{\sum_{1\leq i,j\leq n}\tilde{A}^{\rm rs}_{ij}{\tilde{\Theta}}^{\rm rs}_{iq}{\tilde{\Theta}}^{\rm rs}_{jl}}{\sum_{1\leq i,j\leq n}{\tilde{\Theta}}^{\rm rs}_{iq}{\tilde{\Theta}}^{\rm rs}_{jl}},\quad 1\leq q,l\leq K.\nonumber
\end{equation}
The following theorem provides an upper bound for the deviation of ${\tilde{B}}^{\rm rs}=({\tilde{B}}^{\rm rs}_{ql})_{1 \leq q,l\leq K}$ from $B$.

\begin{theorem}\label{rsamlink}
Suppose that {(\ref{A2})}, {(\ref{A5})} and {(\ref{A6})} hold, then with probability larger than $1-c_6Kn^{-\nu}$ for some $\nu>0$, there exists $c_8>0$ that,
\begin{equation}
\label{4.13}\|{\tilde{B}}^{\rm rs}-B\|_\infty \leq c_8\Big( \frac{\sqrt{K'+r}\sqrt{n\alpha_n}}{{\rm min}\, n_k}+\frac{\sqrt{K'}\sigma_n}{{\rm min}\, n_k}\Big)\left(1+(1-\Psi_n)^{-1}+\frac{2{\rm max}\, n_k}{{\rm min}\, n_k}(1-\Psi_n)^{-2}\right),
\tag{4.13}
\end{equation}
where recall that $\Psi_n$ is defined in (\ref{4.10}).
\end{theorem}

We omit the proof since it is similar to that of Theorem \ref{rprolink}. We can discuss the bound (\ref{4.13}) in a similar way to those in the random projection scheme. For example,
under the special case of SBM in (\ref{4.7}), let $\alpha_n=c_0{\rm log}n/n$ and $p$ be fixed, then the bound (\ref{4.13}) reduces to
the one in (\ref{4.9}). Thus $K=o(({\rm log} n)^{1/3})$ suffices to make sure that the RHS of (\ref{4.13}) vanishes when $n$ goes to infinity.

\section{Extensions}
\label{sec:dcsbm}
Standard SBMs often fail to capture the property of networks with strong degree heterogeneity. As a remedy, in this section we extend our results to degree-corrected stochastic block models (DC-SBMs) coupled with the randomized spherical spectral clustering.
\subsection{Degree-corrected stochastic block models}
Similar to the SBMs, the DC-SBMs \citep{karrer2011stochastic}  are parameterized by the membership matrix $\Theta\in\mathbb M_{n,K}$, and the link probability matrix $B\in[0,1]^{K\times K}$ where $B$ is \emph{not} necessary of full rank and denote ${\rm rank}(B)=K'(K'\leq K)$. To account for the degree heterogeneity, the DC-SBMs additionally introduce the node propensity parameter $\vartheta\in \mathbb R^n_{+}$. With these set-up, the population adjacency matrix is defined as $P:={\rm diag}(\vartheta)\Theta B \Theta^\intercal{\rm diag}(\vartheta)$. To make the parameters identifiable, we follow \citet{lei2015consistency} to assume that $\max_{i\in G_k}\vartheta_i=1$. To facilitate further analysis, let $\phi_k$ be an $n\times1$ vector that is consistent with $\vartheta$ on $G_k$ and zero otherwise. Let $\Omega={\rm diag} (\|\phi_1\|_2,...,\|\phi_{K}\|_2)$, and let $\bar{B}=\Omega {B}\Omega$. The following lemma reveals the eigen-structure of the population matrix $P$.

\begin{lemma}
\label{lem:eigen2}
For a DC-SBM with $K$ communities parameterized by $\Theta\in \mathbb M_{n,K}$, $B\in [0,1]^{K\times K}$ and $\vartheta\in \mathbb R^n_{+}$, we suppose that ${\rm rank}(B)=K'(K'\leq K)$ and the eigenvalue decomposition of $P={\rm diag}(\vartheta)\Theta B \Theta^\intercal{\rm diag}(\vartheta)$ is $U_{n\times K'}\Sigma_{K'\times K'} U^{\intercal}_{K'\times n}$. Denote the eigenvalue decomposition of $\bar{B}$ by $H_{K\times K'}D_{K'\times K'}H_{K'\times K}^\intercal$. For any two vectors $a$ and $b$, ${\rm cos}(a,b)$ is defined to be $a^\intercal b/\|a\|_2\|b\|_2$. Then the following arguments hold.

(a) If $B$ is of full rank, i.e., $K'=K$, then for any $\Theta_{i\ast}=\Theta_{j\ast}$, ${\rm cos}({U}_{i\ast},{U}_{j\ast})=1$, and for $\Theta_{i\ast}\neq\Theta_{j\ast}$, ${\rm cos}({U}_{i\ast},{U}_{j\ast})=0$.

(b) If $B$ is rank deficient, i.e., $K'<K$, then for any $\Theta_{i\ast}=\Theta_{j\ast}$, ${\rm cos}({U}_{i\ast},{U}_{j\ast})=1$, and for any $\Theta_{i\ast}\neq\Theta_{j\ast}$, if $H$'s rows are \emph{not} pairwise proportional such that there exists a deterministic sequence $\{\xi'_n\}_{n\geq 1}<1$ satisfying
\begin{equation}
\label{A7}
\max_{k,l} {\rm cos}({H}_{k\ast},{H}_{l\ast})\leq \xi'_n,
\tag{A7}
\end{equation}
then ${\rm cos}(U_{i\ast},U_{j\ast})={\rm cos}({H}_{g_i\ast},{H}_{g_j\ast})\leq \xi'_n<1$.
\end{lemma}

The following lemma gives an explicit condition on $B$ which suffices for (\ref{A7}).
\begin{lemma}
\label{lem:eigen2condition}
For a DC-SBM with $K$ communities parameterized by $\Theta\in \mathbb M_{n,K}$, $B\in [0,1]^{K\times K}$ and $\vartheta\in \mathbb R^n_{+}$, where suppose that ${\rm rank}(B)=K'\, (K'< K)$ and the eigenvalue decomposition of $P={\rm diag}(\vartheta)\Theta B \Theta^\intercal{\rm diag}(\vartheta)$ is $U_{n\times K'}\Sigma_{K'\times K'} U^{\intercal}_{K'\times n}$. Recall that $\Omega={\rm diag} (\|\phi_1\|_2,...,\|\phi_{K}\|_2)$, and $\bar{B}=\Omega {B}\Omega$. If there exists deterministic positive sequences $\{\eta'_n\}_{n\geq 1}$, $\{\underline{\iota}_n\}_{n\geq 1}$, $\{\overline{\iota}_n\}_{n\geq 1}$ and $\{{\beta}_n\}_{n\geq 1}$ such that
\begin{equation}
\min_{1\leq k<l\leq K} \bar{B}_{kk}\bar{B}_{ll}-\bar{B}_{kl}^2\geq \eta'_n>0  ,\nonumber
\end{equation}
and for any $1 \leq i\leq K'$, $0<\underline{\iota}_n<\Sigma_{ii}<\overline{\iota}_n$, and $0<\min_{1\leq k\leq K} \bar{B}_{kk}\leq \max_{1\leq k\leq K} \bar{B}_{kk}\leq \beta_n$, then (\ref{A7}) holds with $$\xi'_n=\sqrt{1-\frac{\eta'_n}{\overline{\iota}_n \beta_n^2/\underline{\iota}_n}}.$$
\end{lemma}

Compared with Lemma \ref{lem:eigen}, we see that for the DC-SBMs, not the distances but the angles between the rows of true eigenvector $U$ reveal whether the corresponding nodes are in the same community.

\subsection{Randomized spherical spectral clustering}
In light of Lemma \ref{lem:eigen2}, to make the spectral clustering valid on DC-SBMs, we need to normalize the rows of eigenvectors before performing the $k$-means. In this way, the angle-based results in Lemma \ref{lem:eigen2} can be transformed to the distance-based counterpart, and thus making the distance-based $k$-means valid. The resulting algorithm is called spherical spectral clustering; see Algorithm \ref{spectral2}.

\begin{algorithm}[htb]

\renewcommand{\algorithmicrequire}{\textbf{Input:}}

\renewcommand\algorithmicensure {\textbf{Output:} }

\caption{Spherical spectral clustering for $K$ clusters}

\label{spectral2}

\begin{algorithmic}[1]

\REQUIRE ~\\

Cluster number $K$, target rank $K'$, adjacency matrix $A\in \mathbb{R}^{n\times n}$;\\

\ENSURE ~\\
Estimated membership matrix $\hat{{\Theta}}\in\mathbb M_{n,K}$ and centriods $\hat{{X}}\in \mathbb{R}^{K\times K'}$ ;\\
Estimated eigenvectors ${\tilde{U}}=\tilde{{\Theta}}\tilde{{X}}$;\\
~\\
\STATE Find the $K'$ leading eigenvectors $\hat{U}$ of $A$ corresponding to the $K'$ largest eigenvalues of $A$. \\
\STATE Normalize each row of $\hat{U}$ and denote the resulting matrix by $\hat{U}'$, where the rows with Euclidean norm 0's are remained the same.
\STATE Treat each row of $\hat{U}'$ as a point in $\mathbb{R}^{K'}$ and run the Lloyd's algorithm on these points with $K$ clusters. Let
$(\tilde{\Theta},\tilde{X})$ be the solution.
 \\
\end{algorithmic}
\end{algorithm}
The randomized spherical spectral clustering is readily available when we replace the input adjacency matrix $A$ in Algorithm \ref{spectral2} by the randomized counterpart $\tilde{A}$ ($\tilde{A}^{\rm rp}$ or $\tilde{A}^{\rm rs}$). With slight abuse of notation, the output is denoted by
${\tilde{\Theta}}$ (${\tilde{\Theta}}^{\rm rp}$ or ${\tilde{\Theta}}^{\rm rs}$).

\begin{remark}
The spherical spectral clustering algorithms have been studied by several authors; see \citet{lei2015consistency,qin2013regularized}, among others. In particular, \citet{lei2015consistency} remove the zero rows of $\hat{U}$ and use $k$-median instead of $k$-means for technical reasons. Differently, we let the zero rows of $\hat{U}$ be untreated and still use the $k$-means on the normalized vectors. Note that besides $k$-means based algorithms, one could use other clustering algorithms, say subspace clustering \citep{vidal2005generalized,liu2012robust,terada2014strong}, directly on the un-normalized eigenvectors.
\end{remark}

\subsection{Misclassification analysis}
Note that the approximation error bounds $\|\tilde{A}-P\|_2$ ($\tilde{A}$ can be $\tilde{A}^{\rm rs}$ or $\tilde{A}^{\rm rp}$; see (\ref{4.1}) and (\ref{4.10})) only make use of the low-rank nature of $P$, hence they remain the same under the DC-SBMs. The following theorem provides the misclassification error rate of randomized spherical spectral clustering on DC-SBMs, where output ${\tilde{\Theta}}$ represents  ${\tilde{\Theta}}^{\rm rp}$ and ${\tilde{\Theta}}^{\rm rs}$ respectively when $\tilde{A}=\tilde{A}^{\rm rp}$ and $\tilde{A}=\tilde{A}^{\rm rs}$.

\begin{theorem}\label{rpromis2}
For a DC-SBM with $K$ communities parameterized by $\Theta\in \mathbb M_{n,K}$, $B\in [0,1]^{K\times K}$ with ${\rm rank}(B)=K' \leq K$ and $\vartheta\in \mathbb R^n_{+}$. Let $\tilde{\vartheta}$ be an $n\times 1$ vector such that the $i$th element is $\vartheta_i/\|\phi_{g_i}\|_2$, where recall that $\phi_k$ is an $n\times 1$ vector that consistent with $\vartheta$ on $G_k$ and zero otherwise. The following results hold for the output ${\tilde{\Theta}}$ of the randomized spherical spectral clustering.

(a) For $K'=K$, suppose that there exists an absolute constant $c_9>0$ such that,
\begin{equation}
\label{A8}\frac{1}{\min_i \tilde{\vartheta}_i^2}\cdot\frac{K'\|\tilde{A}-P\|_2^2}{\gamma_n^2 {\rm min}\, n_k}\leq c_9,\tag{A8}
\end{equation}
then there exist subsets $S_k\in G_k$ for $k=1,...,K$ such that
\begin{equation}
\label{5.1}L_1({\tilde{\Theta}}, \Theta)\leq\sum_{k=1}^K \frac{|S_k|}{n_k}\leq c_9^{-1}\frac{1}{\min_i \tilde{\vartheta}_i^2}\cdot\frac{K'\|\tilde{A}-P\|_2^2}{\gamma_n^2 {\rm min}\, n_k}.
\tag{5.1}
\end{equation}
Moreover, for $G=\cup _{k=1}^K(G_k\backslash S_k)$, there exists a $K\times K$ permutation matrix $J$ such that
\begin{equation}
{\tilde{\Theta}}_{G\ast}J=\Theta_{G\ast}.\nonumber
\end{equation}

(b) For $K'<K$, suppose that (\ref{A7}) holds, $\max_i\Sigma_{ii}<\overline{\iota}_n$, and $\min_{1\leq k\leq K} B_{kk}>0$. Also suppose there exists an absolute constant $c_{10}>0$ such that,
\begin{equation}
\label{A9}\frac{\overline{\iota}_n}{\min_i \tilde{\vartheta}_i^2 \min \bar{B}_{kk}}\cdot\frac{K'\|\tilde{A}-P\|_2^2}{(1-\xi'_n)\gamma_n^2 {\rm min}\, n_k}\leq c_{10},\tag{A9}
\end{equation}
then there exist subsets $S_k\in G_k$ for $k=1,...,K$ such that
\begin{equation}
\label{5.2}L_1({\tilde{\Theta}}, \Theta)\leq\sum_{k=1}^K \frac{|S_k|}{n_k}\leq c_{10}^{-1}\frac{\overline{\iota}_n}{\min_i \tilde{\vartheta}_i^2 \min \bar{B}_{kk}}\cdot\frac{K'\|\tilde{A}-P\|_2^2}{(1-\xi'_n)\gamma_n^2 {\rm min}\, n_k}.\tag{5.2}
\end{equation}
Moreover, for $G=\cup _{k=1}^K(G_k\backslash S_k)$, there exists a $K\times K$ permutation matrix $I$ such that
\begin{equation}
{\tilde{\Theta}}_{G\ast}I=\Theta_{G\ast}.\nonumber
\end{equation}
\end{theorem}

(\ref{A8}) and (\ref{A9}) are technical conditions that ensure the bound (\ref{5.1}) and (\ref{5.2}) valid. (\ref{5.1}) and (\ref{5.2}) can be made explicitly by incorporating the bound of $\|\tilde{A}^{\rm rp}-P\|_2$ or $\|\tilde{A}^{\rm rs}-P\|_2$ coupled with the corresponding assumptions; see Theorem \ref{rproappro} and \ref{rsamappro}. Note that $\min_i \tilde{\vartheta}_i$ reflects the degree heterogeneity in some sense. Larger $\min_i \tilde{\vartheta}_i$ indicates less degree heterogeneity and thus better clustering performance.

\section{Related work and discussion}
\label{sec:related}
{In this section, we review and discuss the literature that is closely related to the current work. We classify them into three groups: spectral clustering, randomization techniques, and iterative methods for fast eigen-decomposition.}

The community detection is one of the fundamental problems in network analysis. The SBMs and their variants have been useful tools for modeling networks with communities and thus being widely studied \citep{abbe2017community}. In particular, a multitude of researches focus on spectral clustering and its variants, see \citet{arroyo2021overlapping,chin2015stochastic,fishkind2013consistent,joseph2016impact,lei2015consistency,li2020community,lyzinski2014perfect,
paul2020spectral,qin2013regularized,rohe2011spectral,tang2021asymptotically,su2019strong,yang2020simultaneous,yun2014accurate,yun2016optimal}, and references therein, among which weak (strong) consistency, namely the fraction (number) of misclustered nodes decreases to zero as $n$ grows, are well-established. Compared with most of these works, the current work has novelty in terms of both algorithms and also theoretics. In respect of algorithms, the randomized spectral clustering algorithms can deal with networks with up to millions number of nodes, {showing the advantage over original spectral clustering with full eigenvalue decomposition.} In respect of theoretics, the approximation error bound $\|\tilde{A}-P\|_2$ is optimal under mild conditions though we use randomized adjacency matrix $\tilde{A}$. As a by-product, we generalize the common assumption ${\rm rank}(B)=K$ in SBMs and DC-SBMs to ${\rm rank}(B)\leq K$, which is of independent interest and rarely mentioned in the works of literature except \citet{tang2021asymptotically,fishkind2013consistent}, and a few others.

There are also various prior works on spectral clustering using randomized methods, see \citet{liao2020sparse,sakai2009fast,sinha2018k,tremblay2016compressive,tremblay2020approximating,wang2019scalable,yan2009fast}, among others. For example, \citet{sakai2009fast} developed fast spectral clustering algorithms by using random projection and random sampling techniques in order to reduce the data dimensionality and cardinality. \citet{yan2009fast} provided a general framework for fast spectral clustering where a distortion-minimizing local transformation is first applied to the data to reduce the dimensionality. \citet{tremblay2016compressive} proposed the compressive spectral clustering using the randomized techniques in graph signal processing. {Compared with this line of works}, the merits of this work lie in that we study the effect of randomization from the statistical point of view\textendash under the framework of SBMs and DC-SBMs. The current methods can obtain optimal error for $\|\tilde{A}-P\|_2$ under mild conditions, indicating that the optimization error induced by random projection or random sampling are dominated by the statistical error induced by the randomness of networks from SBMs and DC-SBMs. It should be noted that the structure of SBMs and DC-SBMs facilitates bounding the approximation error in the random sampling regime. The resulting bound is tighter than those obtained by simply combining the optimization error bound $\|\tilde{A}^{\rm rs}-A\|_2$ in \citet{achlioptas2007fast} and the statistical error bound $\|{A}-P\|_2$ in \citet{lei2015consistency}. Note that \citet{li2020network} also studied the deviation of $\tilde{A}^{\rm rs}$ from $P$ but in the context of network cross-validation. It turns out that $K\leq n/{\rm log}n$ is additionally required therein to ensure that the concentration bound of $\tilde{A}^{\rm rs}$ meets that of the full adjacency matrix $A$, provided that $p$ is fixed.

Iterative methods are widely used for partial eigen-decomposition and there are fruitful works in this line; see \citet{allen2016lazysvd,baglama2005augmented,calvetti1994implicitly,lehoucq1995analysis}, among others. {We illustrate the merits of this work as follows.} Actually, in the random sampling scheme, we use the iterative methods of \citep{calvetti1994implicitly,rspectra} as our baseline method and study how sampling could be used to further accelerate the partial eigen-decomposition. While for the random projection scheme, it has the following advantages \citep{halko2011finding}. First, the random projection-based methods are scalable because the matrix-vector operations can be done via multi-threading and distributed computing, which has been exploited in the R package (\textsf{Rclust}) of this work. Second, the random projection-based methods have low communication costs as it only requires few passes over the input matrix. Further, the communication costs could be reduced by considering single-pass version \citep{tropp2017randomized}. At last, our experiments show that the randomized methods are faster than iterative methods while achieving satisfactory performance provided that the network's scale is super-large, say millions of nodes.

\section{Numerical results}
\label{sec:simu}
In this section, we empirically compare the finite sample performance of the randomized spectral clustering, namely, the random projection and the random sampling, with the original spectral clustering, where we use uniform sampling in the random sampling scheme for computational convenience. We will start with a simple SBM model to test the effect of $n,K,\alpha$ on the approximation error, misclassification error, estimation error for $B$, respectively. Then we extend our model setting to more complex models. At last, we test the effect of hyper parameters, including the power parameter $q$ and the oversampling parameter $r$ in the random projection scheme, and the sampling parameter $p$ in the random sampling scheme.

\subsection{Theoretical bounds evaluation}
\label{sub:sim1}
{To be consistent with Section \ref{sec:theo}, we use the following three metrics to evaluate the theoretical performance of each method. The first one is the spectral derivation of the ``approximated'' adjacency matrix $\hat{A}$ from the population adjacency matrix $P$, namely, $\|\hat{A}-P\|_2$, where $\hat{A}$ can be $\tilde{A}^{\rm rs}$, $\tilde{A}^{\rm rp}$ or ${A}$. The second metric is the sum of the fractions of misclustered nodes within each true cluster, namely,
\begin{align}
\underset{J\in E_K}{\rm min}\,\underset{1\leq k\leq K}{\sum}\;(2n_k)^{-1}\|(\hat{{\Theta}}J)_{G_{k\ast}}-\Theta_{G_{k\ast}}\|_0,\nonumber
\end{align}
where $\hat{{\Theta}}$ can be ${\tilde{\Theta}}^{\rm rp}$, ${\tilde{\Theta}}^{\rm rs}$ or $\tilde{{\Theta}}$.
The third metric is the derivation of the estimated link probability matrix $\hat{{B}}$ from the true link probability matrix $B$, namely, $\|{\hat{B}}-B\|_\infty$, where ${\hat{B}}$ can be $\tilde{{B}}^{\rm rp}$, $\tilde{{B}}^{\rm rs}$, or the counterpart corresponding to the original spectral clustering.  Throughout this subsection, the SBMs parameterized by $(\Theta,B)$ were homogeneously generated in the following way,
\begin{equation}
P=\Theta B\Theta^{\intercal}=\Theta(\alpha_n\lambda I_K+\alpha_n(1-\lambda)1_K1_K^{\intercal})\Theta^{\intercal},\nonumber
\end{equation}
where $1_K$ represents a $K$ dimensional vector of 1's and $\lambda$ is a constant, and the community sizes are balanced to be $n/K$. To see how the above mentioned metrics change with $n$, $K$, $\alpha_n$, we conduct the following four experiments.}

\begin{figure*}[!htbp]{}
\centering
\subfigure[]{\includegraphics[height=4.8cm,width=5cm,angle=0]{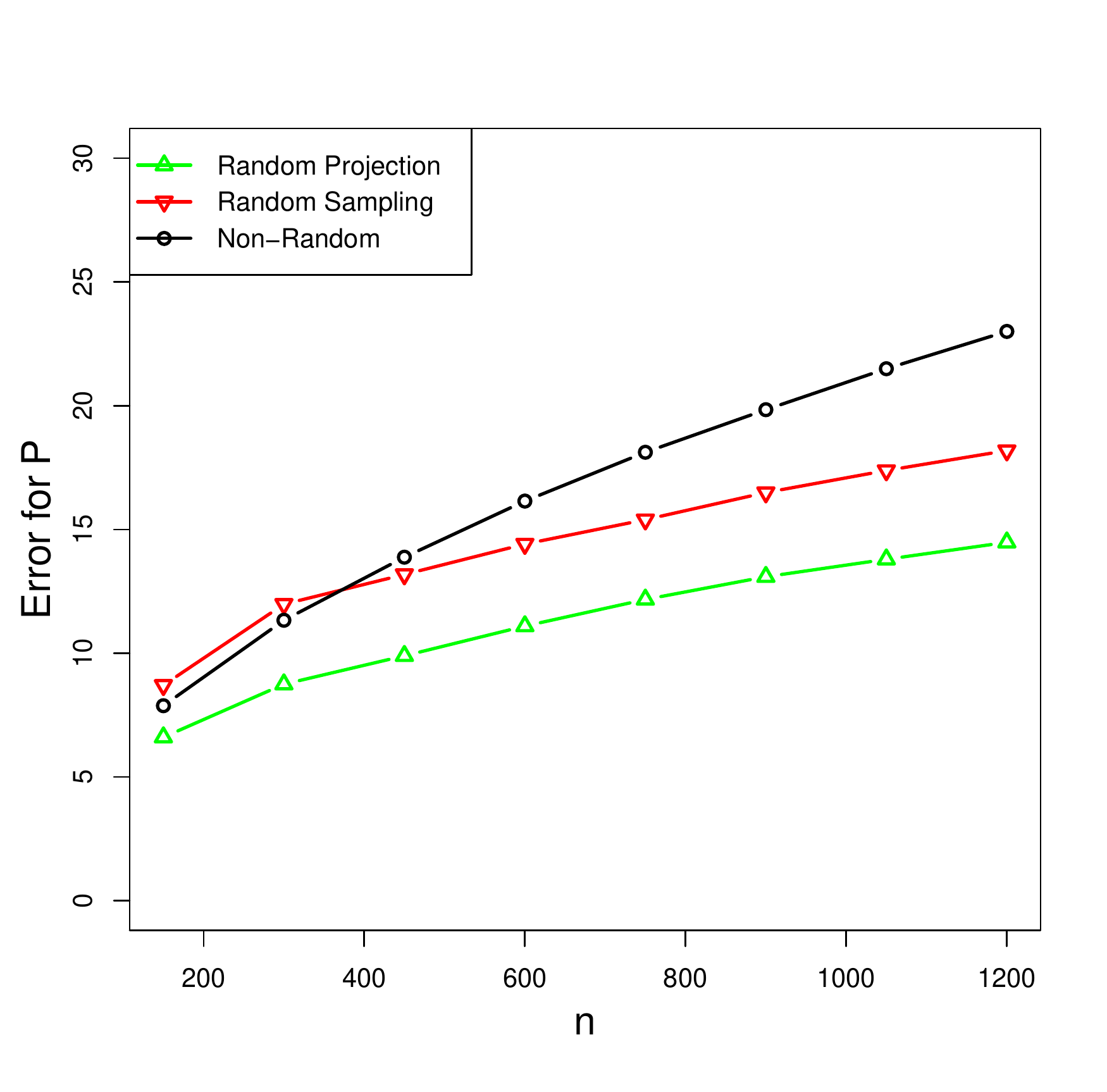}}
\subfigure[]{\includegraphics[height=4.8cm,width=5cm,angle=0]{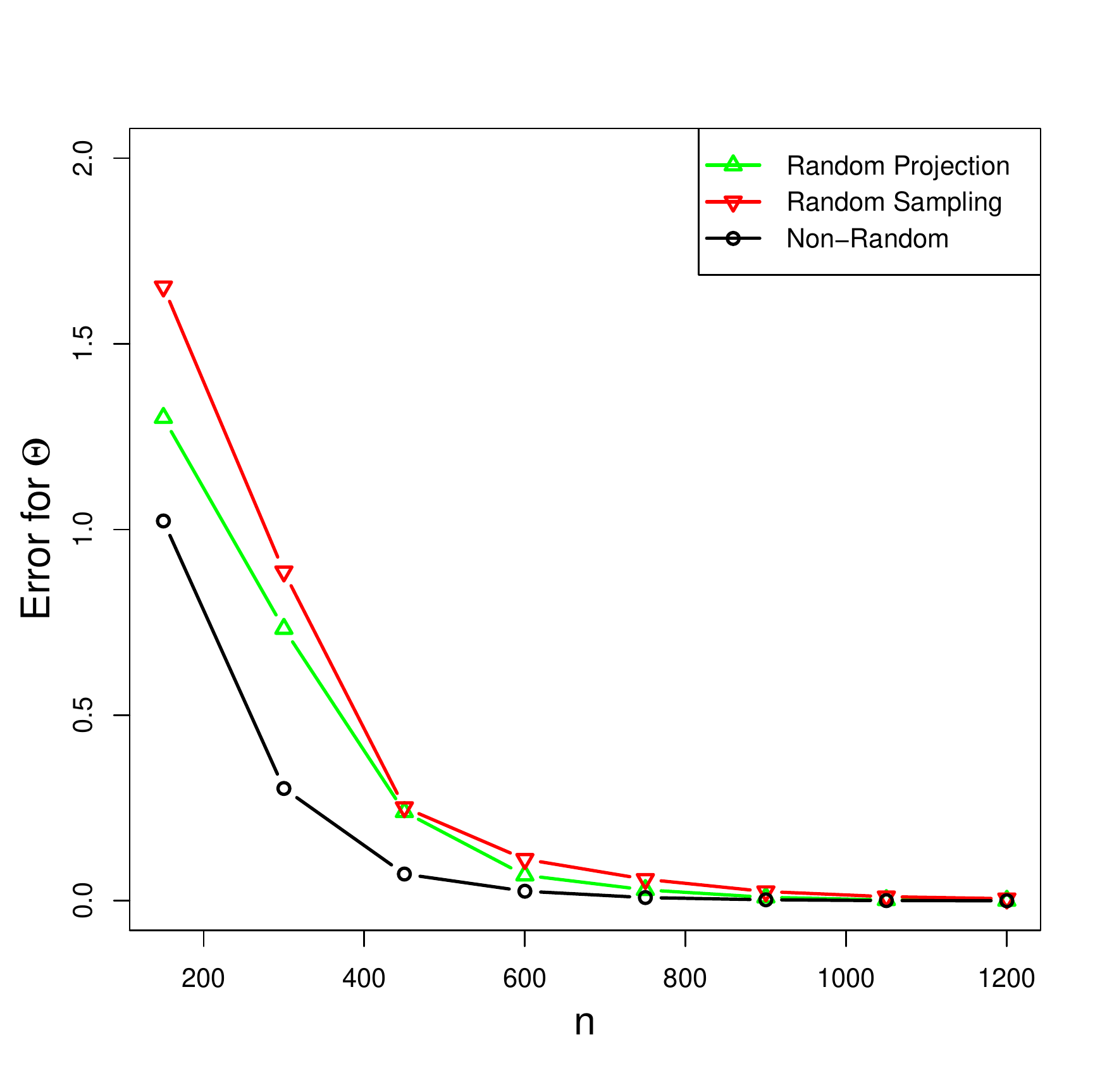}}
\subfigure[]{\includegraphics[height=4.8cm,width=5cm,angle=0]{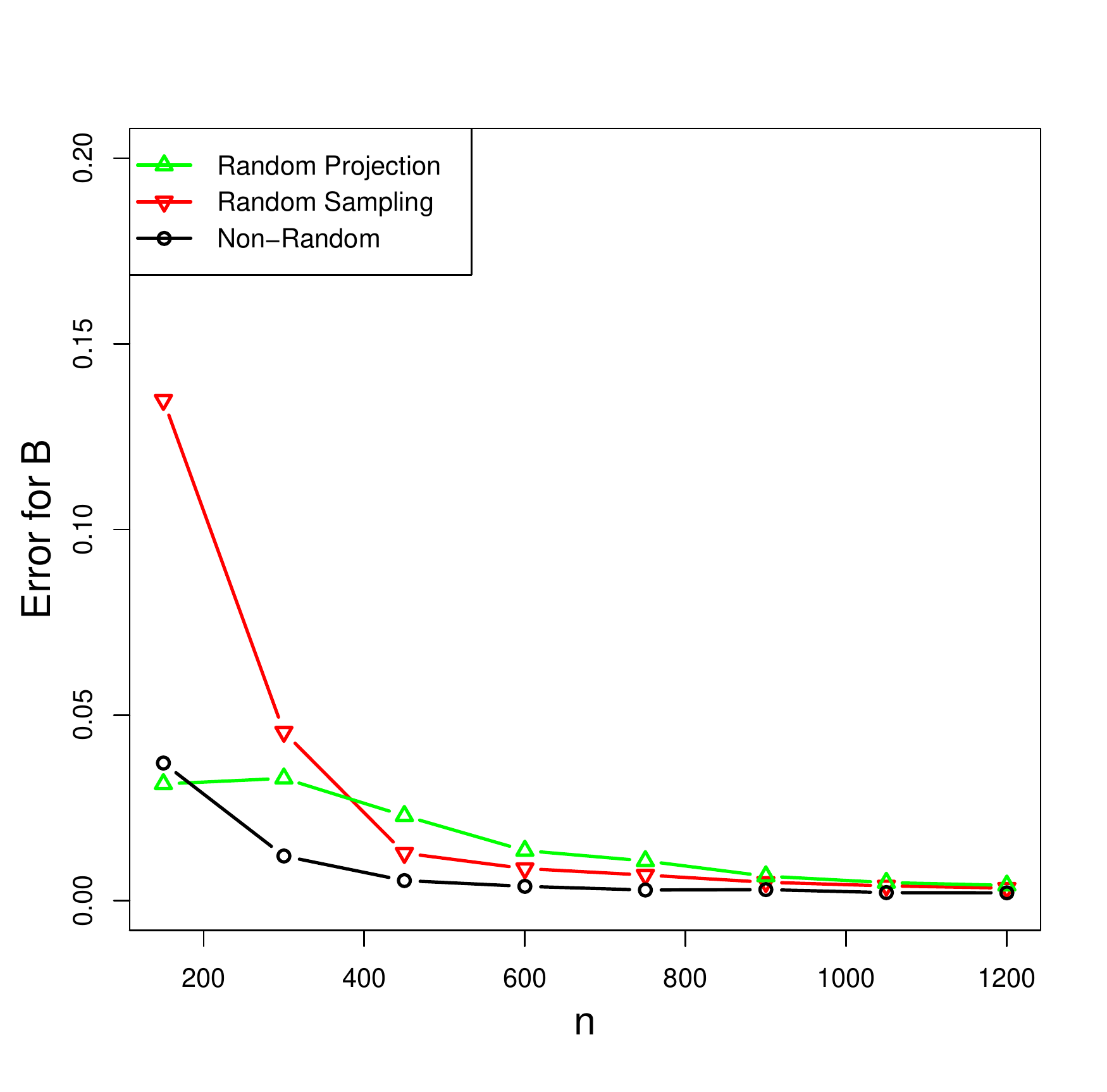}}
\caption{The average effect of $n$ on the three metrics over 20 replications. (a), (b), (c) correspond to the approximation error for $P$, the misclassification error for $\Theta$, and the estimation error for $B$, respectively. The other parameters $K=3, \alpha_n=0.2, \alpha_n(1-\lambda)=0.1$, $r=10$, $q=2$, $p=0.7$, and $\Omega$ had i.i.d. standard Gaussian entries, respectively. }\label{effectofn}
\end{figure*}

\paragraph{Experiment 1.} In this experiment, we aim to evaluate the effect of $n$ on the three metrics. To that end, we let
$n$ vary while keeping the other parameters fixed at $K=3, \alpha_n=0.2, \alpha_n(1-\lambda)=0.1, q=2,r=10,p=0.7$. The random test matrix in the random projection scheme was generated with i.i.d. standard Gaussian entries, respectively.  Figure \ref{effectofn} shows the average results of 20 replications, where ``non-random'' refers to the original spectral clustering. Recall that the error bound for $P$ increases with order $O(\sqrt{n})$, the error bound for $\Theta$ decreases with order $O(1/n)$, and the error bound for $B$ vanishes as $n$ goes to infinity. As expected, from Figure \ref{effectofn} we can see that the randomized methods perform worse than the original spectral clustering when $n$ is small, say $n<600$, but they become almost identical when $n$ becomes large, say $n>800$, which is actually the focus of this paper (see Figure \ref{effectofn}(b) and (c)). As for the approximation error, we see that the random projection and the random sampling perform better than the original spectral clustering (see Figure \ref{effectofn}(a)), which is partially because of the constants' effects.

\begin{figure*}[!htbp]{}
\centering
\subfigure[]{\includegraphics[height=4.8cm,width=5cm,angle=0]{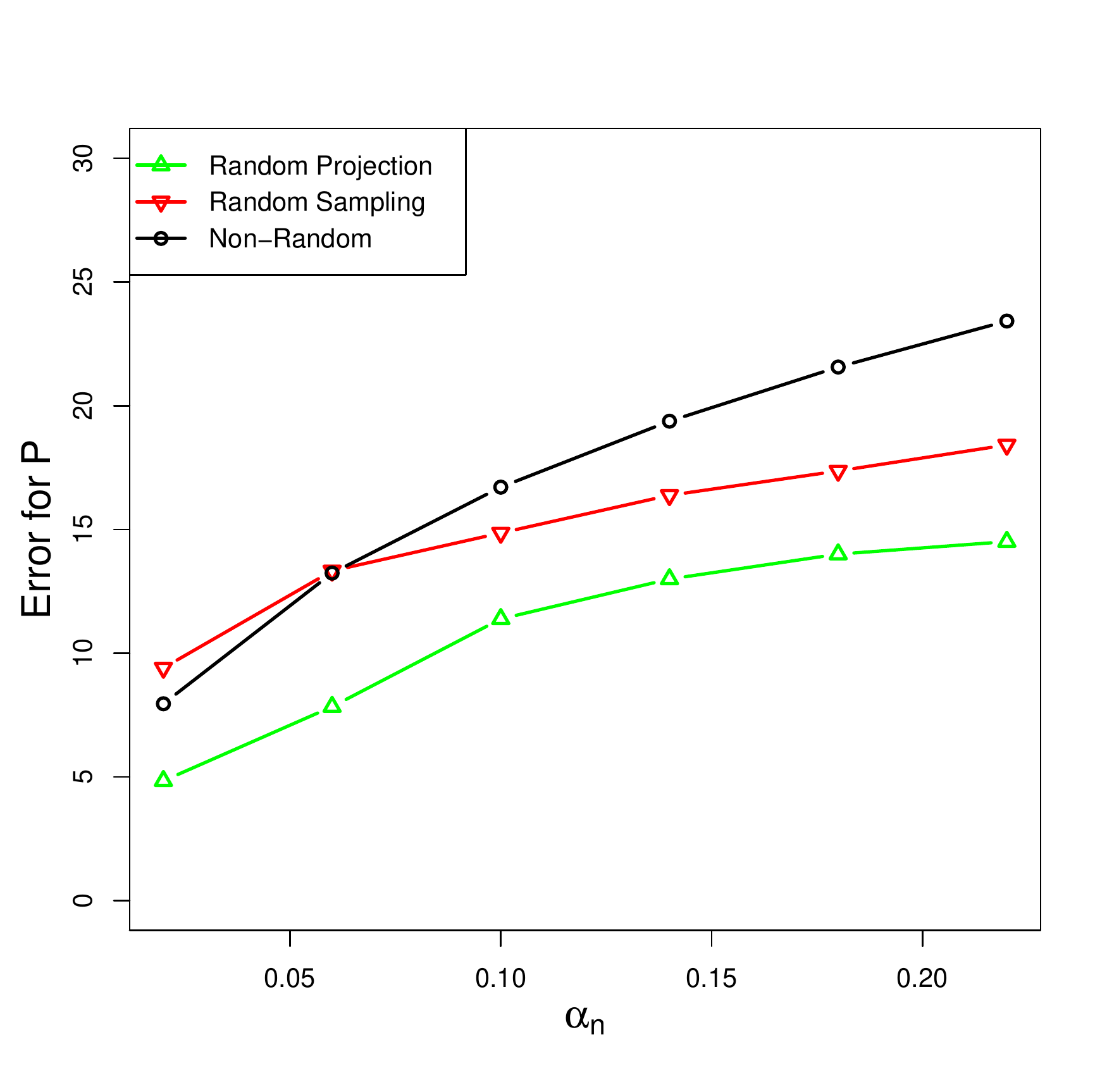}}
\subfigure[]{\includegraphics[height=4.8cm,width=5cm,angle=0]{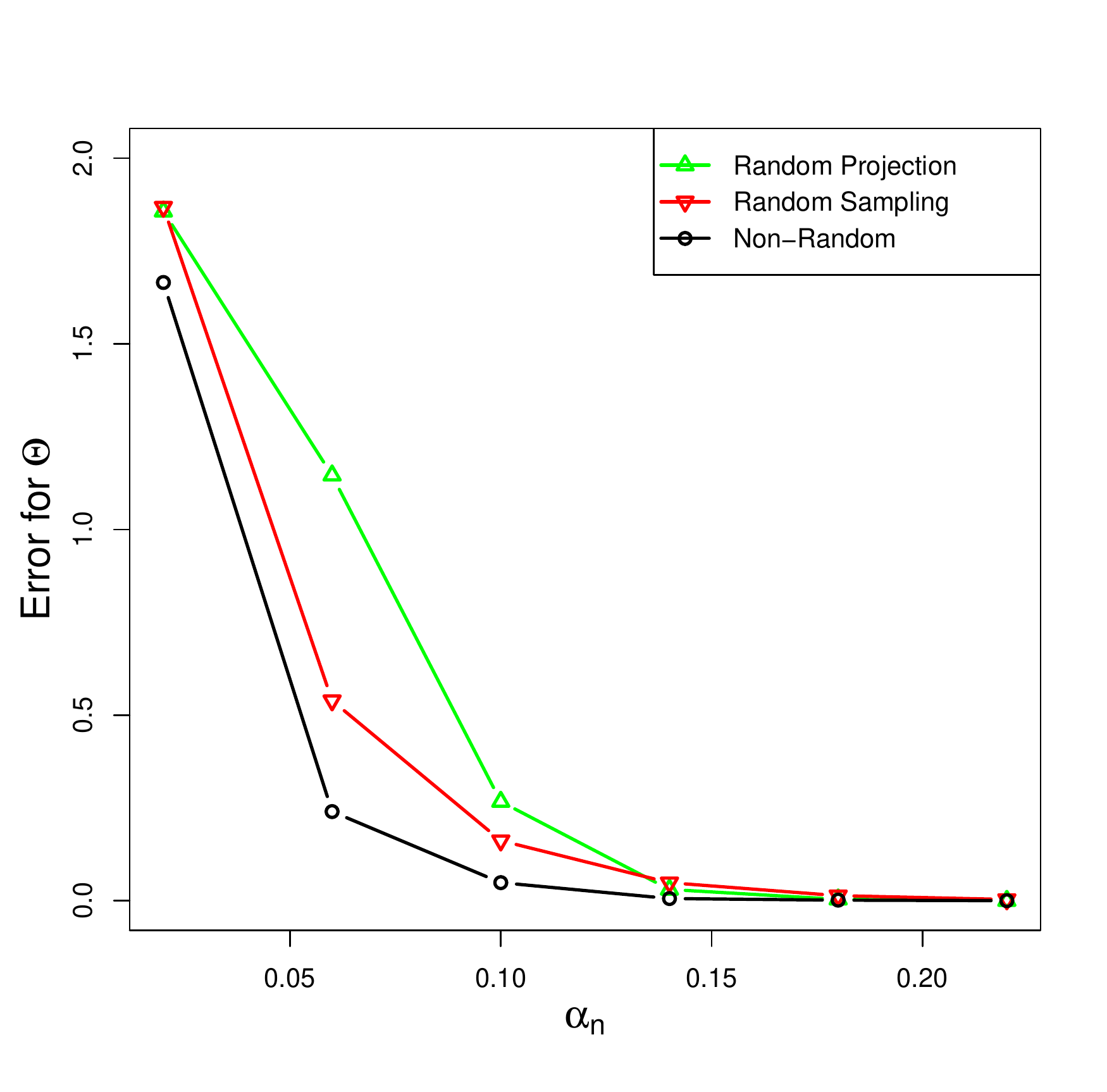}}
\subfigure[]{\includegraphics[height=4.8cm,width=5cm,angle=0]{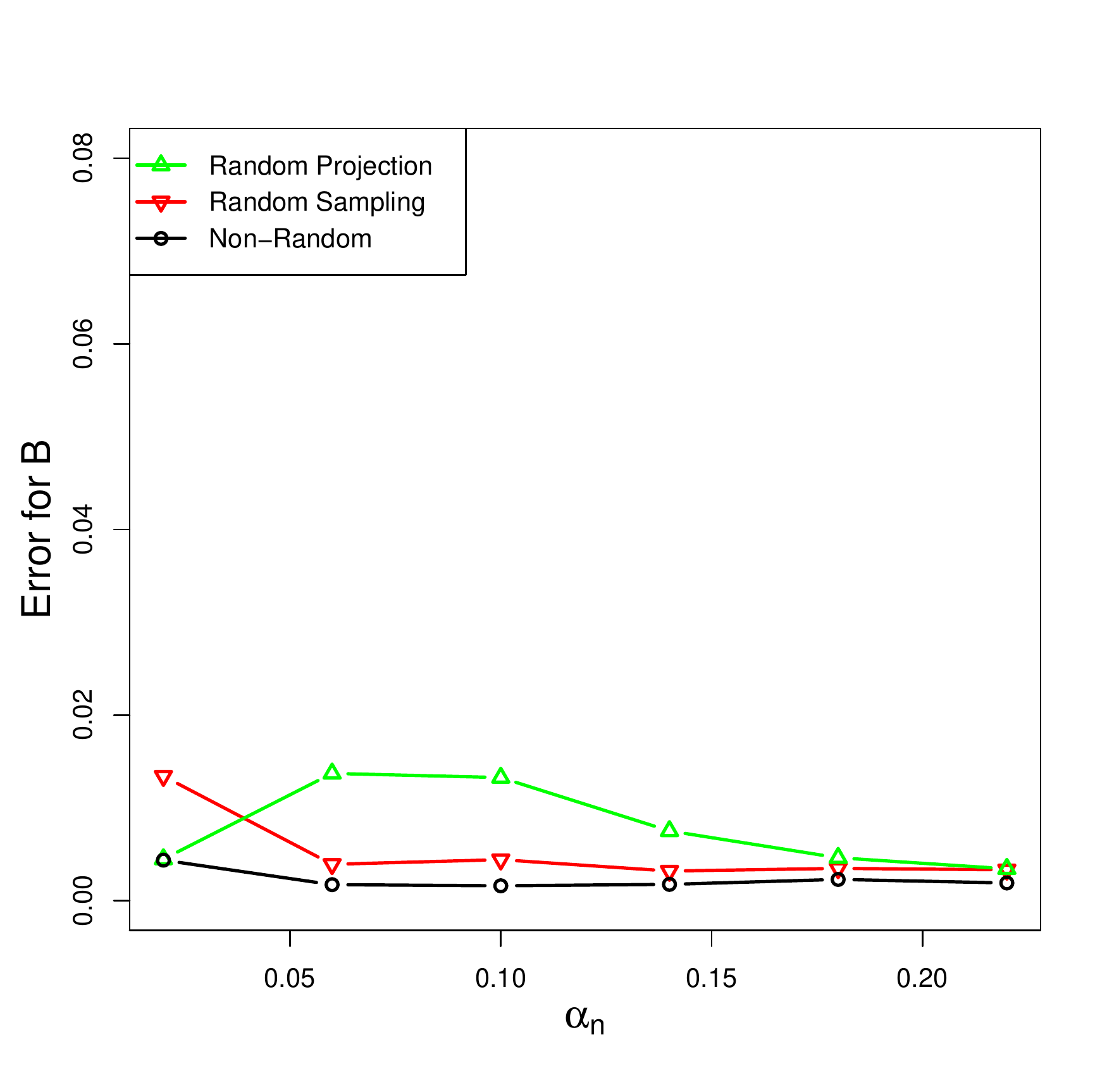}}
\caption{The average effect of $\alpha$ on the three metrics over 20 replications. (a), (b), (c) correspond to the approximation error for $P$, the misclassification error for $\Theta$, and the estimation error for $B$, respectively. The other parameters $n=1152, K=3, \lambda=0.5$, $r=10$, $q=2$, $p=0.7$, and $\Omega$ had i.i.d. standard Gaussian entries, respectively.}\label{effectofalpha}
\end{figure*}

\paragraph{Experiment 2.} In this experiment, we evaluate the effect of $\alpha_n$ on the three metrics. We fix the sample size for the moment, and focus on the influence of the maximum link probability $\alpha$. Specifically, we let $\alpha$ vary and the between cluster probability was set as $\alpha(1-0.5)$ varying with $\alpha$. The sample size $n$ was fixed at 1152. The other parameters were the same as those in Experiment 1. Figure \ref{effectofalpha} displays the average results of 20 replications. By the theoretical results, we know that the error bound for $P$ increases with order $O(\sqrt{\alpha})$, the error bound for $\Theta$ decreases with order $O(1/\alpha_n)$, and the error bound for $B$ decreases ultimately after some increase at the beginning as $\alpha$ increases. The empirical results in Figure \ref{effectofalpha} coincide with the theoretical results in some sense. {The error for $P$ increases slowly with $alpha_n$, while the error for $\Theta$ and $B$ both decrease eventually with $\alpha_n$. In addition, the gap between the randomized and the original spectral clustering in Figure \ref{effectofalpha}(b) and (c) closes as $\alpha$ increases.}

\begin{figure*}[!htbp]{}
\centering
\subfigure[]{\includegraphics[height=4.8cm,width=5cm,angle=0]{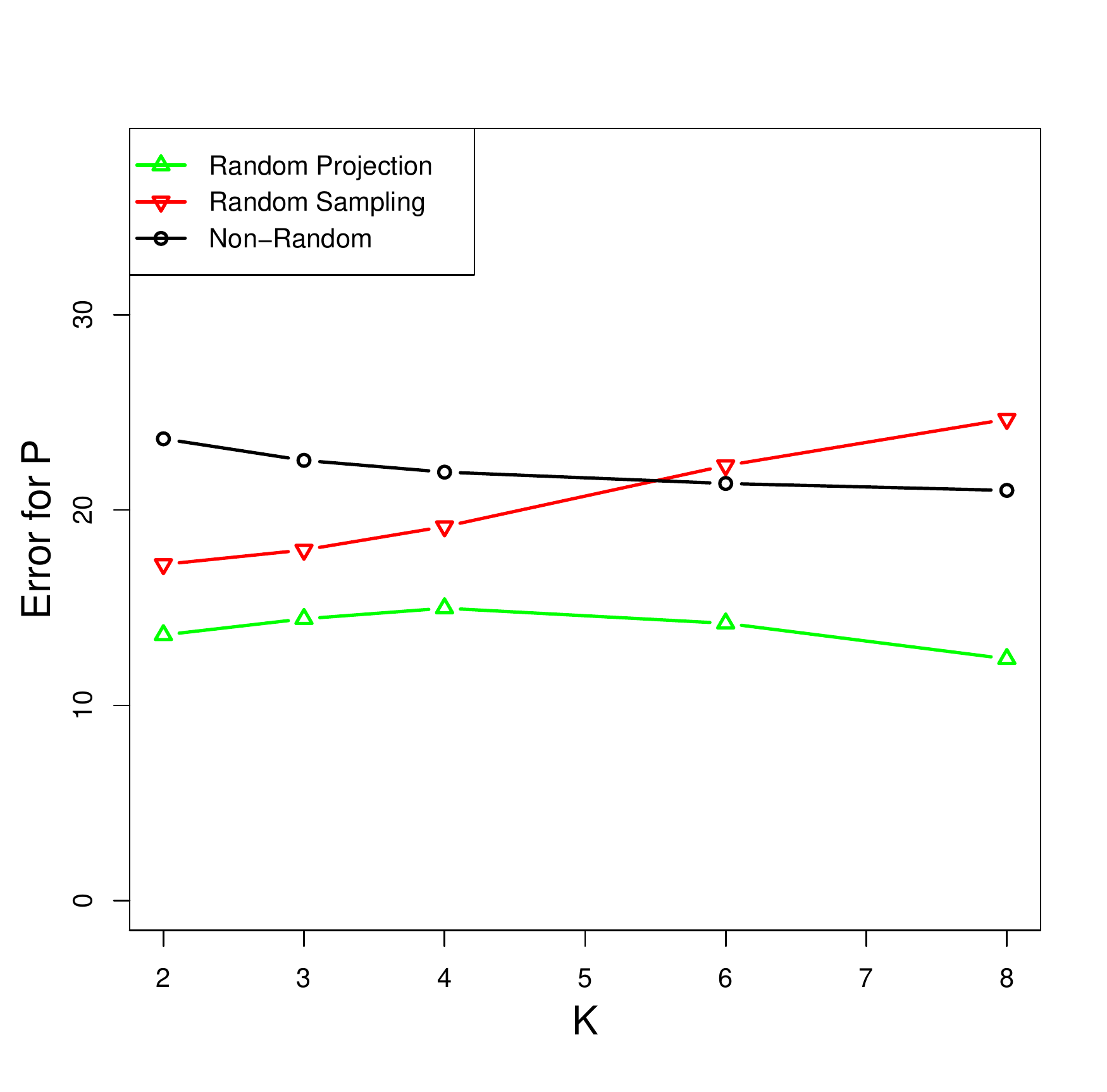}}
\subfigure[]{\includegraphics[height=4.8cm,width=5cm,angle=0]{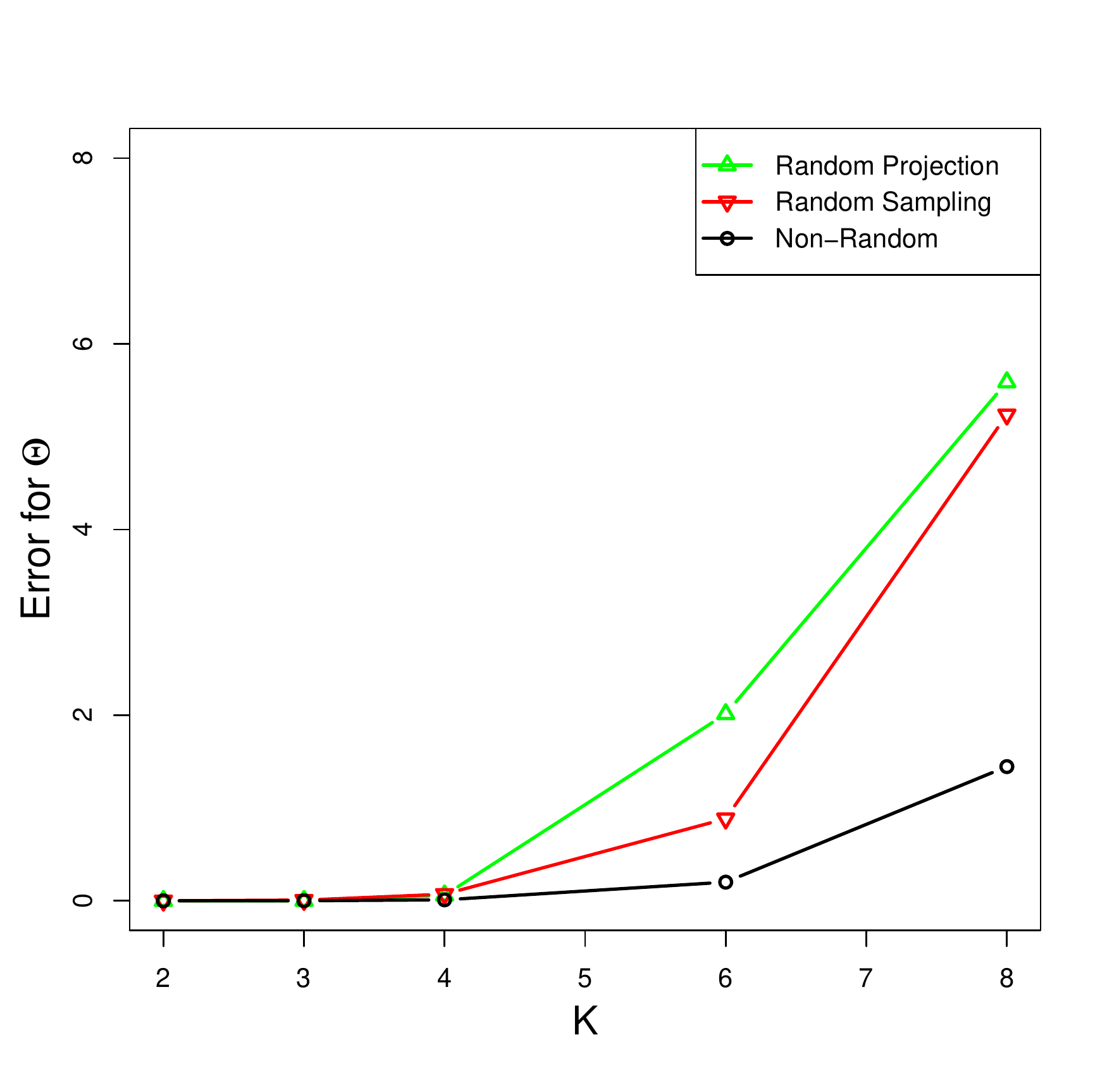}}
\subfigure[]{\includegraphics[height=4.8cm,width=5cm,angle=0]{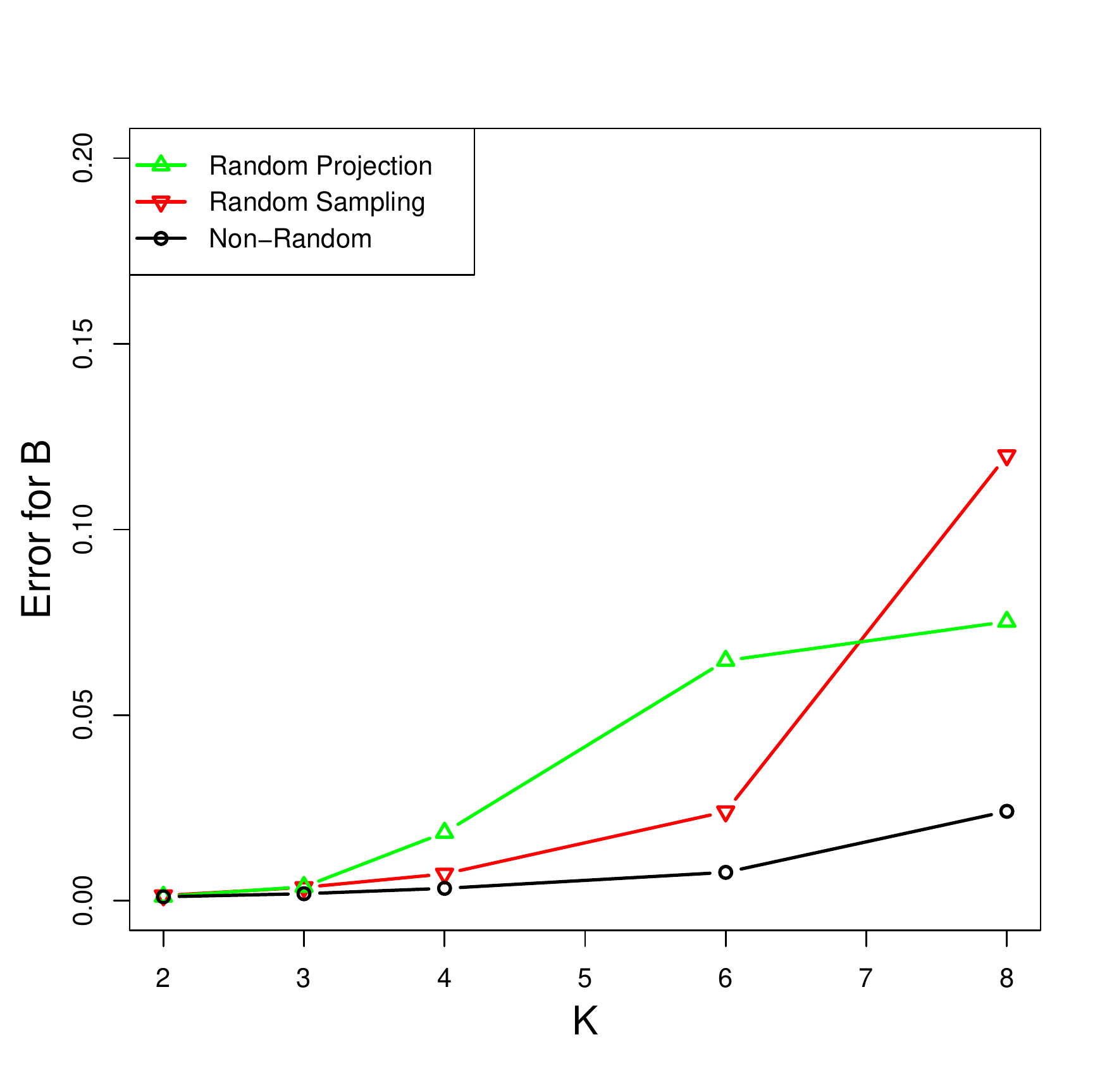}}
\caption{The average effect of $K$ on the three metrics over 20 replications. (a), (b), (c) correspond to the approximation error for $P$, the misclassification error for $\Theta$, and the estimation error for $B$, respectively. The other parameters $n=1152, \alpha_n=0.2, \alpha_n(1-\lambda)=0.1$, $r=10$, $q=2$, $p=0.7$, and $\Omega$ had i.i.d. standard Gaussian entries, respectively. }\label{effectofk}
\end{figure*}

\paragraph{Experiment 3.} In this experiment, we test the effect of $K$ on the three metrics. Specifically, we let $K$ vary, the within cluster probability $\alpha=0.2$, and the between cluster probability $\alpha(1-0.5)=0.1$, respectively. The
other parameters were the same as those in Experiment 2. The average results of 20 replications are shown in Figure \ref{effectofk}. The theoretical bounds indicate that the error bound for $\Theta$ increases with order $O(K^3)$, and the error bound for $B$ increases with $K$. {As expected, the empirical results support the theoretical findings (see Figure \ref{effectofk}(b) and (c)). The error for $\Theta$ and $B$ both increases with $K$.} While for the approximation error, recall that our randomized $\tilde{A}$ attains the minimax optimal rate which does not rely on $K$ (see Theorem 3.6 of \citet{gao2015rate}).  Empirically, from Figure \ref{effectofk}(a) we see that the approximation error for $P$ changes slowly as $K$ increases, which is partially due to the randomness throughout the experimental procedure.

\begin{figure*}[!htbp]{}
\centering
\subfigure[]{\includegraphics[height=4.8cm,width=5cm,angle=0]{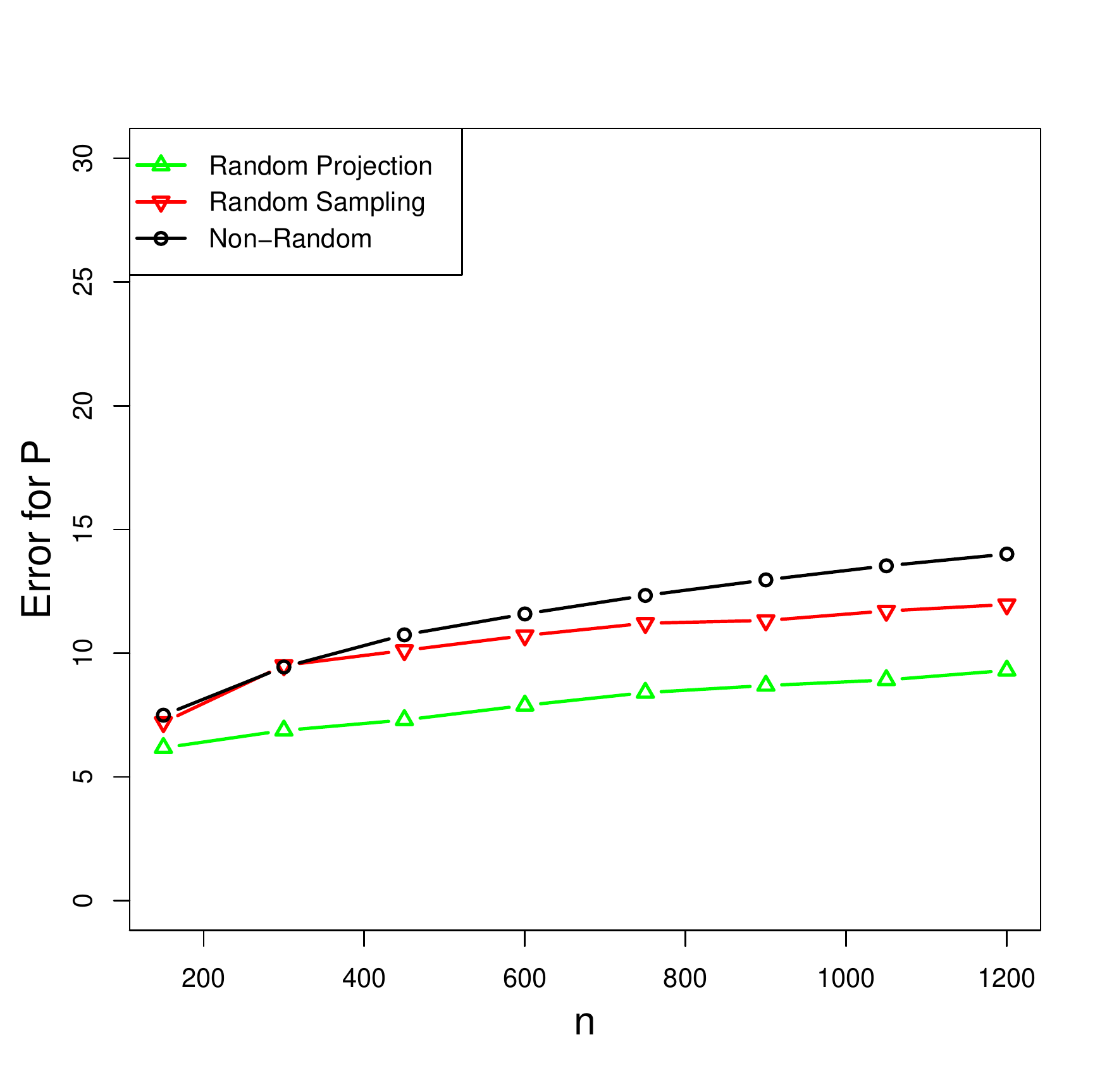}}
\subfigure[]{\includegraphics[height=4.8cm,width=5cm,angle=0]{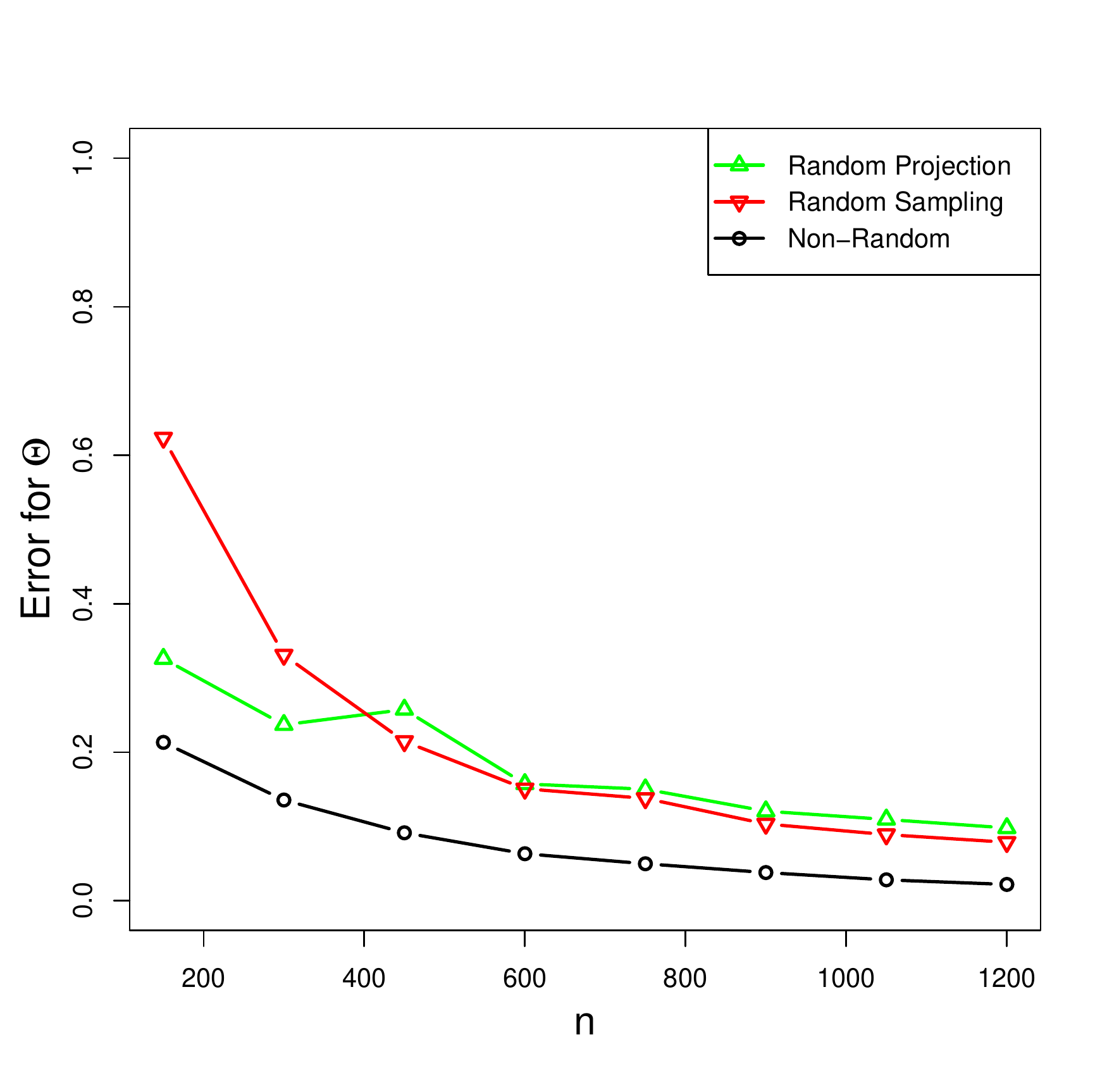}}
\subfigure[]{\includegraphics[height=4.8cm,width=5cm,angle=0]{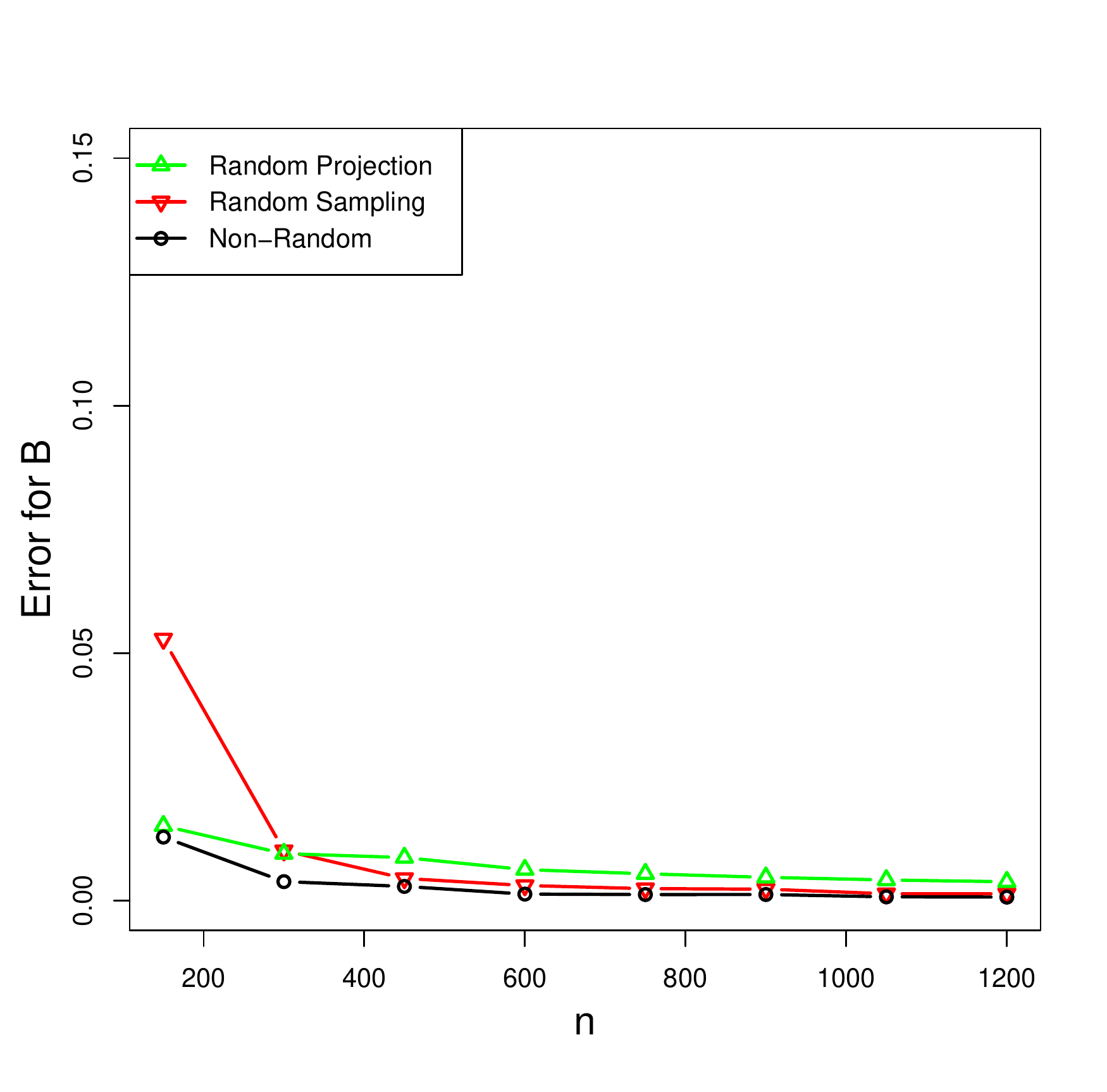}}
\caption{The average effect of $n$ and $\alpha_n$ on the three metrics over 20 replications. (a), (b), (c) correspond to the approximation error for $P$, the misclassification error for $\Theta$, and the estimation error for $B$, respectively. The within cluster probability $\alpha_n=2/\sqrt{n}$ and the between cluster probability $\alpha_n(1-0.5)=1/\sqrt{n}$.
The other parameters $K=2$, $r=10$, $q=2$, $p=0.7$, and $\Omega$ had i.i.d. standard Gaussian entries, respectively. }\label{effectofnalpha}
\end{figure*}

\paragraph{Experiment 4.} In the above three experiments, we fixed all the other parameters except the one that we pay attention to. Indeed, in view of the theoretical bounds, all the parameters can vary with $n$. To see the so-called high-dimensional performance of each method, in this experiment, we consider a simple setting that the within cluster and between cluster probabilities decrease with $n$ according to $\alpha_n=2/\sqrt{n}$ and $\alpha_n(1-0.5)=1/\sqrt{n}$, respectively. In such a setting, to ensure the decreasing trend of the misclustered error, $K$ should be of smaller order than $n^{1/6}$, which is rather small for $n$ smaller than, say, 1000. Hence we set $K=2$ for simplicity. The
other parameters were the same as those in Experiment 2. Figure \ref{effectofnalpha} shows the average curves for each method in terms of three metrics. {As expected, the misclassification error and the error for $B$ both decrease as $n$ increases, showing the high-dimensional feature of the theoretics. In addition, the performance of randomized methods become close to that of the original spectral clustering as $n$ increases.}

\subsection{Model extensions}
Besides the simple SBMs we considered in subsection \ref{sub:sim1}, we here consider the following six more complex models.
\begin{itemize}
  \item\textbf{ Model 1} (Full-rank SBM with random $B$): $K=3$, and the elements of $B$ are generated randomly according to $B_{ii}\sim {\rm Uniform}(0.2,0.3)$ and $B_{ij}\sim {\rm Uniform}(0.01,0.1)$, the community sizes are balanced.
  \item\textbf{ Model 2} (Full-rank SBM with random $B$ and unblanced communities): The parameter set-up is identical to that of Model 1 except that the proportions of the number nodes within of each community over that of the whole nodes are $\frac{1}{6}, \frac{1}{2},\frac{1}{3}$, respectively.
  \item\textbf{ Model 3} (Rank-deficient SBM): $K=3$, and the community sizes are balanced. {The link probability matrix $B:=CC^\intercal$ where
      \begin{equation*}
      C:= \left[\begin{matrix}
     \frac{2{\rm sin}\,0}{3} &  \frac{2{\rm cos}\,0}{3}\\
    \frac{ {\rm sin}\,\frac{\pi}{5}}{2}  &  \frac{ {\rm cos}\,\frac{\pi}{5}}{2}\\
     \frac{5 {\rm sin}\,\frac{2\pi}{5}}{6}  &  \frac{5 {\rm cos}\,\frac{2\pi}{5}}{6}
      \end{matrix}\right].
      \end{equation*}}

 \item\textbf{ Model 4} (Full-rank DC-SBM): $K=3$, and the elements of $B$ are generated randomly according to $B_{ii}\sim {\rm Uniform}(0.4,0.6)$ and $B_{ij}\sim {\rm Uniform}(0.01,0.2)$. Within each true cluster $k$, $\vartheta_{i\in G_k}$'s are i.i.d. 0.2 with probability (w.p.) 0.8, and 1 w.p. 0.2. $\vartheta$'s are then normalized such that its maximum value is 1 within each true cluster.
  \item\textbf{Model 5} (Full-rank DC-SBM with more heterogeneity): Except for the node propensity parameter $\vartheta$, the parameter set-up is identical to that of Model 4. The $\vartheta$'s are generated as follows.  Within each true cluster $k$, $\vartheta_{i\in G_k}$'s are i.i.d. with its element being 0.1 w.p. 0.4, being 0.2 w.p. 0.4, and being 1 with probability 0.2. $\vartheta$'s are then normalized such that its maximum value is 1 within each true cluster.
  \item\textbf{Model 6} (Rank-deficient DC-SBM):  The parameter set-up is identical to that of Model 4 except the formulation of $B$. Particularly, $B$ is the same with that in Model 3.
  %\item\textbf{Model 8} (Rank-deficient disassortative SBM):  The parameter set-up is identical to that of Model 1 except that the SBM is disassortative in that $B$ is not diagonally dominant. In particular \begin{equation*}
%      B:=CC^\intercal= \left[\begin{matrix}
%     0.4 & 0\\
%      0.5&  0.05\\
%     0  &  0.4
%      \end{matrix}\right]\cdot \left[\begin{matrix}
%     0.4 & 0.5& 0\\
%     0  &0.05&  0.4
%      \end{matrix}\right]=\left[\begin{matrix}
%     0.16 & 0.2&0\\
%      0.2&  0.2525&0.02\\
%     0  &  0.02&0.16
%      \end{matrix}\right].
%      \end{equation*}
%

\end{itemize}

\begin{figure*}[!htbp]{}
\centering
\small (I) Approximation error of model 1, model 2 and model 3.\\
\subfigure[Model 1]{\includegraphics[height=4.8cm,width=5cm,angle=0]{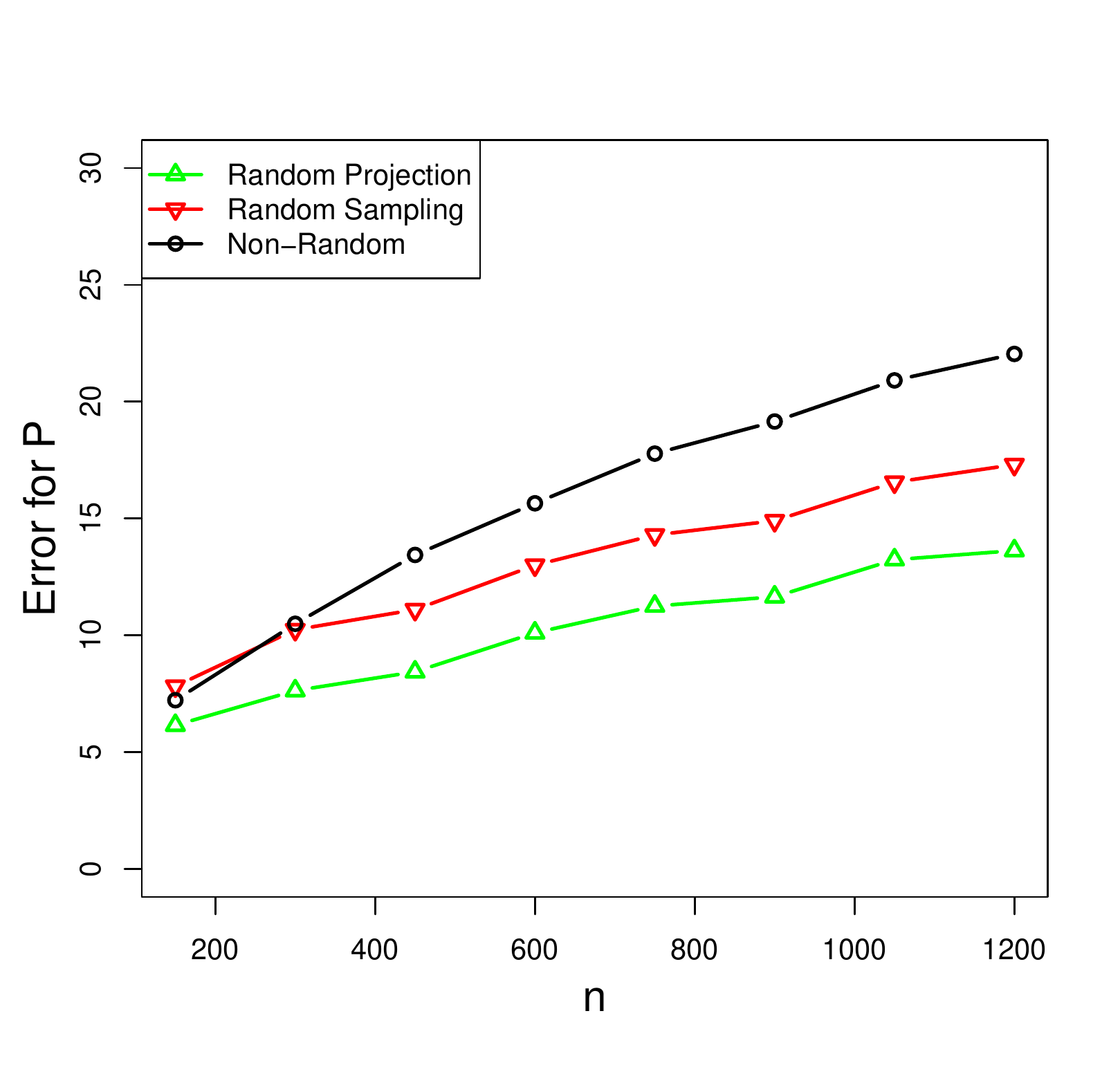}}
\subfigure[Model 2]{\includegraphics[height=4.8cm,width=5cm,angle=0]{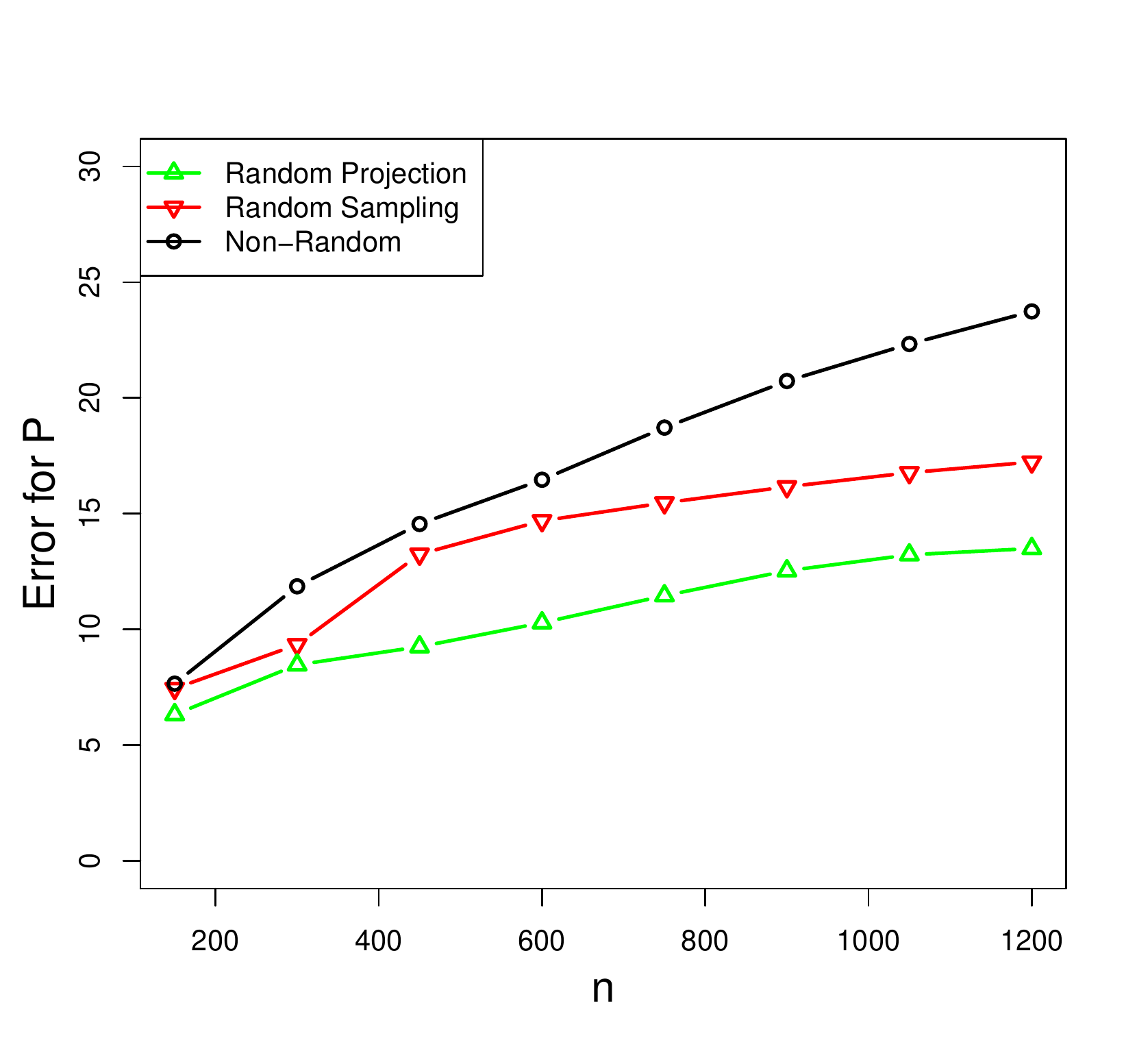}}
\subfigure[Model 3]{\includegraphics[height=4.8cm,width=5cm,angle=0]{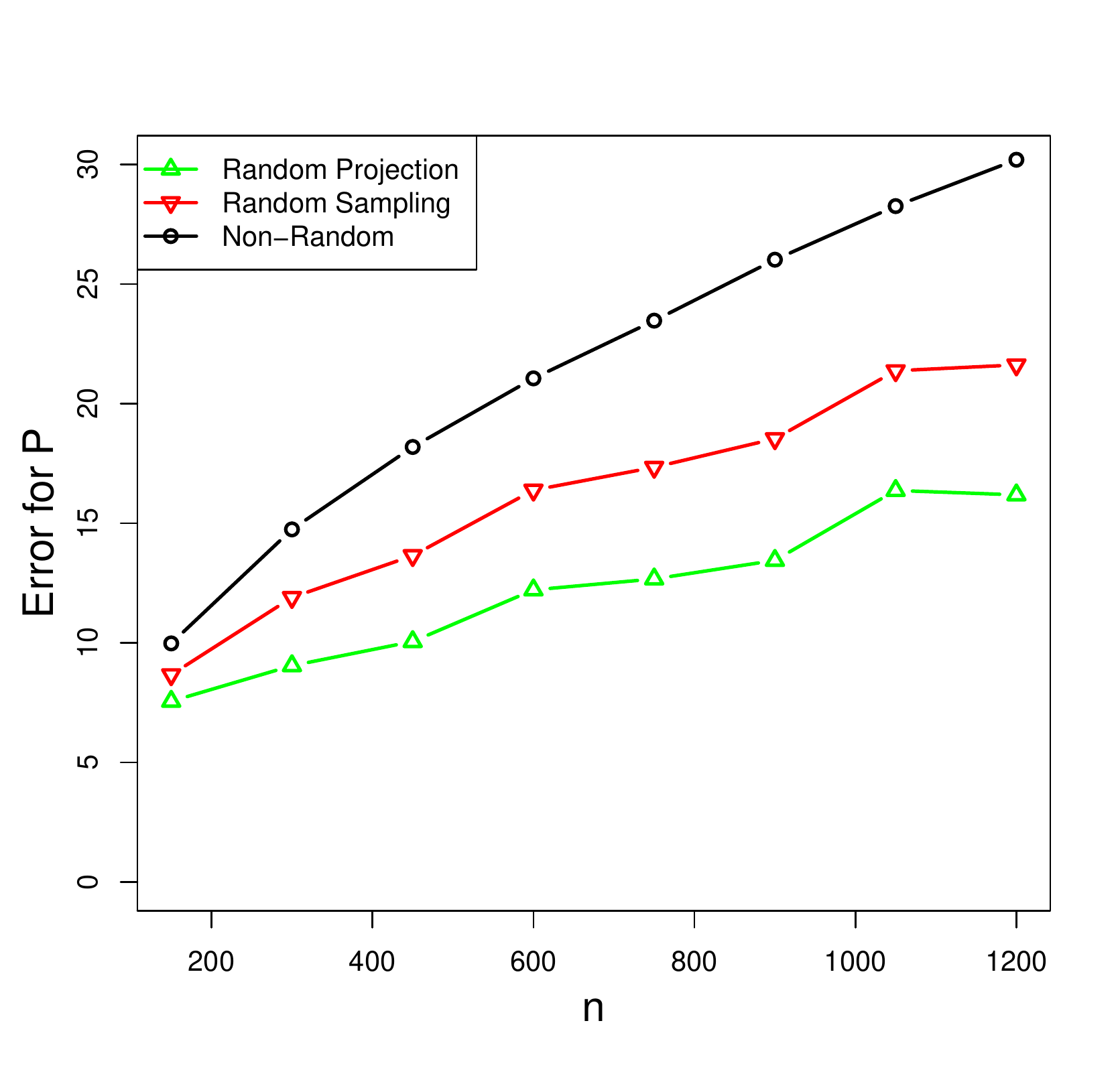}}\vspace{1cm}\\
(II) Misclassification error of Model 1, Model 2 and Model 3.\\
\subfigure[Model 1]{\includegraphics[height=4.8cm,width=5cm,angle=0]{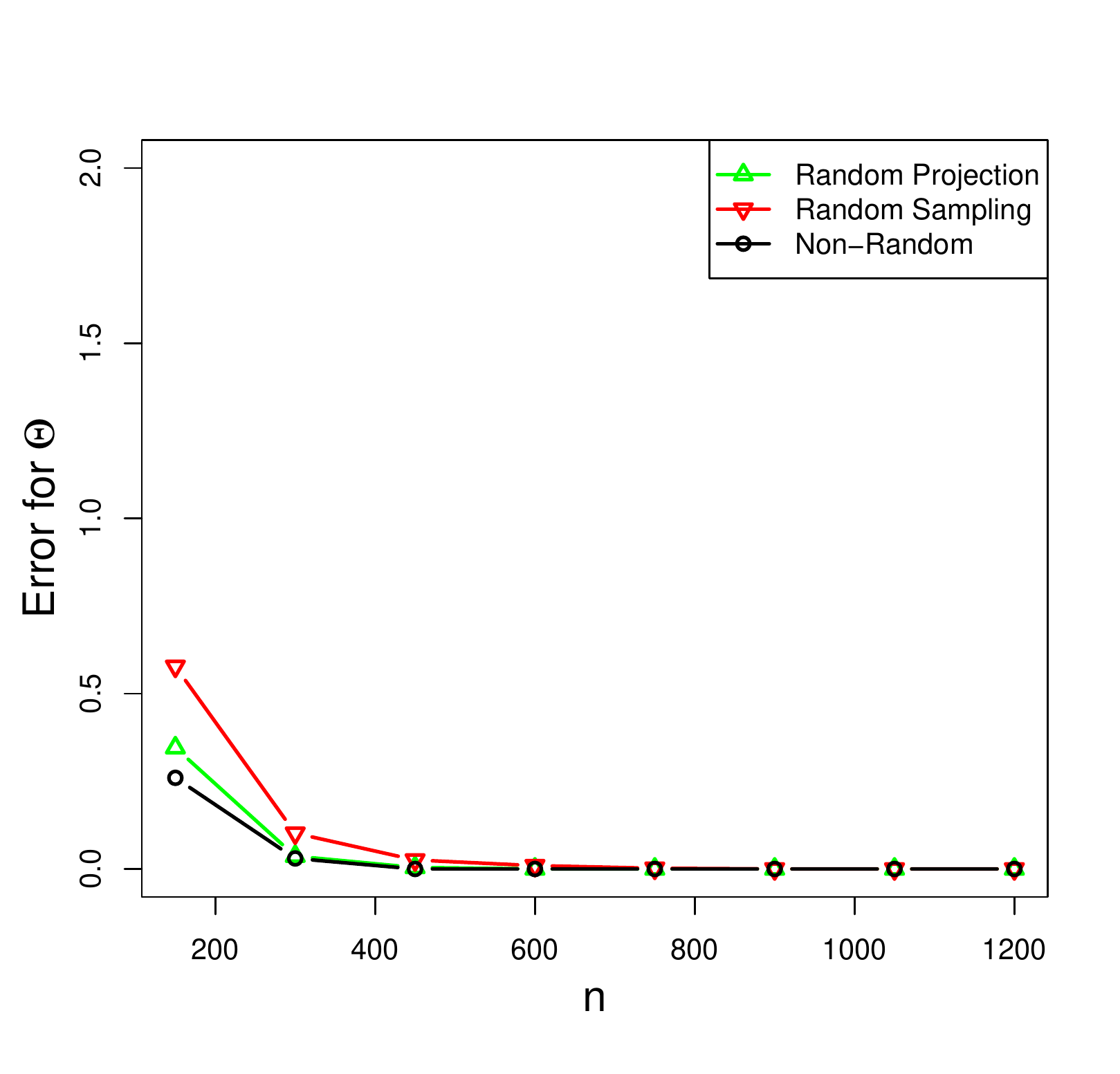}}
\subfigure[Model 2]{\includegraphics[height=4.8cm,width=5cm,angle=0]{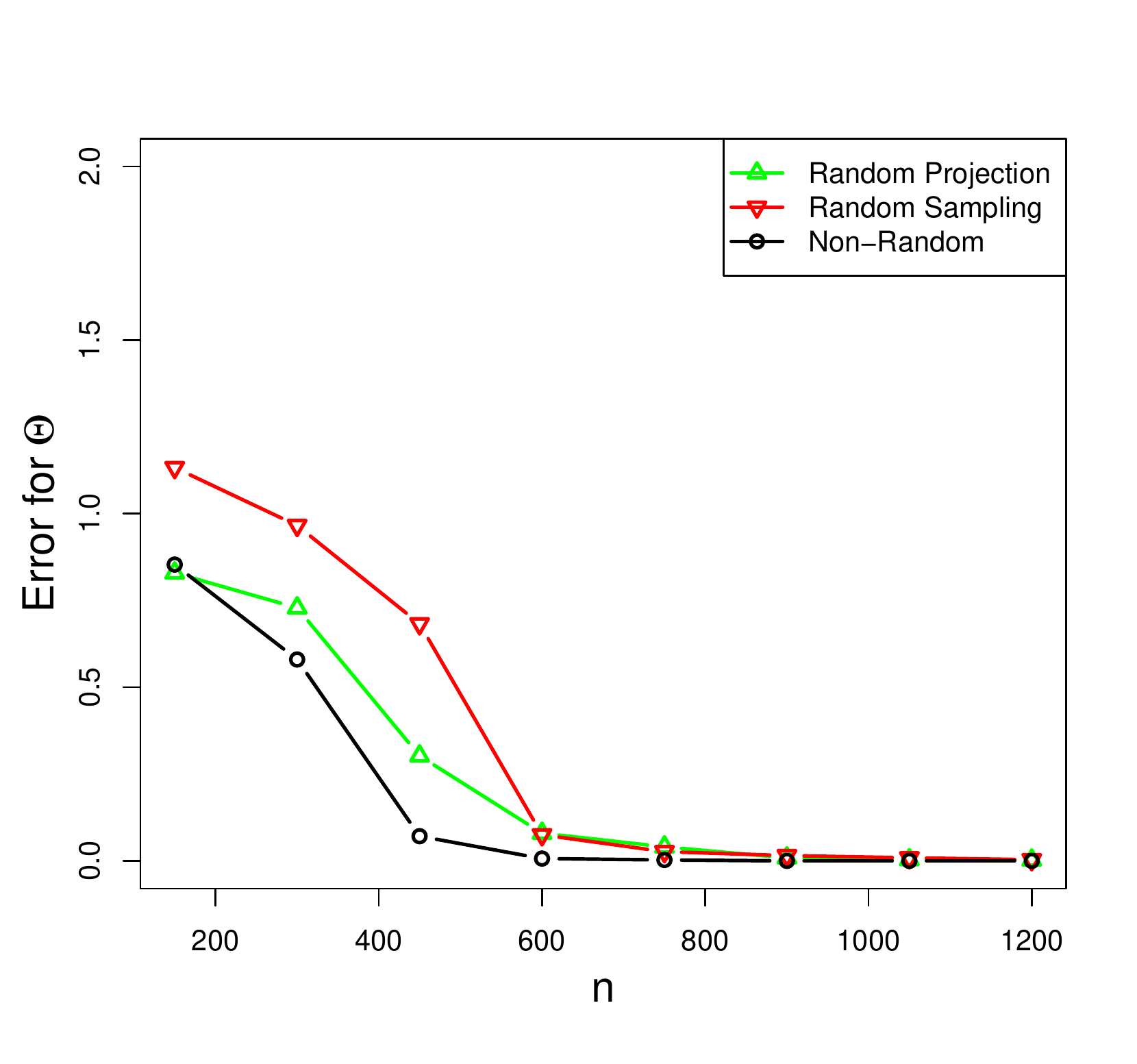}}
\subfigure[Model 3]{\includegraphics[height=4.8cm,width=5cm,angle=0]{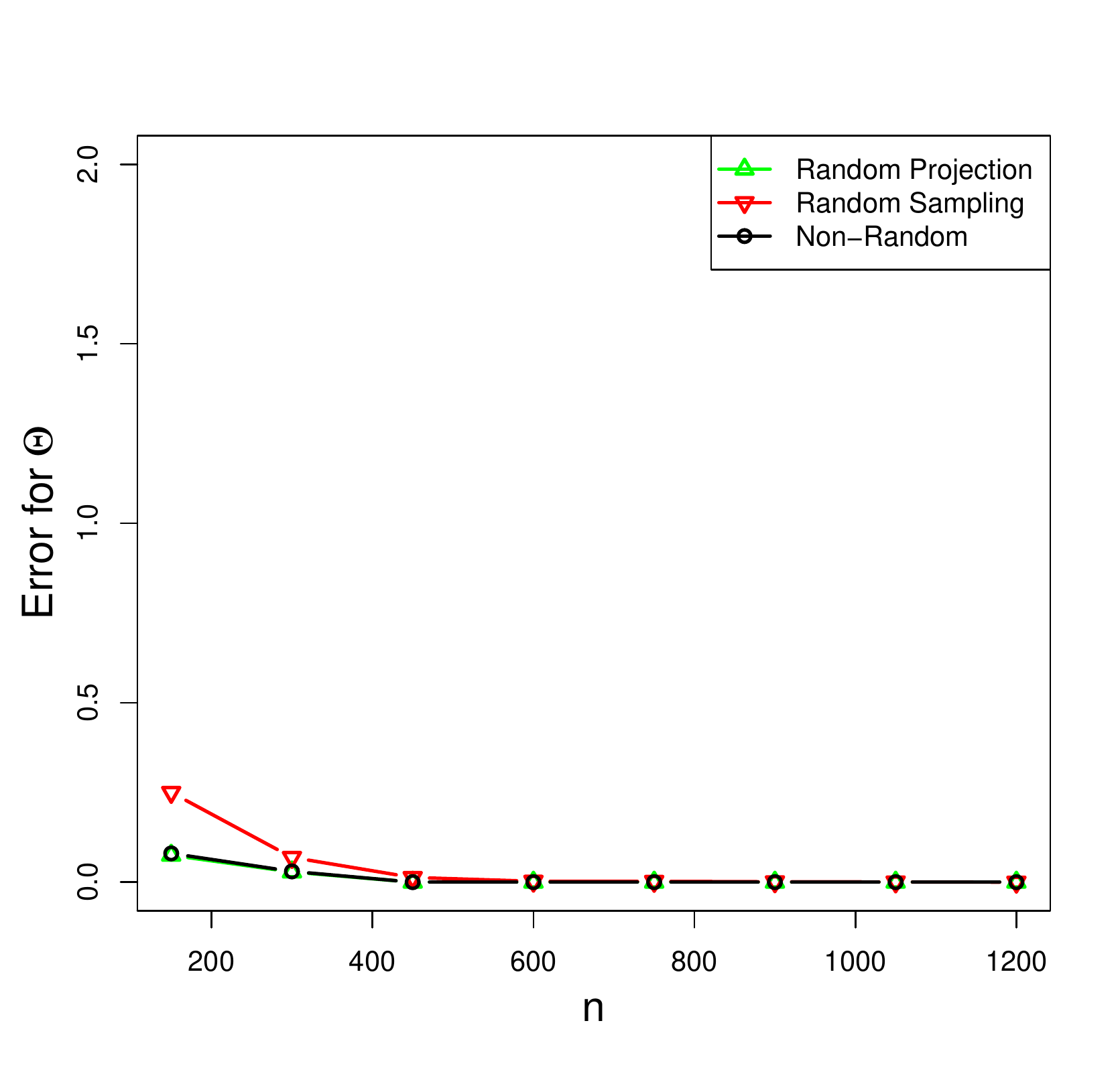}}
\vspace{1cm}\\
(III) Estimation error for link probability matrix of model 1, model 2 and model 3.\\
\subfigure[Model 1]{\includegraphics[height=4.8cm,width=5cm,angle=0]{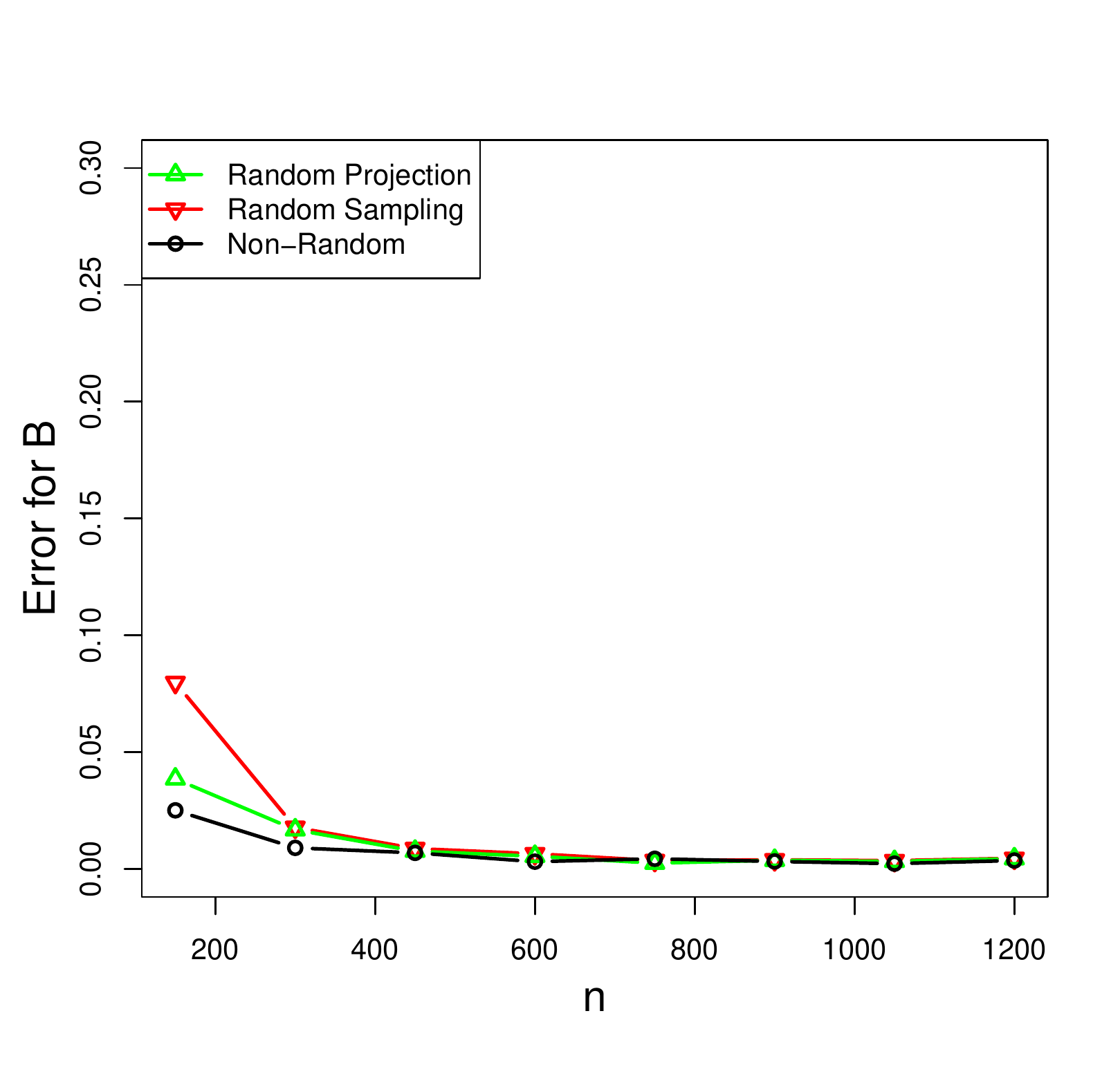}}
\subfigure[Model 2]{\includegraphics[height=4.8cm,width=5cm,angle=0]{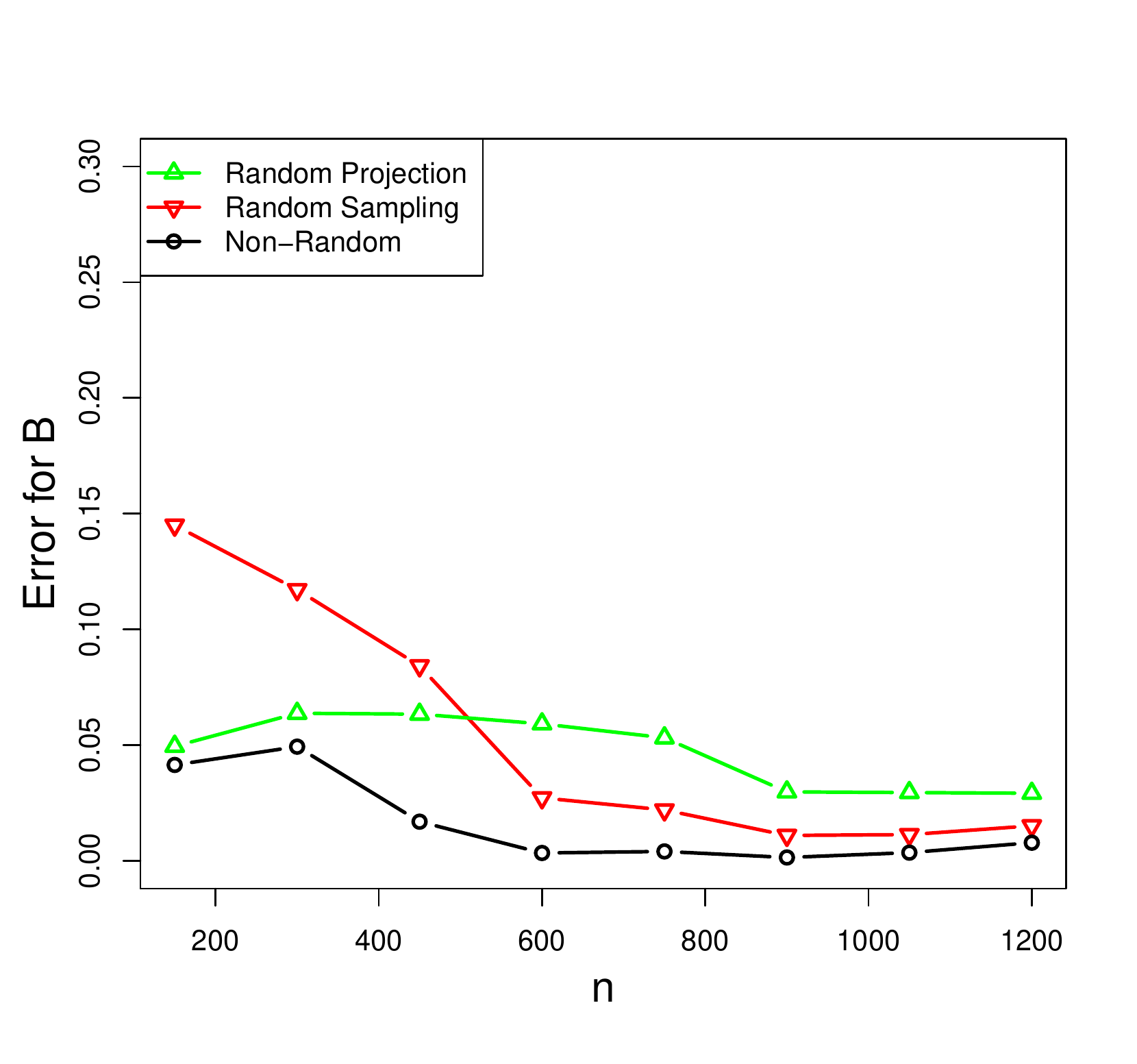}}
\subfigure[Model 3]{\includegraphics[height=4.8cm,width=5cm,angle=0]{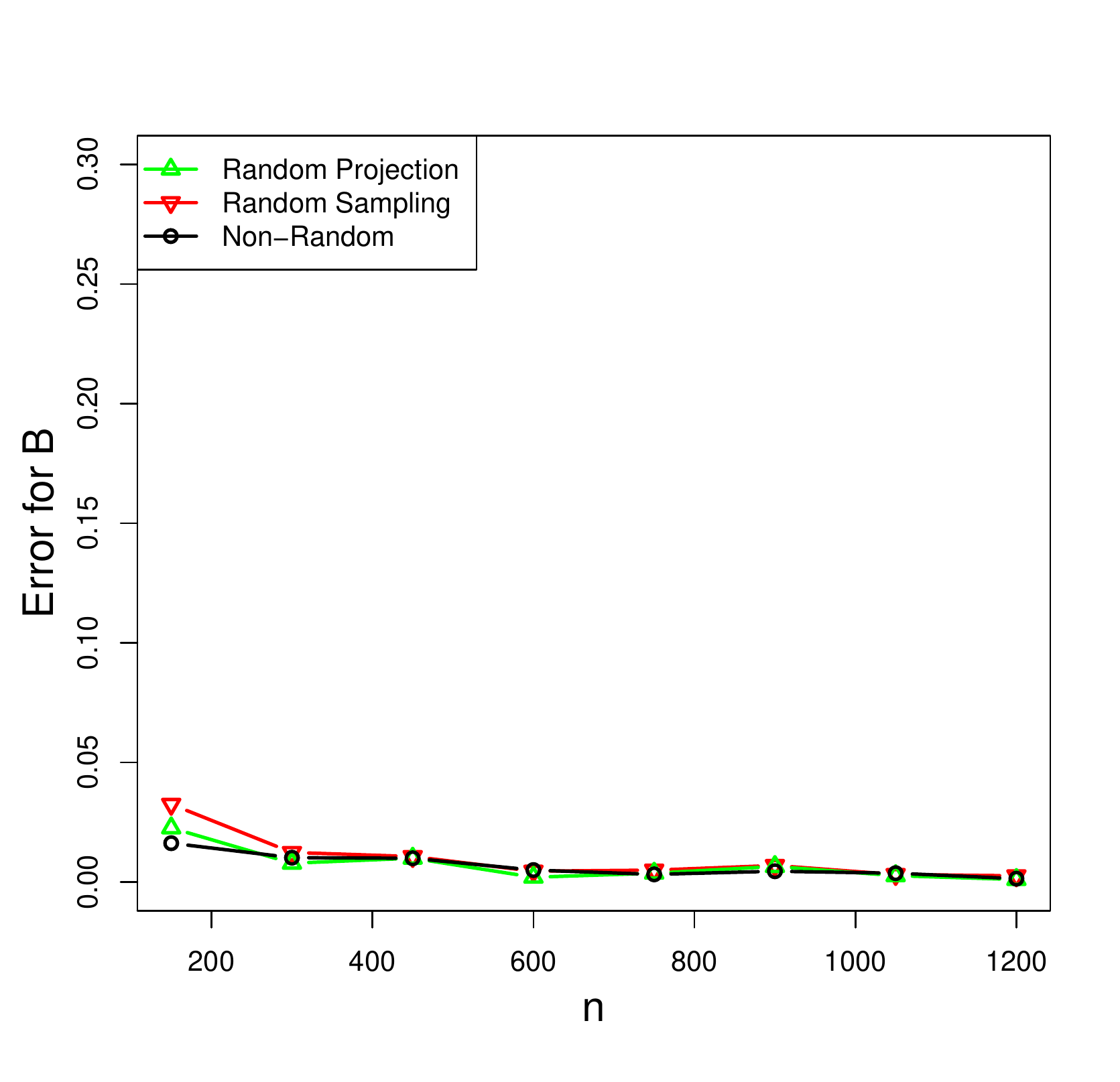}}

\caption{Averaged results of each method over 20 replications on Models 1-3. Each column corresponds to a model. The first, second and third row corresponds to the approximation error, the misclassification error, and the estimation error for link probability matrix, respectively.}\label{model123}
\end{figure*}

\begin{figure*}[!htbp]{}
\centering
\small (I) Approximation error of Model 4, Model 5 and Model 6.\\
\subfigure[Model 4]{\includegraphics[height=4.8cm,width=5cm,angle=0]{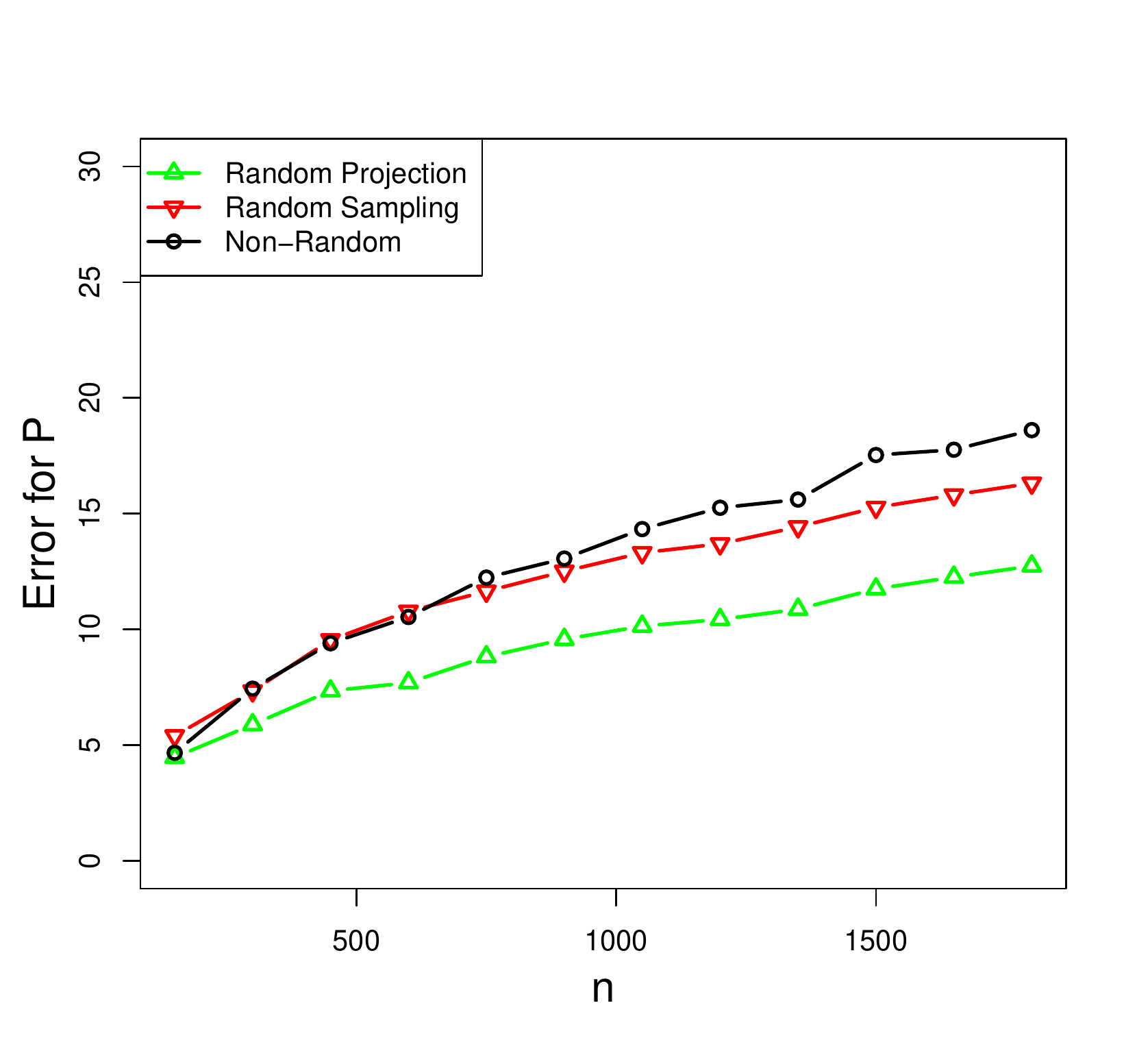}}
\subfigure[Model 5]{\includegraphics[height=4.8cm,width=5cm,angle=0]{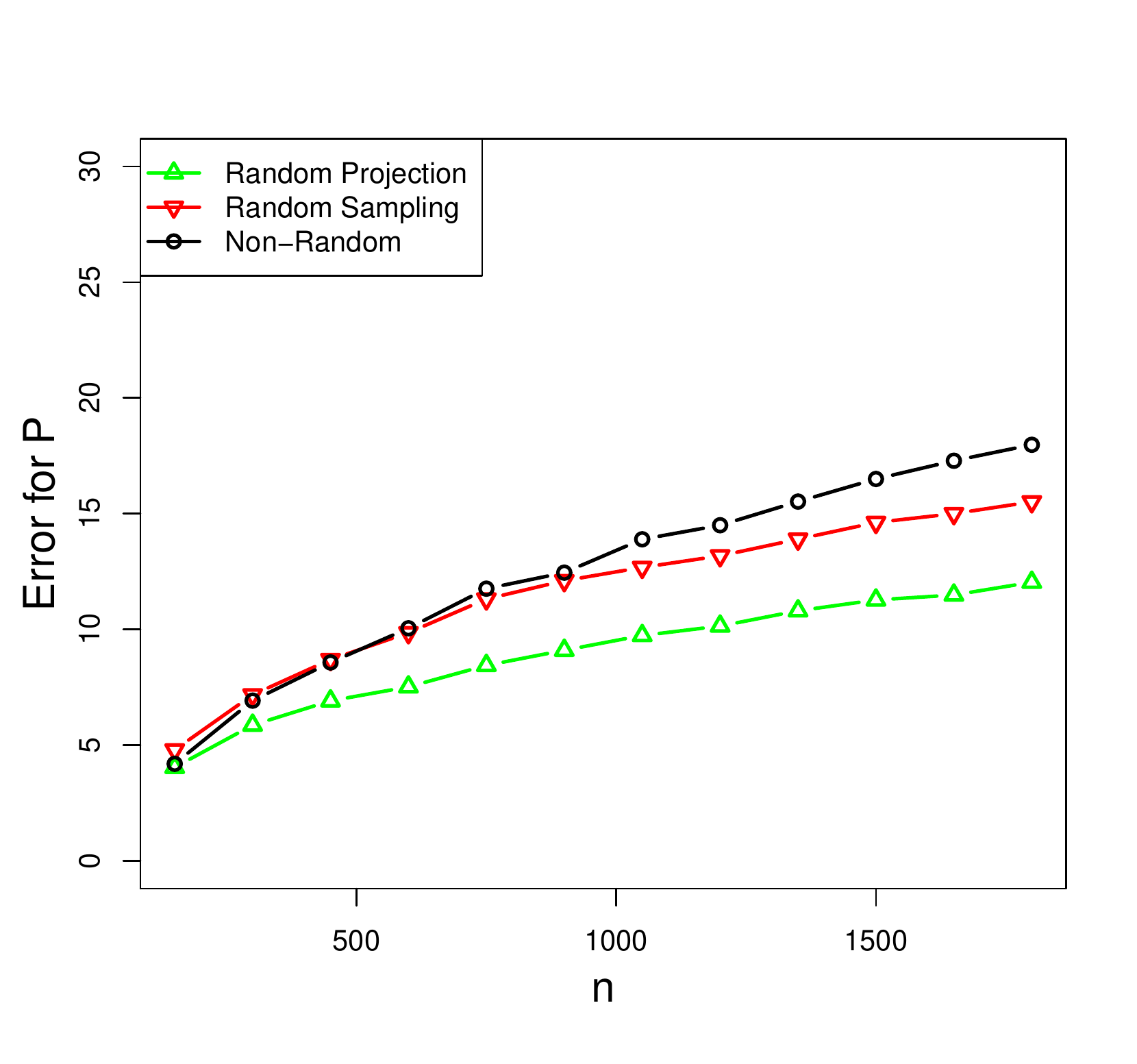}}
\subfigure[Model 6]{\includegraphics[height=4.8cm,width=5cm,angle=0]{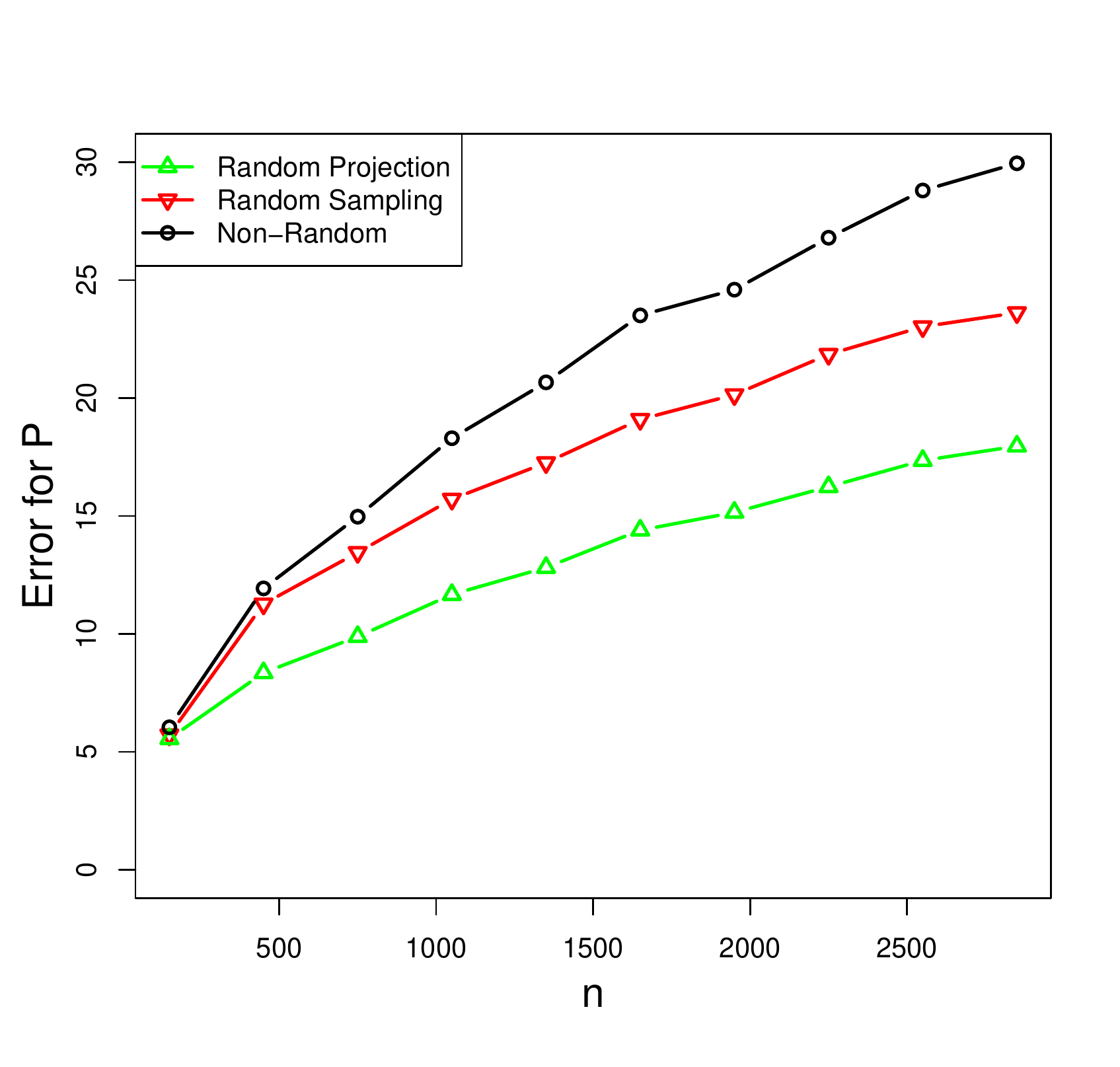}}\vspace{1cm}\\
(II) Misclassification error of Model 4, Model 5 and Model 6.\\
\subfigure[Model 4]{\includegraphics[height=4.8cm,width=5cm,angle=0]{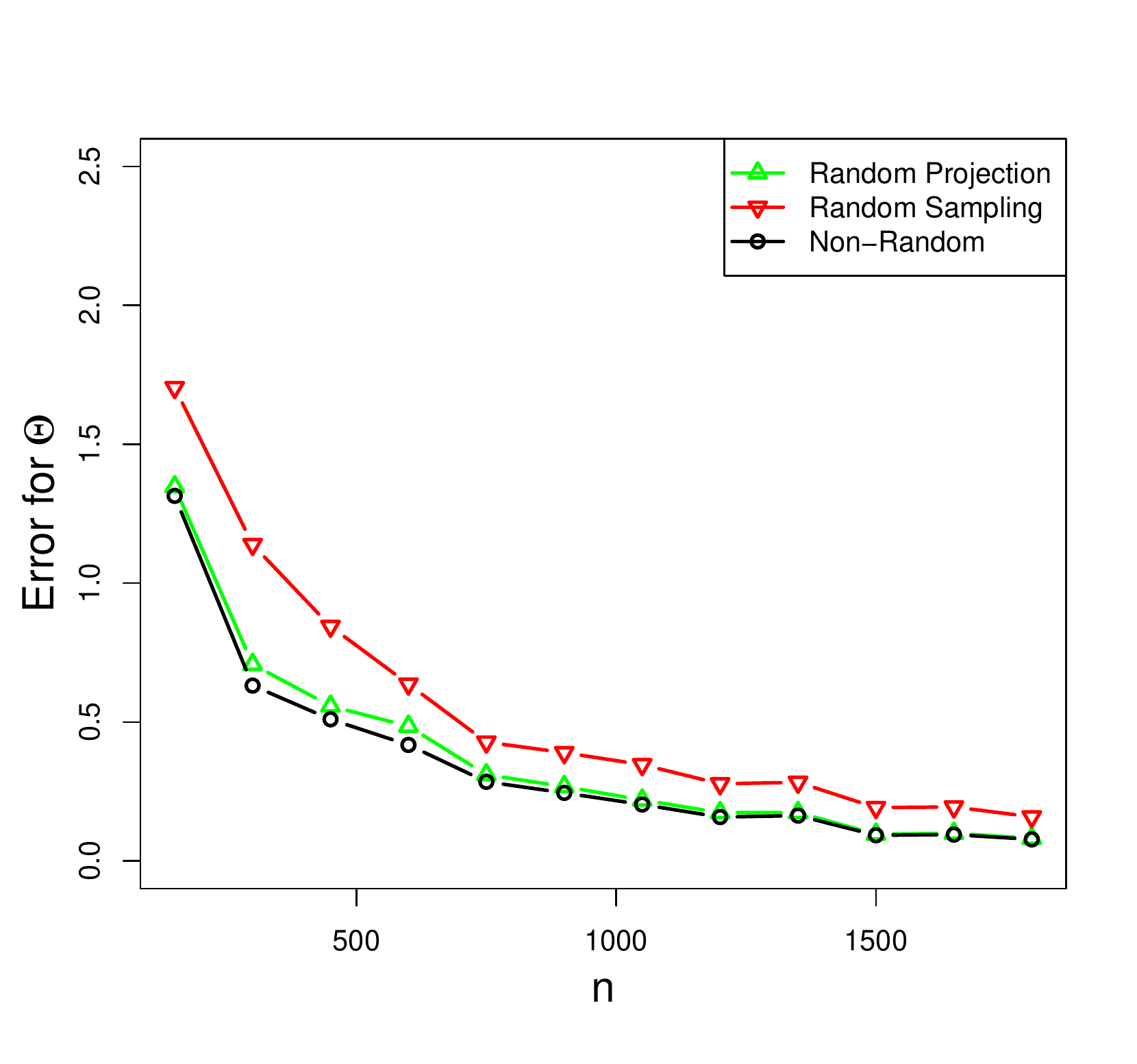}}
\subfigure[Model 5]{\includegraphics[height=4.8cm,width=5cm,angle=0]{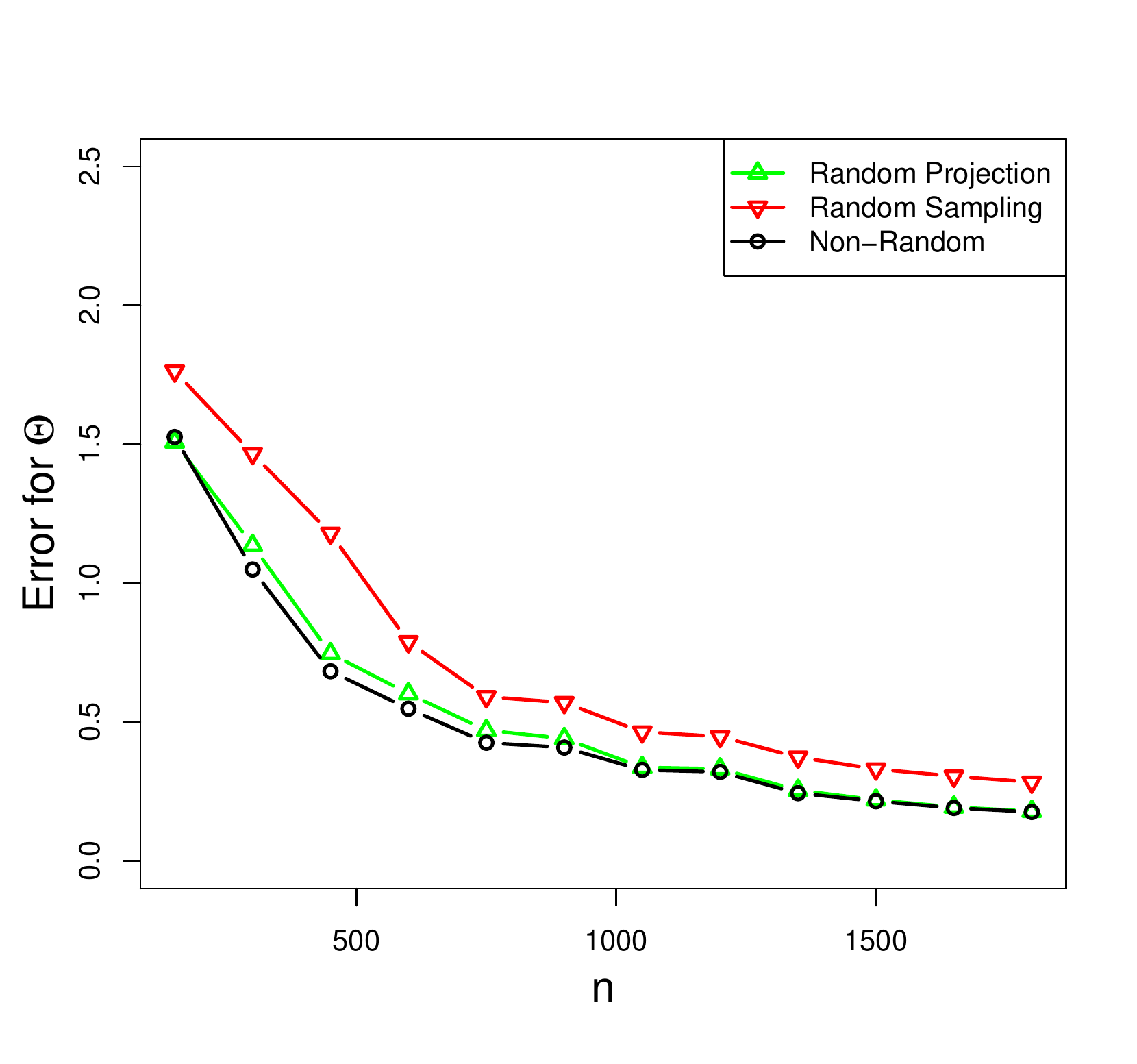}}
\subfigure[Model 6]{\includegraphics[height=4.8cm,width=5cm,angle=0]{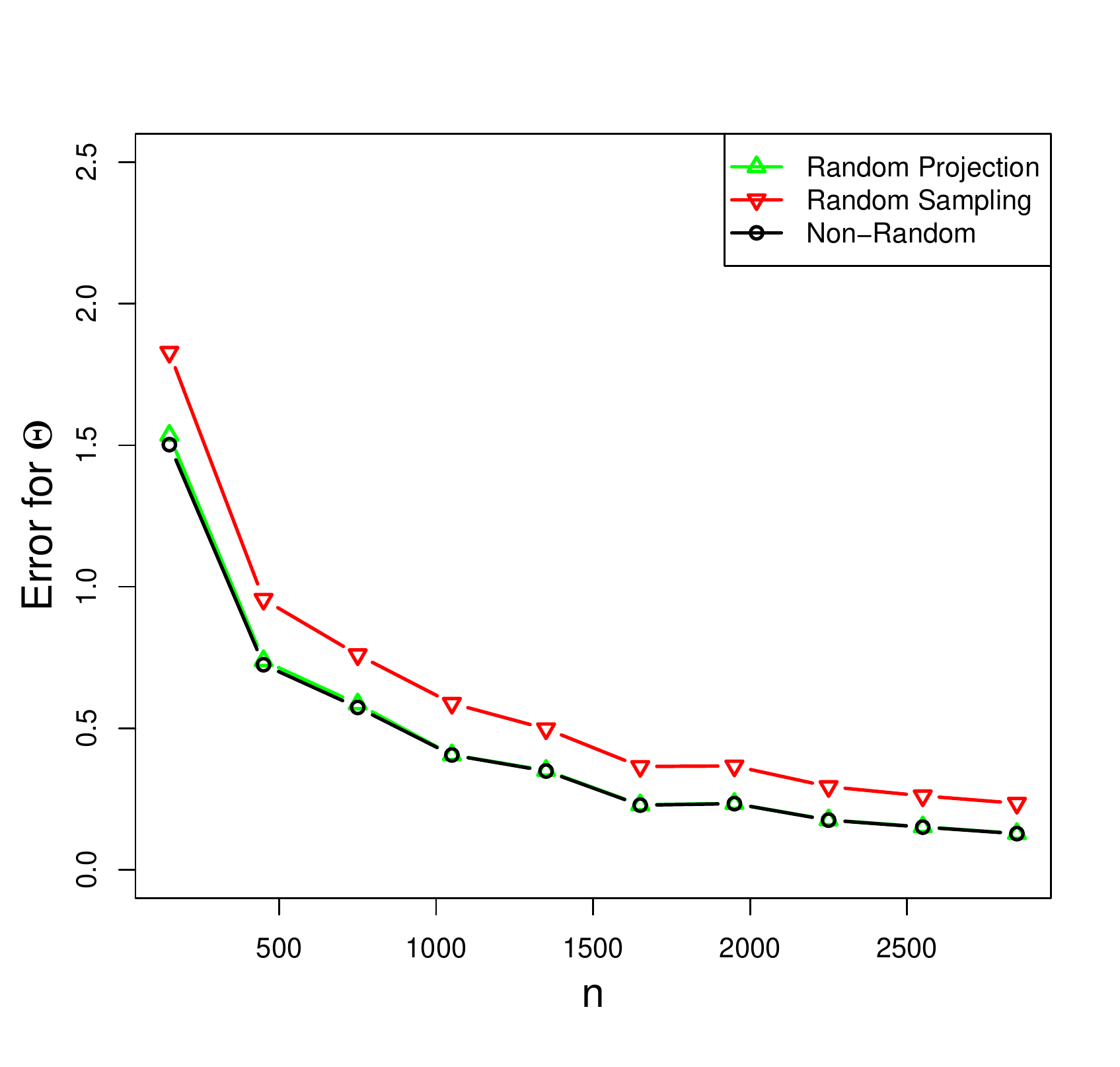}}
\caption{Averaged results of each method over 20 replications on Models 4-6. Each column corresponds to a model. The first and second row corresponds to the approximation error and the misclassification error, respectively.}\label{model456}
\end{figure*}

Figure \ref{model123} and \ref{model456} display the averaged results over 20 replications of model 1-3 and model 4-6, respectively. It can be seen that for all models we tested, the clustering performance of the randomized spectral clustering algorithms become close to that of the original spectral clustering as the sample size $n$ increases, coinciding with theoretical results.

{In the above experiments, we only considered assortative networks where nodes tend to be connected with those in the same community, which is mainly because that we require the link probability matrix to be diagonally dominant in some sense (Lemma \ref{lem:eigencondition} and \ref{lem:eigen2condition}). For the disassortative networks, it would be our future work to study their clustering methods specifically in the rank deficient setting, though it is suggest that the absolute eigenvalue would help finding the proper communities in full rank setting \citep{rohe2011spectral}.}

%\begin{figure*}[!htbp]{}
%\centering
%\small (I) Approximation error of model 7, model 8 and model 9.\\
%\subfigure[Model 7]{\includegraphics[height=4.8cm,width=5cm,angle=0]{m4_app.pdf}}
%\subfigure[Model 8]{\includegraphics[height=4.8cm,width=5cm,angle=0]{m8_app.pdf}}
%\subfigure[Model 9]{\includegraphics[height=4.8cm,width=5cm,angle=0]{m6_app.pdf}}\vspace{1cm}\\
%(II) Misclassification error of model 7, model 8 and model 9.\\
%\subfigure[Model 7]{\includegraphics[height=4.8cm,width=5cm,angle=0]{m4_mis.pdf}}
%\subfigure[Model 8]{\includegraphics[height=4.8cm,width=5cm,angle=0]{m8_mis.pdf}}
%\subfigure[Model 9]{\includegraphics[height=4.8cm,width=5cm,angle=0]{m6_mis.pdf}}
%\caption{Averaged results of each method over 20 replications on models 7-9. Each column corresponds to a model. The first and second row corresponds to the approximation error and the misclassification error, respectively.}\label{model789}
%\end{figure*}

\subsection{Additional experiments}
\begin{figure*}[!htbp]{}
\centering
(I) Effect of the oversampling parameter $r$\\
\subfigure[${\rm F}_1$]{\includegraphics[height=4.8cm,width=5cm,angle=0]{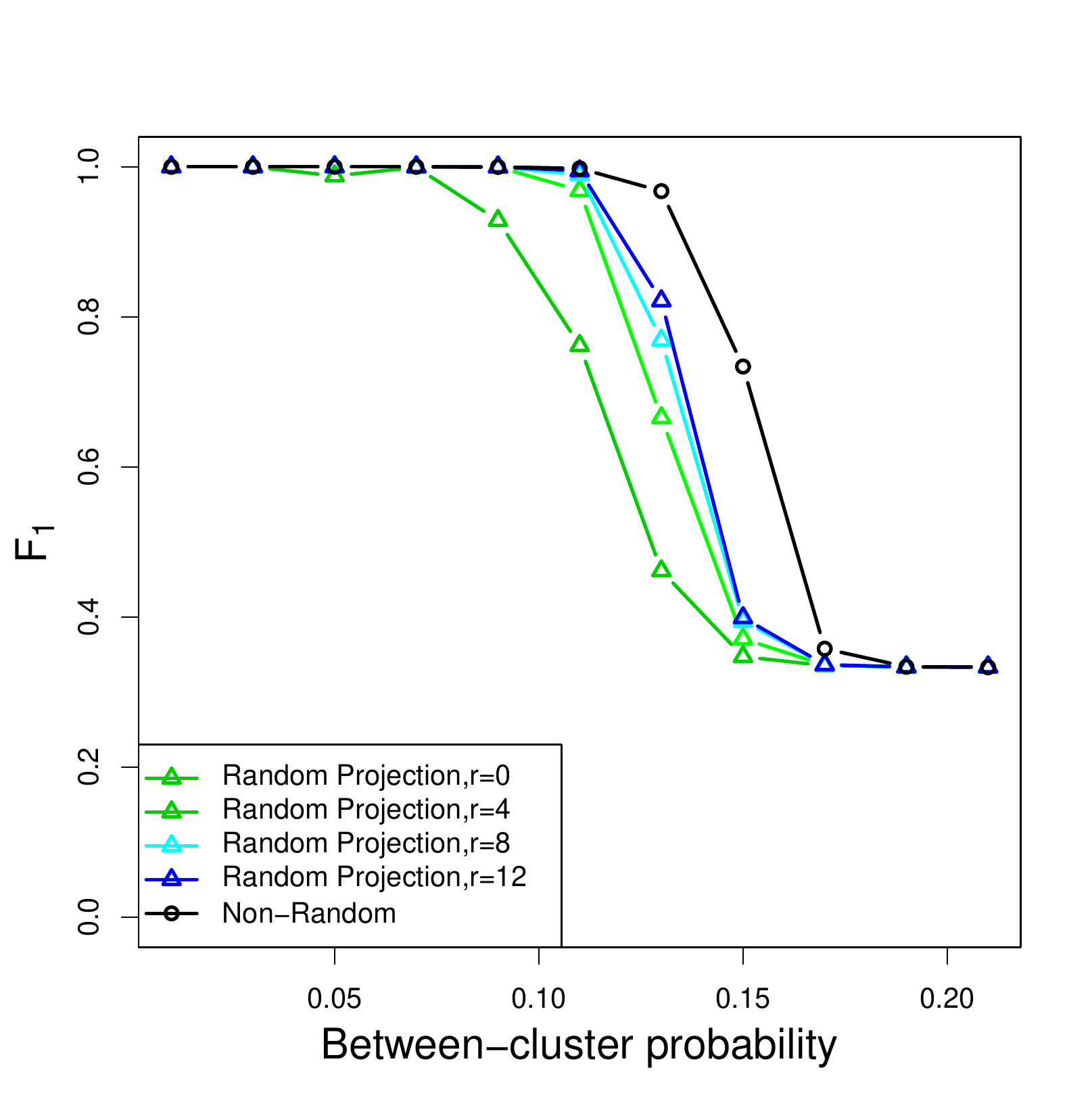}}
\subfigure[NMI]{\includegraphics[height=4.8cm,width=5cm,angle=0]{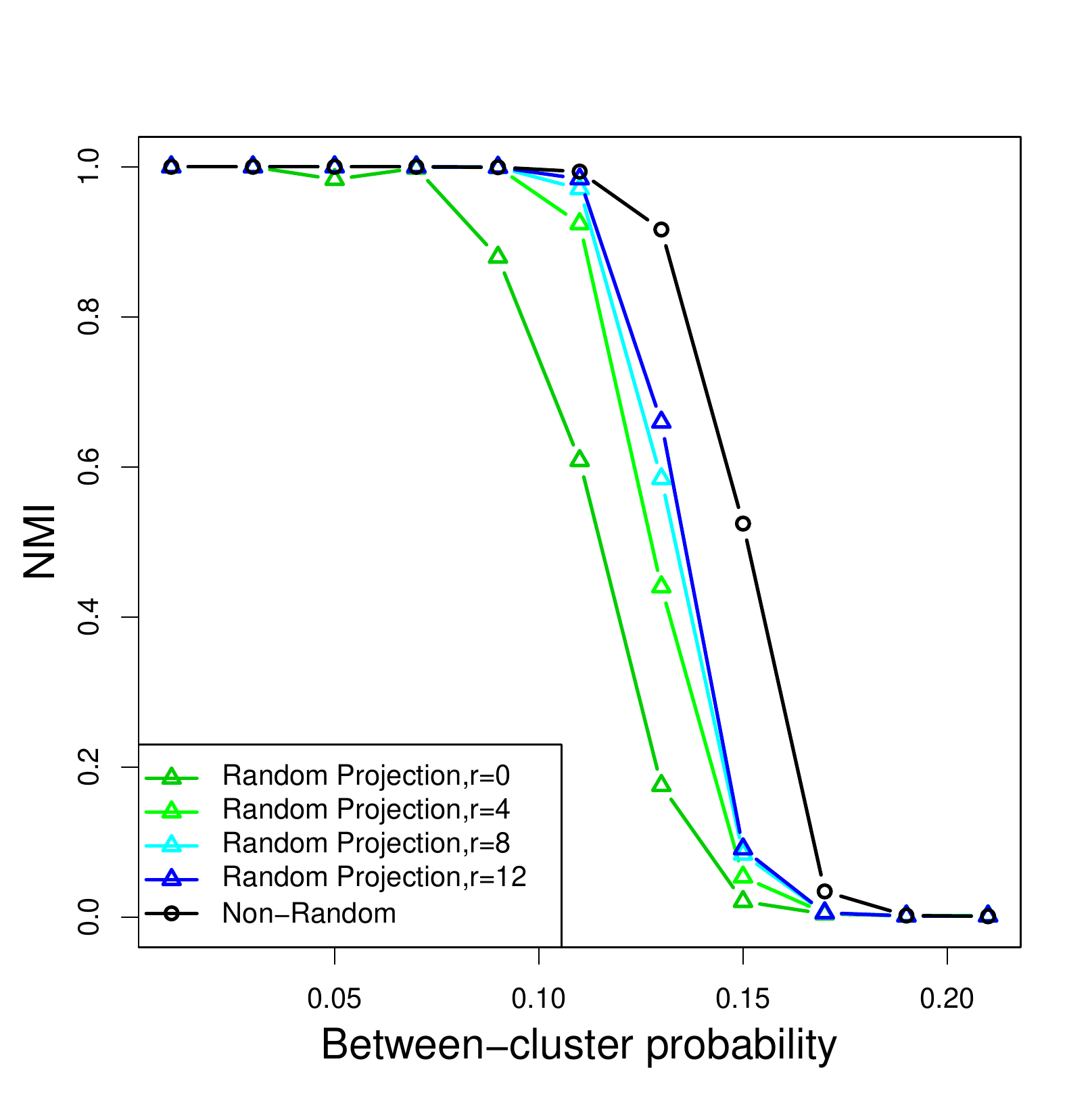}}
\subfigure[ARI]{\includegraphics[height=4.8cm,width=5cm,angle=0]{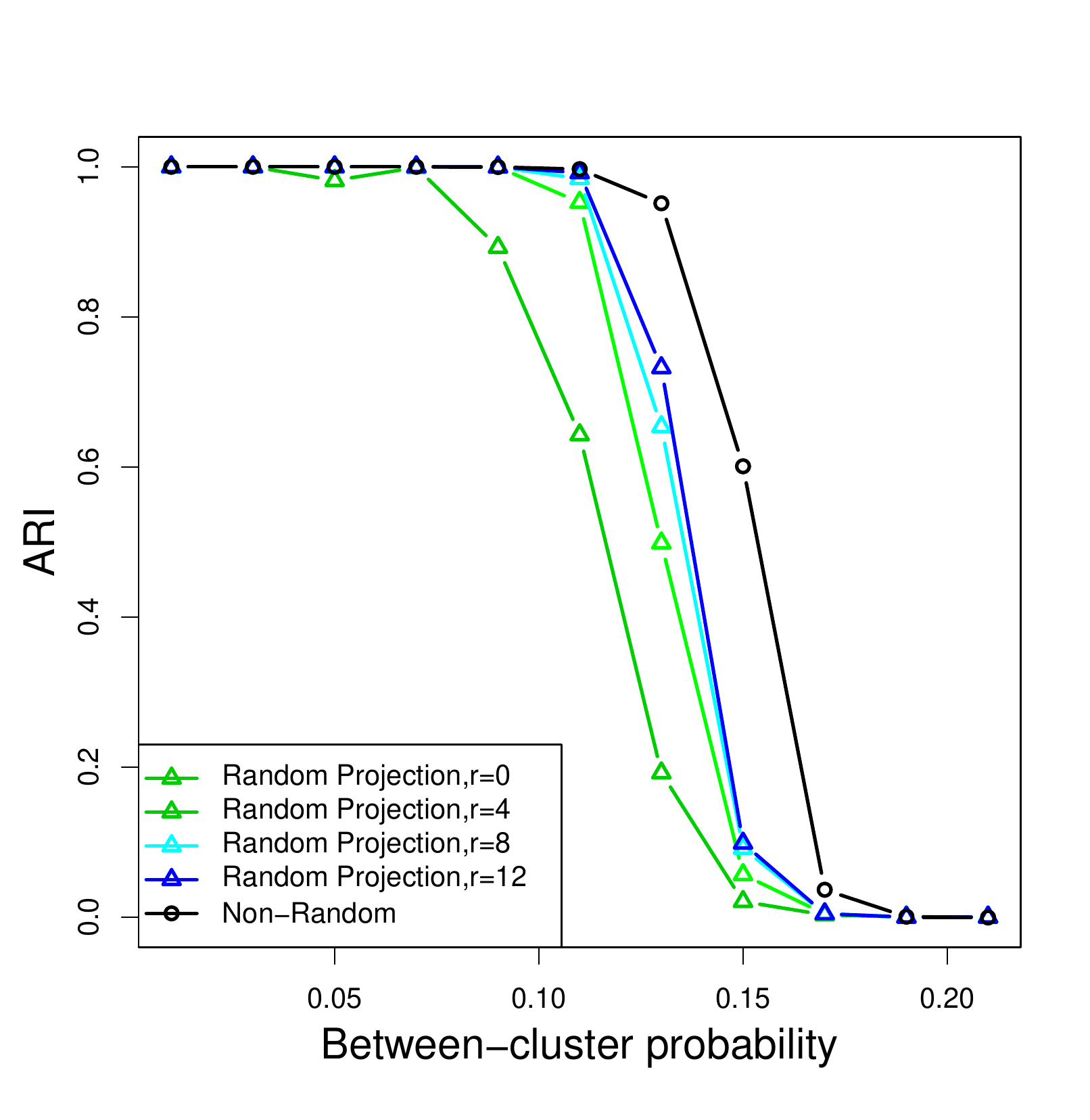}}\\
(II) Effect of the power parameter $q$\\
\subfigure[${\rm F}_1$]{\includegraphics[height=4.8cm,width=5cm,angle=0]{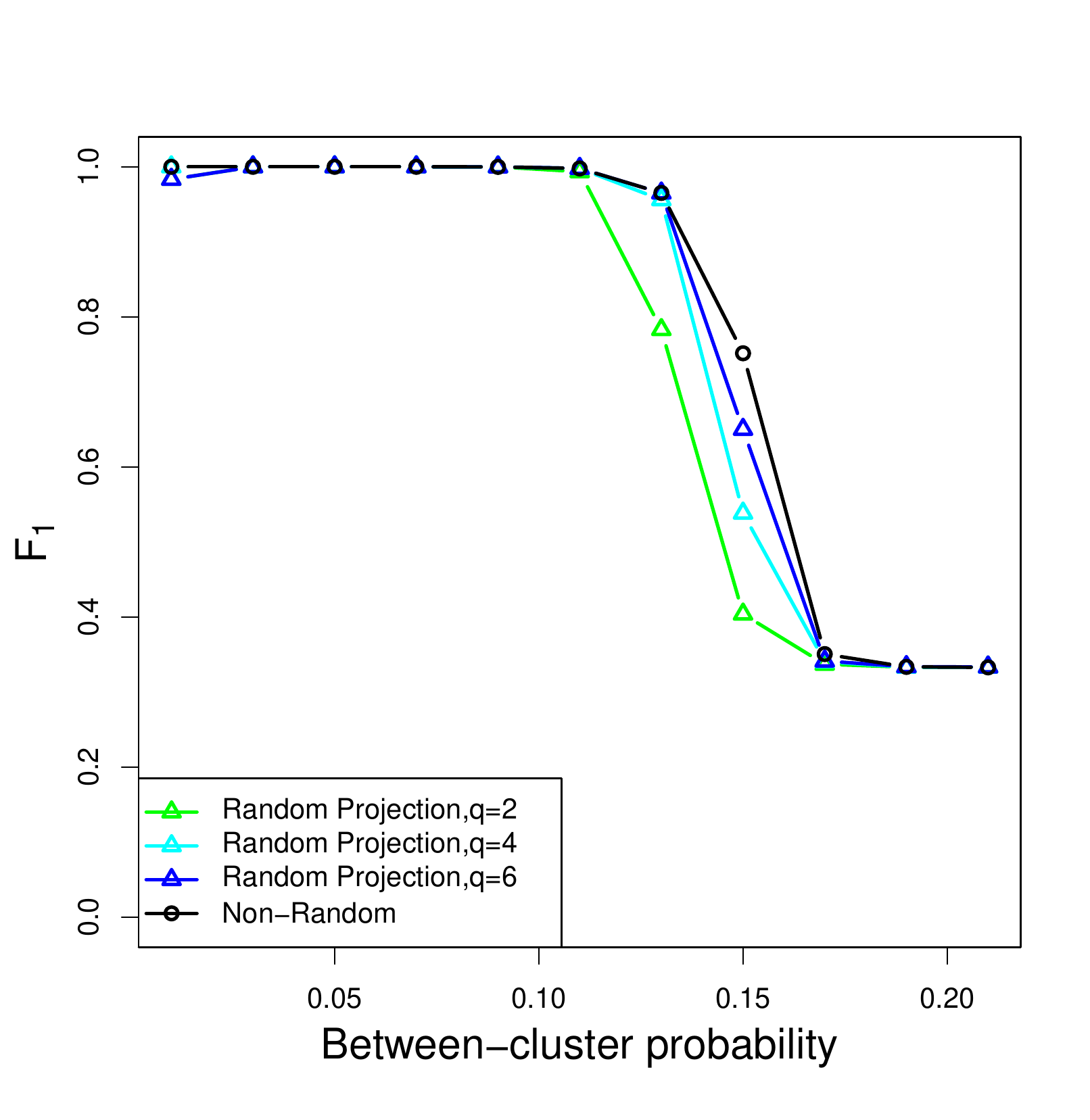}}
\subfigure[NMI]{\includegraphics[height=4.8cm,width=5cm,angle=0]{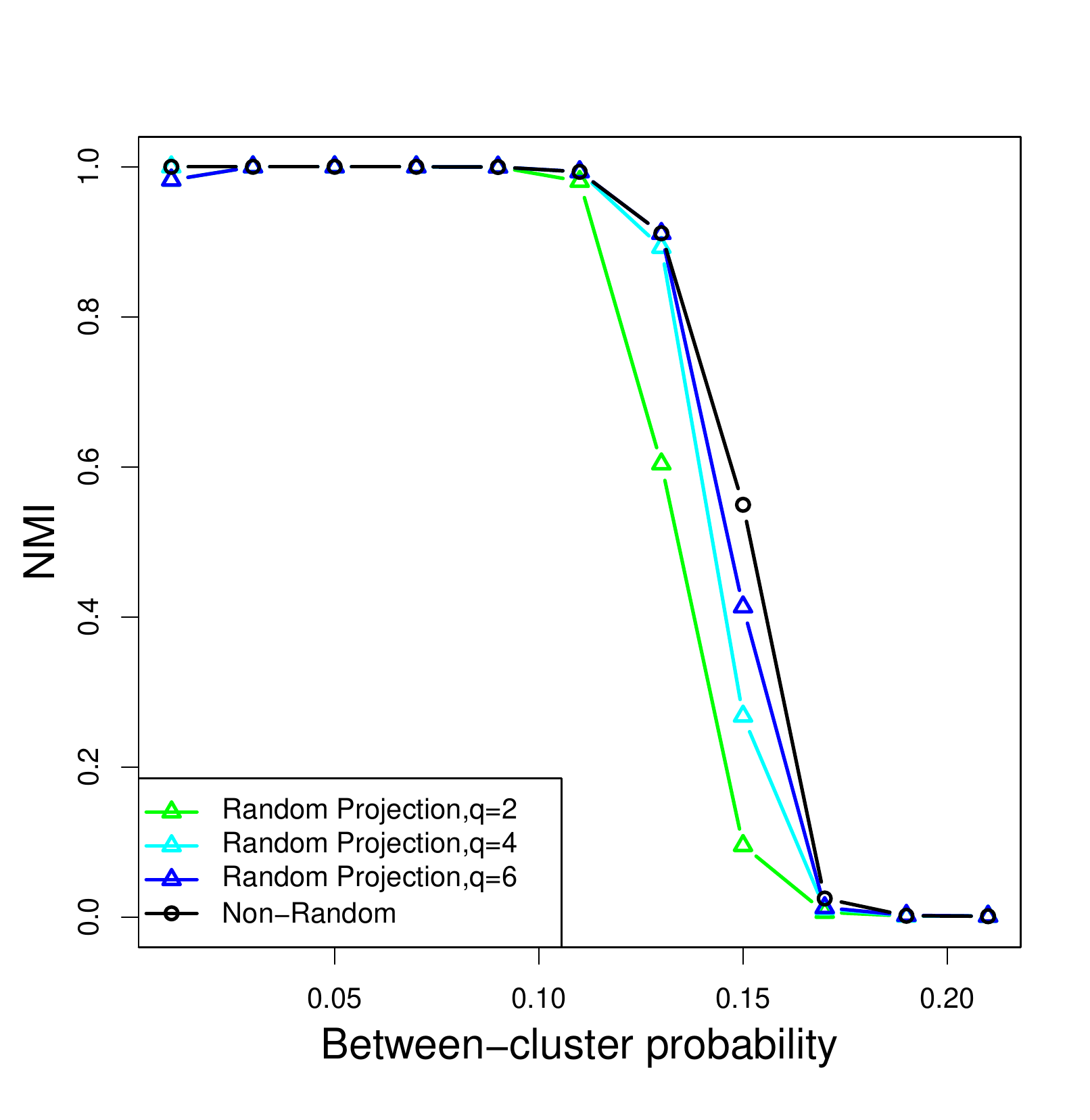}}
\subfigure[ARI]{\includegraphics[height=4.8cm,width=5cm,angle=0]{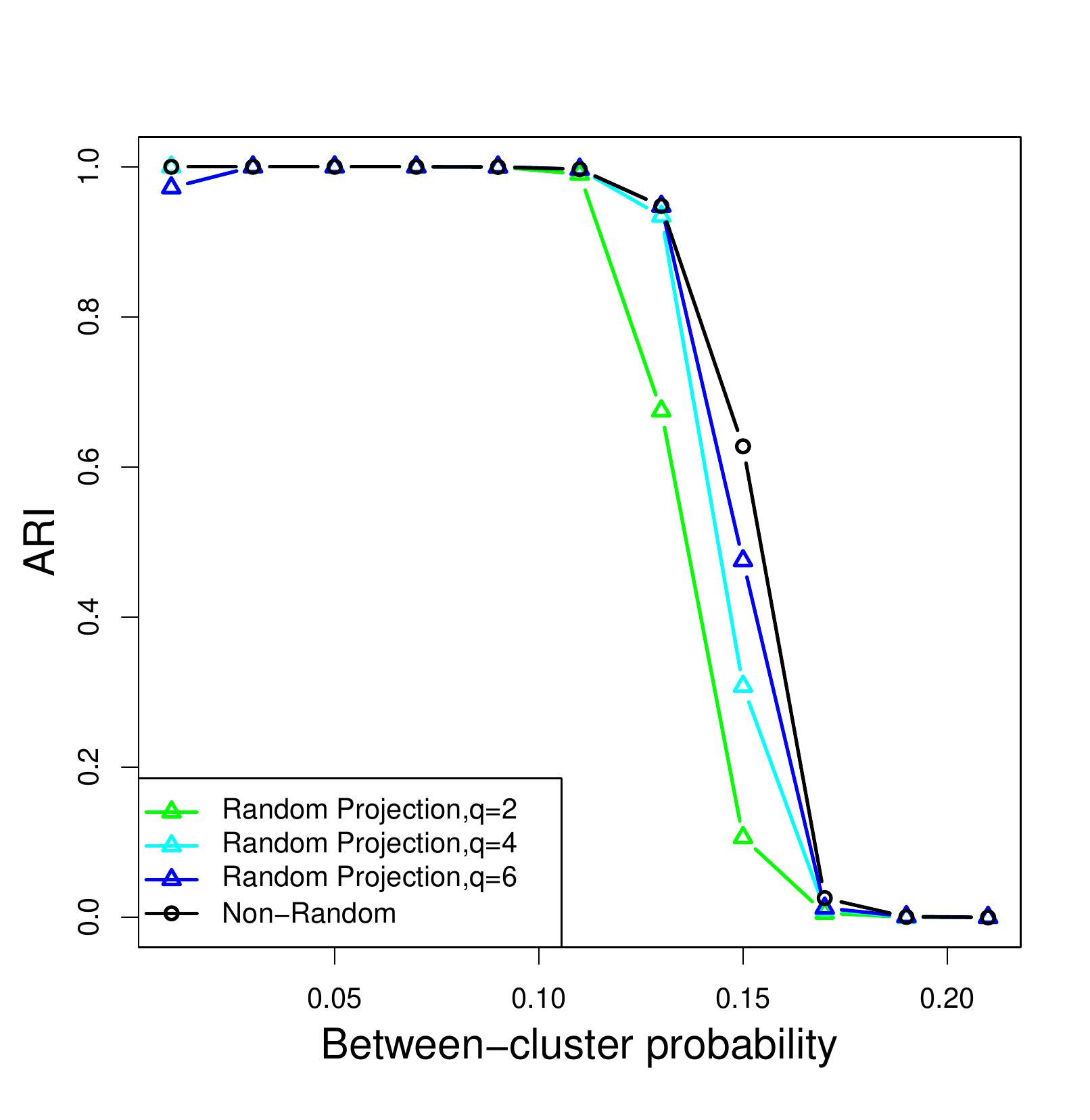}}\\
(III) Effect of the test matrix $\Omega$\\
\subfigure[${\rm F}_1$]{\includegraphics[height=4.8cm,width=5cm,angle=0]{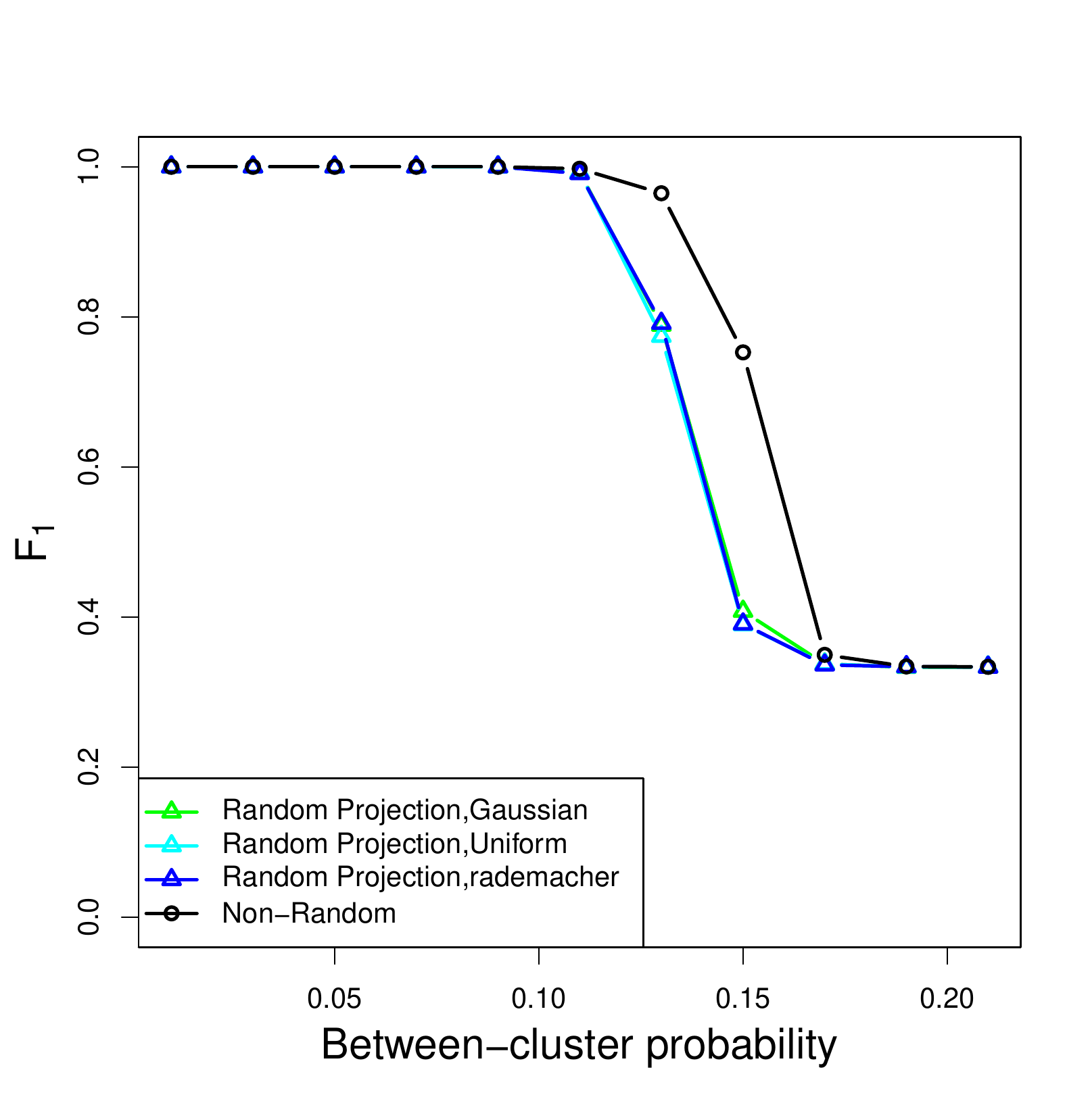}}
\subfigure[NMI]{\includegraphics[height=4.8cm,width=5cm,angle=0]{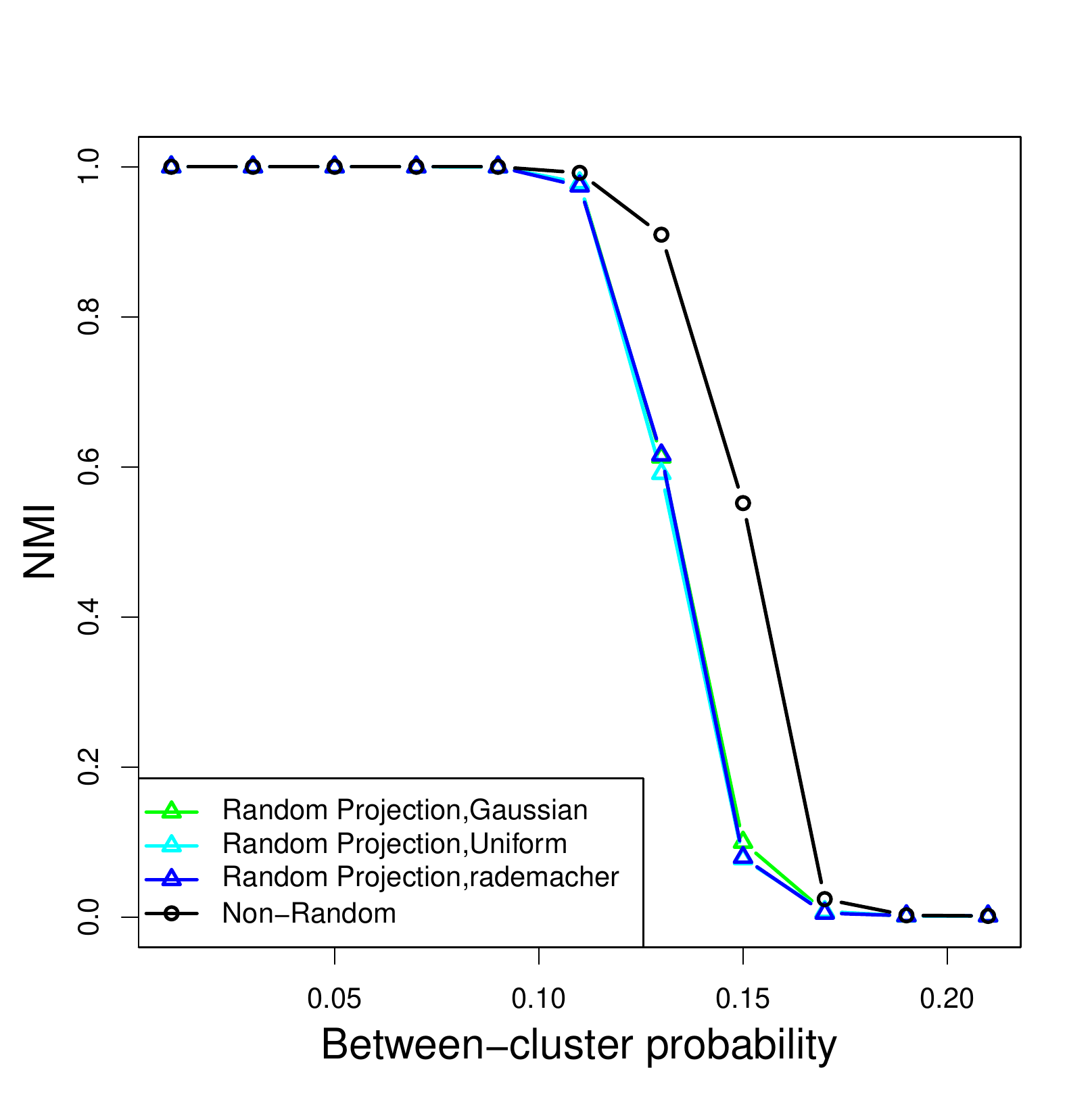}}
\subfigure[ARI]{\includegraphics[height=4.8cm,width=5cm,angle=0]{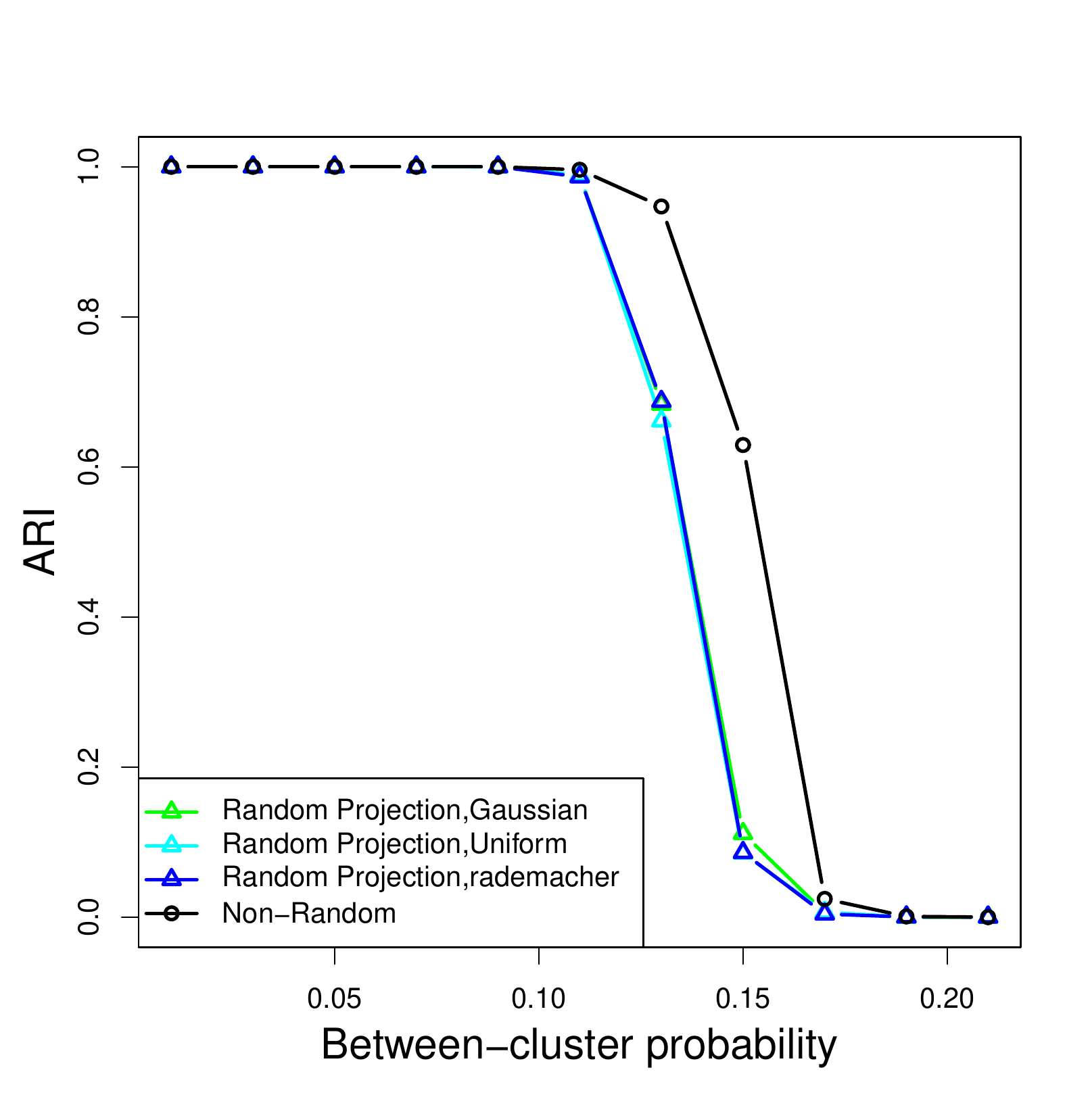}}
\caption{Effects of the parameters $r,q,\Omega$ in the random projection scheme. Each row corresponds to the effect of one parameter with the others fixed. Each column corresponds to a measure for the clustering performance. The other parameters are fixed at $n=1152$, $K=3$, and the within cluster probability $\alpha=0.2$. }\label{senrp}
\end{figure*}

\begin{figure*}[!htbp]{}
\centering
\subfigure[${\rm F}_1$]{\includegraphics[height=4.8cm,width=5cm,angle=0]{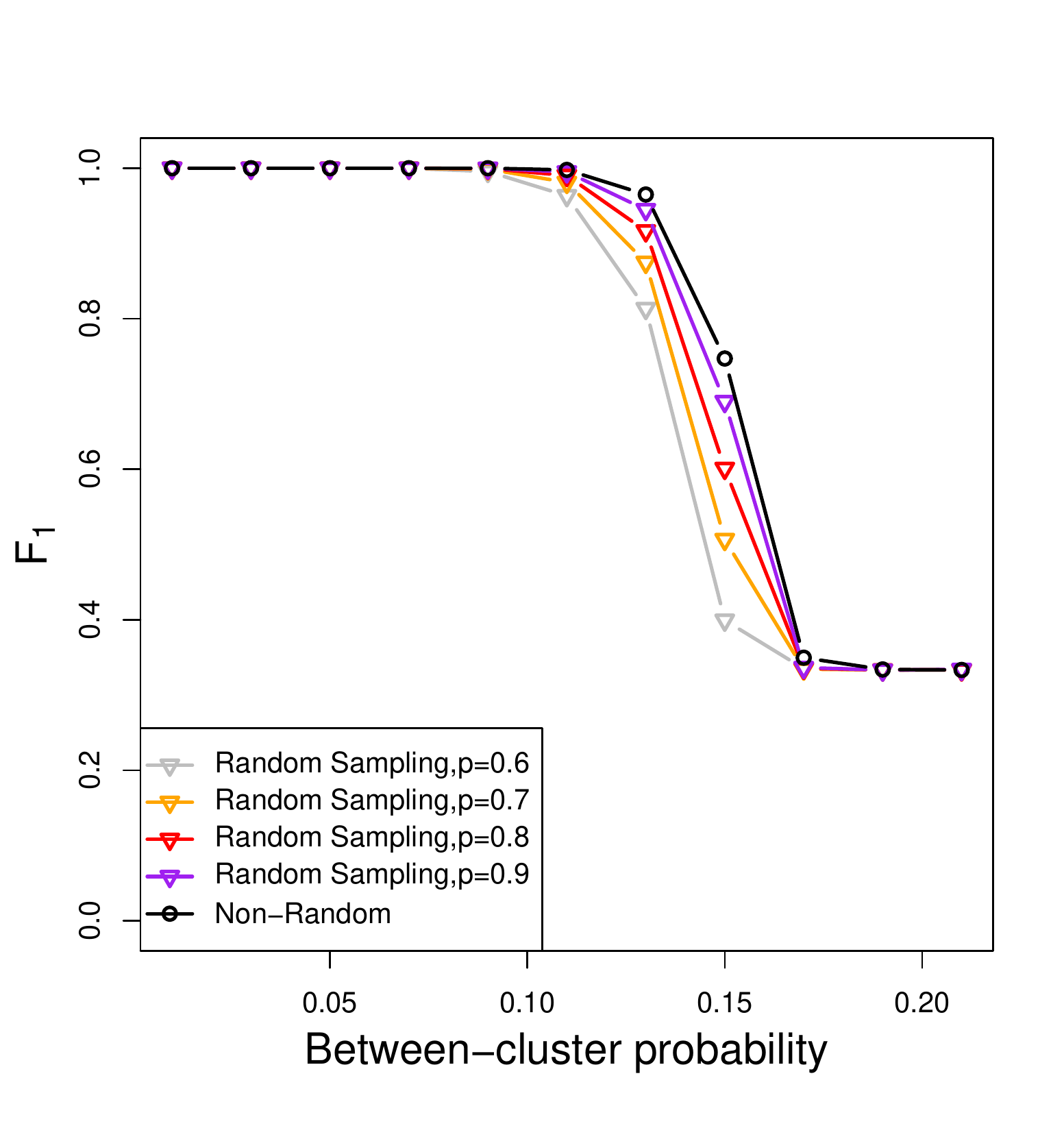}}
\subfigure[NMI]{\includegraphics[height=4.8cm,width=5cm,angle=0]{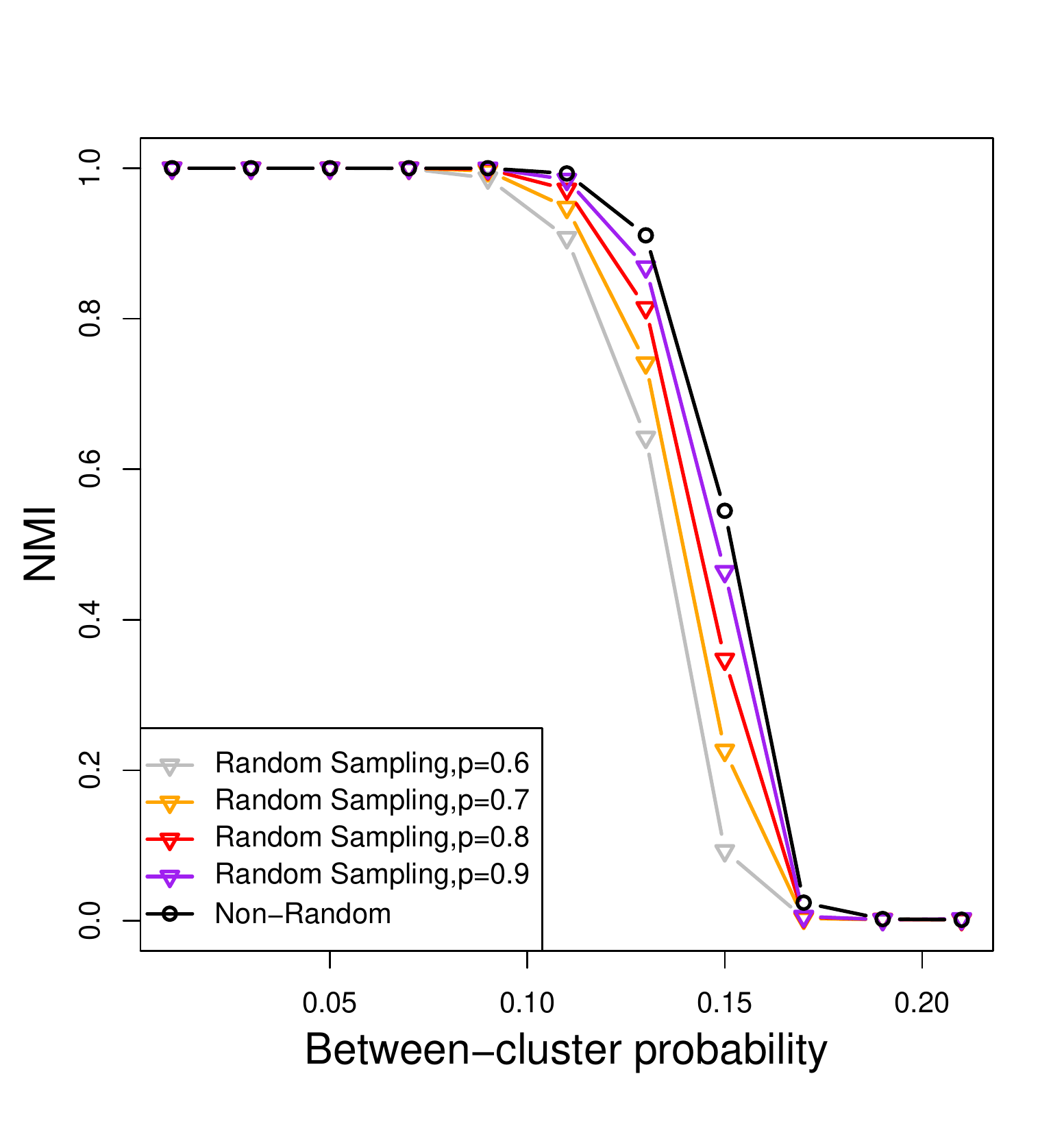}}
\subfigure[ARI]{\includegraphics[height=4.8cm,width=5cm,angle=0]{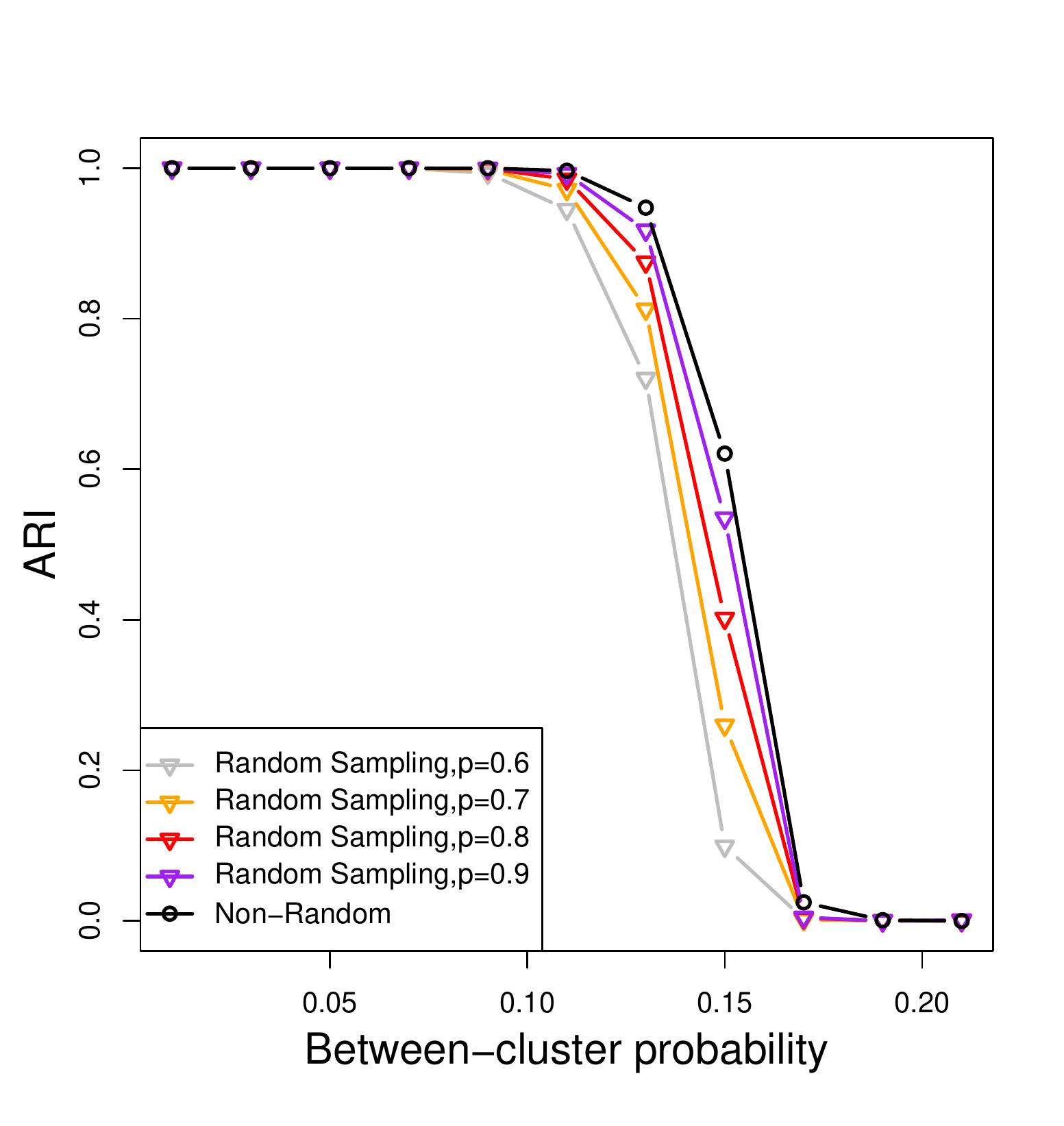}}\\
\caption{Effect of the parameters $p$ within the random sampling scheme. Each column corresponds to a measure for the clustering performance. The other parameters are fixed at $n=1152$, $K=3$, and the within cluster probability $\alpha=0.2$.  }\label{senrs}
\end{figure*}

To see how hyper parameters $r,q,p$ and the distribution of test matrix affect the performance of corresponding method, we here conduct another series of experiments. Specifically, to remove the computational cost of finding the best permutation matrix over the permutation matrix set, we use $\rm F_1$ score ($\rm F_1$), Normalized Mutual Information (NMI), and Adjusted Rand Index (ARI) \citep{hubert1985comparing,manning2010introduction} to justify the clustering performance of each method. These indexes measure the similarity of two clusters, and here we refer to the estimated and true clusters, from different perspectives. The larger these indexes, the better the clustering algorithm performs. The parameters were basically set as $n=1152$, $K=3$, and the within cluster probability $\alpha=0.2$. To see the effect of other parameters, we varied the oversampling parameter $r\in\{0, 4, 8, 12\}$, the power parameter $q\in\{2, 4, 6\}$, the sampling rate $p\in\{0.6, 0.7, 0.8, 0.9\}$, and the distribution of test matrix $\Omega$ was generated as i.i.d. Gaussian (standard), uniform (from -1 to 1), and Rademacher (take values $+1$ and $-1$ with equal probability). And for each setting, we let the between cluster probability vary. Figure \ref{senrp} and \ref{senrs} show the averaged results of the random projection scheme and the random sampling scheme, respectively. As expected, larger $r$, $q$ and $p$ lead to better clustering performance but at the cost of computational efficiency. One should choose these parameters according to the problem at hand. In addition, among the distribution of $\Omega$ we tested, it has little effect on the resulting clustering performance of random projection.

\section{Real data examples}
\label{sec:real}
In this section, we numerically evaluate the merits of randomized spectral clustering in terms of accuracy and efficiency. Specifically, we first compare the clustering accuracy of each method on four small-scale real networks, using the original spectral clustering as the baseline method. After that, we examine the computational efficiency as well as the relative error of each method on four large-scale networks, where we compare randomized methods with several iterative methods.

\subsection{Accuracy evaluation}
In this subsection, we test the effectiveness of randomized spectral clustering on four network datasets, including the European email network \citep{leskovec2007graph,yin2017local}, the political blog network \citep{adamic2005political}, the statistician coauthor network and the statistician citation network \citep{ji2016coauthorship}, where the first two datasets have ground truth community assignments and the last two have no ground truth community assignment. Table \ref{tablesmall} shows the basic statistics about the networks, where for the first two networks, the target rank is set as the number of true clusters, while for the last two networks, the target rank follows \citet{ji2016coauthorship}. For the datasets with ground truth labels, we computed $\rm F_1$, NMI, and ARI
\citep{hubert1985comparing,manning2010introduction} between the estimated clusters and the true clusters for each of the three methods, namely, the random projection, the random sampling, and the original spectral clustering, respectively. While for the datasets without ground truth labels, we computed $\rm F_1$, NMI, and ARI between the clusters estimated by the randomized spectral clustering and the clusters estimated by the original spectral clustering. Our aim is to show that randomized algorithms perform comparably to the original spectral clustering. Hence for the datasets with ground truth labels, the smaller gap of $\rm F_1$, NMI, and ARI between randomized and original spectral clustering indicate the better match between these methods. While for the datasets without ground truth labels, larger $\rm F_1$, NMI, and ARI indicate the better match. For the random projection scheme, the oversampling parameter $r=10$, the power parameter $q=2$, and the random test matrix has i.i.d. Gaussian entries. And for the random sampling scheme, we test two cases, namely, $p=0.7$ and $0.8$. Table \ref{table} summarizes the average performance of these methods over 20 replications with the standard deviations in the parentheses. From Table \ref{table} we see that all the methods perform very similarly to each other in terms of $\rm F_1$, NMI, and ARI, and the results are rather stable, which shows the effectiveness of randomized methods.

\begin{table*}[!htbp]
\centering
\footnotesize
\caption{ A summary of the four small-scale undirected networks.
}\vspace{0.5cm}
\def\arraystretch{1.5}
\begin{tabular}{lccc}
\hline
{Networks}&{No. of nodes}&{No. of edges}&{Target rank}\\
\hline
European email network&986&16,064&42\\
Political blog network& 1,222&16,714&2\\
Statisticians coauthor network&2,263&4,388&3\\
Statisticians citation network&2,654&20,049&3\\

\hline
\end{tabular}
\label{tablesmall}
\end{table*}

%\begin{figure*}[!htbp]{}
%\centering
%\subfigure[European email network]{\includegraphics[height=8cm,width=8cm,angle=0]{email_clustertrue22.pdf}}
%\subfigure[Political blog network]{\includegraphics[height=8cm,width=8cm,angle=0]{blog_clustertrue22.pdf}}
%\subfigure[Statisticians coauthor network]{\includegraphics[height=8cm,width=8cm,angle=0]{coauthor22.pdf}}
%\subfigure[Statisticians citation network]{\includegraphics[height=8cm,width=8cm,angle=0]{citation22.pdf}}
%\caption{The four networks used in the real data analysis. Each network corresponds to the largest connected components of the original networks. (a) The European email network with $n=986$ and $K=42$. (b) The political blog network with $n=1222$ and $K=2$. (c) The statisticians coauthor network with $n=2263$. (d) The statisticians citation network with $n=2654$.}\label{network}
%\end{figure*}

\begin{table}[!htbp]
\centering
\footnotesize
\caption{The clustering performance of each method on four real network datasets based on randomized spectral clustering algorithms. For the European email network and political blog network, the performance is evaluated based on a known ground truth. For the statisticians coauthor citation networks, the performance is evaluated based on the original spectral clustering.
}\vspace{1cm}
\def\arraystretch{1.5}
\begin{tabular}{lccc}
\hline
Methods&$\rm F_1$ &NMI&ARI\\
\hline
(a) European email network\\
Random Projection&0.165(0.007)&0.558(0.006)&0.100(0.009)\\
Random Sampling ($p=0.7$)&0.126(0.007)&0.417(0.010)&0.059(0.008)\\
Random Sampling ($p=0.8$)&0.131(0.005)&0.436(0.010)&0.064(0.006)\\
Non-Random&0.154(0.006)&0.571(0.005)&0.088(0.007)\\
(b) Political blog network\\
Random Projection&0.641(0.004)&0.178(0.004)&0.079(0.006)\\
Random Sampling ($p=0.7$)&0.642(0.003)&0.177(0.007)&0.077(0.007)\\
Random Sampling ($p=0.8$)&0.641(0.004)&0.177(0.008)&0.077(0.009)\\
Non-Random&0.641(0.004)&0.178(0.004)&0.079(0.006)\\
(c) Statisticians coauthor network (No true labels)\\
Random Projection (relative)&0.981(0.012)&0.646(0.197)&0.715(0.246)\\
Random Sampling (relative) ($p=0.7$)&0.970(0.011)&0.480(0.148)&0.593(0.193)\\
Random Sampling (relative) ($p=0.8$)&0.973(0.011)&0.544(0.142)&0.639(0.190)\\
(d) Statisticians citation network (No true labels)\\
Random Projection (relative)&0.990(0.021)&0.881(0.166)&0.926(0.140)\\
Random Sampling (relative) ($p=0.7$)&0.981(0.019)&0.759(0.125)&0.863(0.120)\\
Random Sampling (relative) ($p=0.8$)&0.981(0.022)&0.770(0.163)&0.861(0.149)\\
\hline
\end{tabular}
\label{table}
\end{table}

\subsection{Efficiency evaluation}
In this subsection, we examine the computational efficiency of randomized methods for partial eigenvalue decomposition on four large-scale real undirected networks, including
DBLP collaboration network, Youtube social network, Internet topology network, LiveJournal social network \citep{yang2015defining,leskovec2005graphs}. These four networks are large-scale with up to millions of nodes and tens of millions of edges. Table \ref{table1} shows the basic statistics about the networks, where the target rank corresponds to a network is $k$ if there exits a large gap between the $k$-th and $(k+1)$-th largest (in absolute value) approximated eigenvalues. We compare the performance of our methods with iterative methods, including the implicitly restarted Lanczos algorithm \citep{calvetti1994implicitly} (\textsf{svds} in R package \textsf{RSpectra} \citep{rspectra}), the augmented implicitly restarted Lanczos bidiagonalization algorithms \citep{baglama2005augmented} (\textsf{irlba} and \textsf{partial\_eigen} in R package \textsf{irlba} \citep{irlba}). In addition, we also compare our implementation with the randomized methods implemented in \textsf{svdr} in R package \textsf{irlba} \citep{irlba}. Note that the full eigenvalue decomposition always fails in such large-scale data setting.

Table \ref{table2} shows the median computational time of each method over 20 replications, where all computations are done on a machine with {Intel Core i9-9900K CPU 3.60GHz, 32GB memory, and 64-bit WS operating-system, and R version 4.0.4} is used for all computations. For the random projection-based method, the power parameter is 2 and the oversampling parameter is 10. For the random sampling-based method, the sampling probability is 0.7. We can see from Table \ref{table2} that the our methods shows great advantage over compared methods especially when the network scale is large. In particular, the random sampling-based method is efficient no matter the sampling time is included or not.

\begin{figure*}[!htbp]{}
\centering
\subfigure[DBLP, $K=3$]{\includegraphics[height=3.2cm,width=5cm,angle=0]{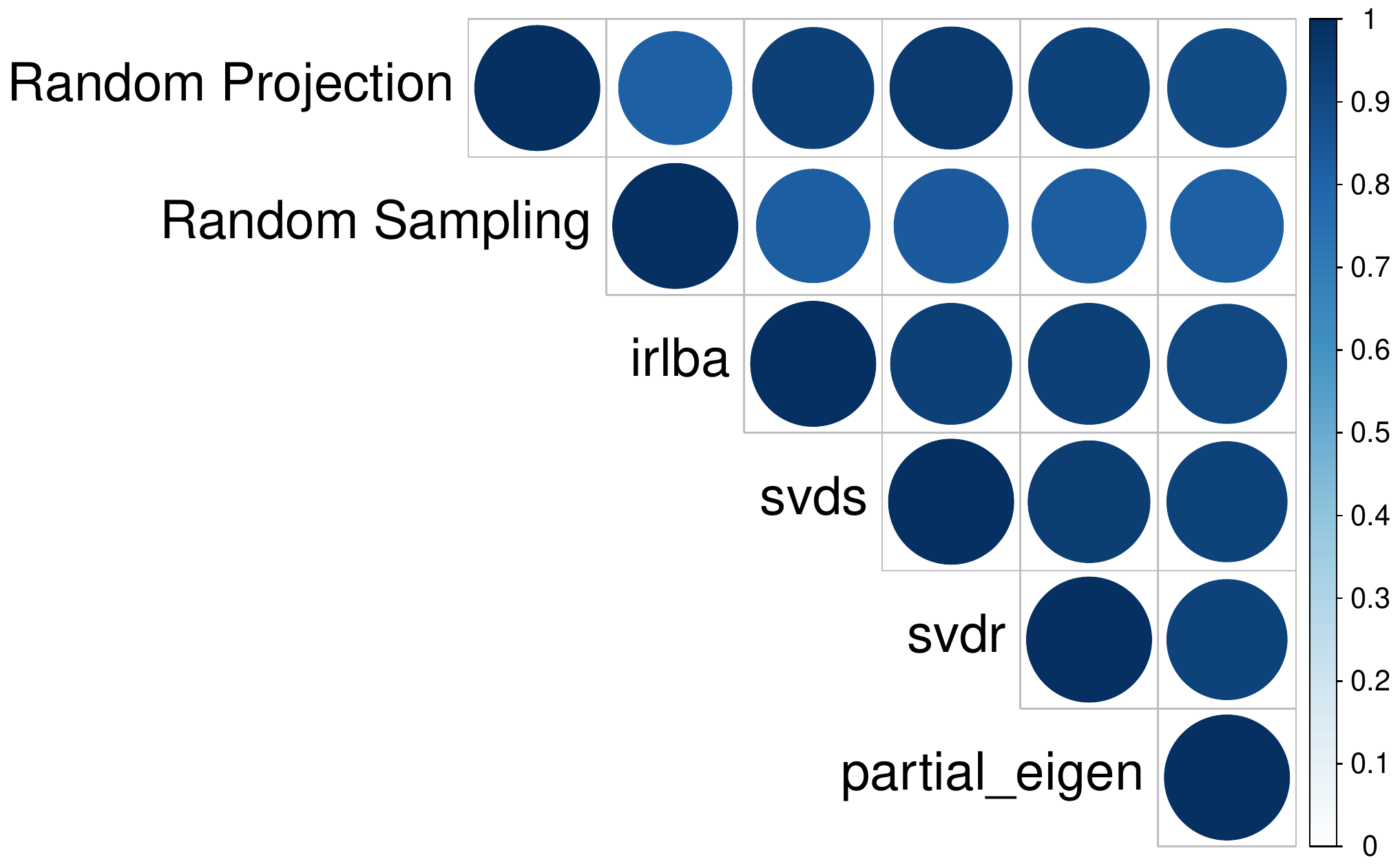}}\hspace{1cm}
\subfigure[Youtube, $K=7$]{\includegraphics[height=3.2cm,width=5cm,angle=0]{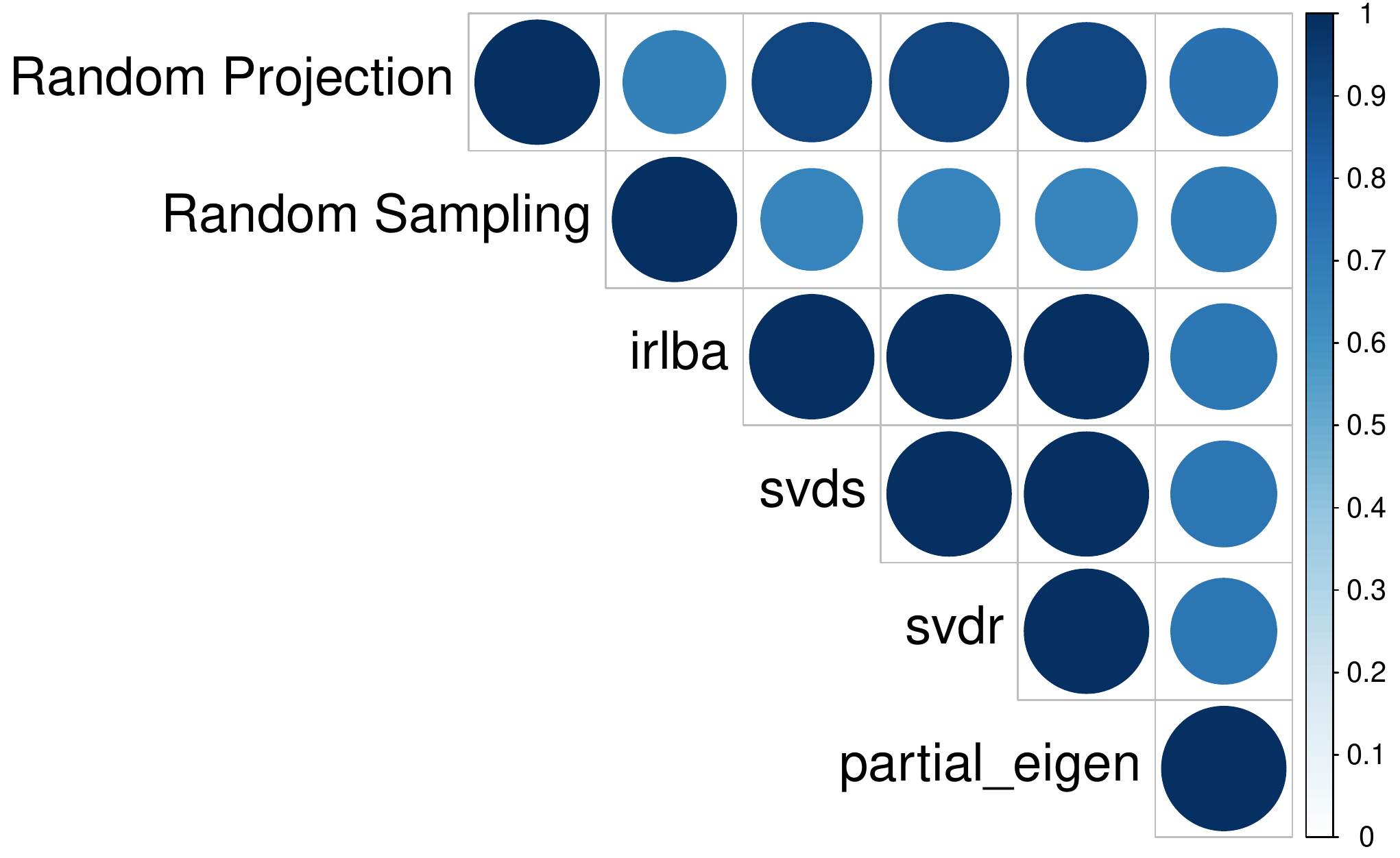}}\\
\subfigure[Internet,$K=4$]{\includegraphics[height=3cm,width=5cm,angle=0]{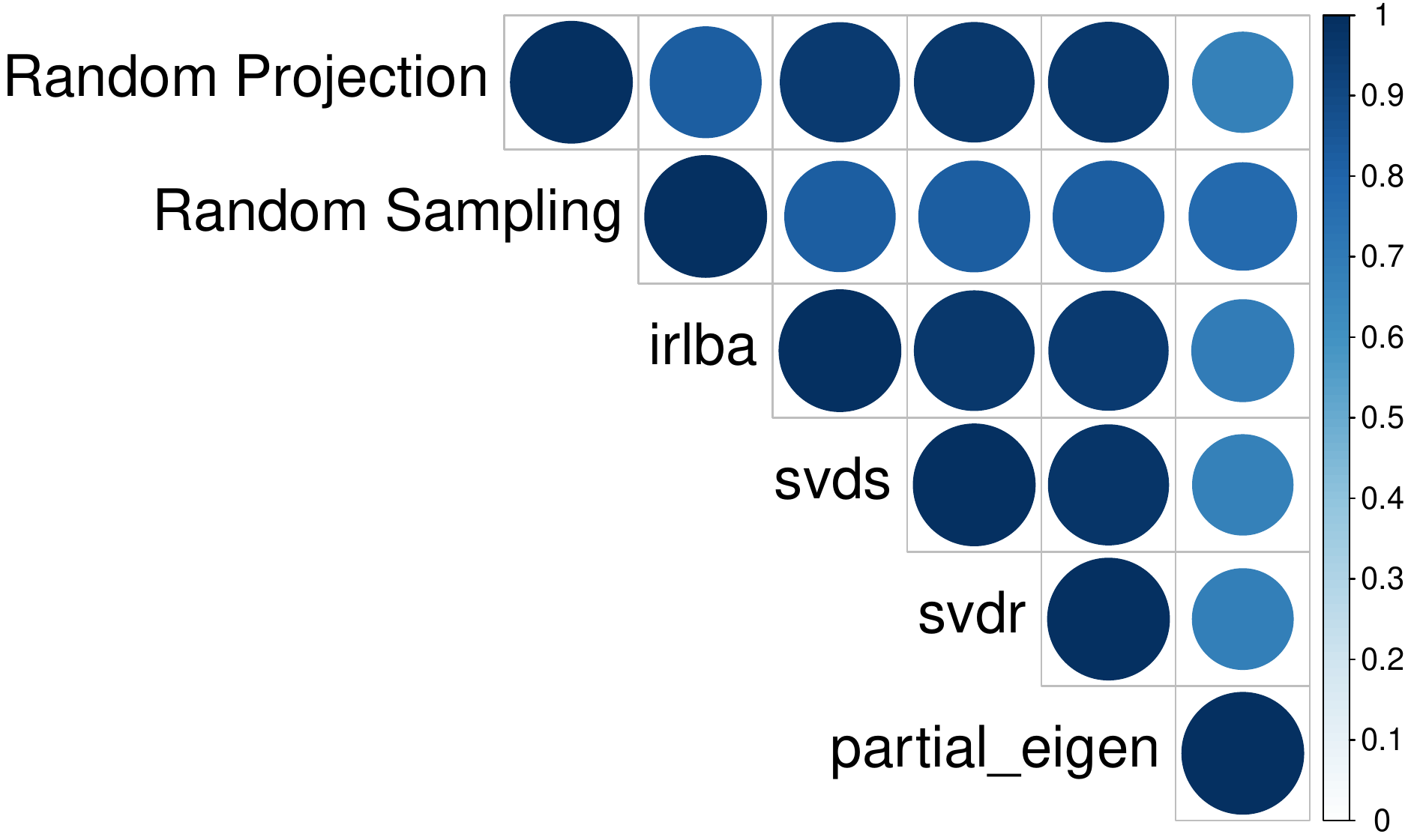}}\hspace{1cm}
\subfigure[LiveJournal, $K=4$]{\includegraphics[height=3cm,width=5.1cm,angle=0]{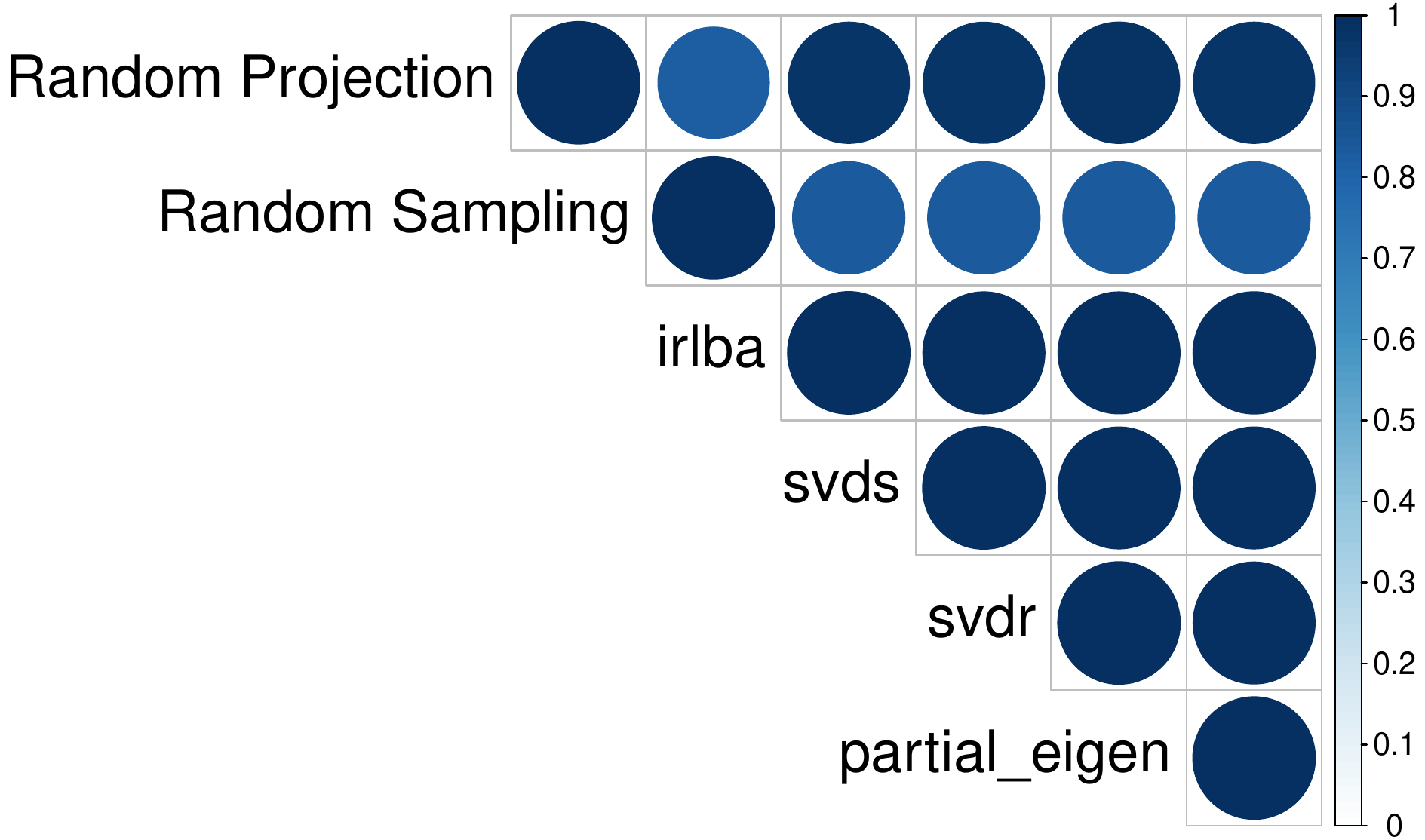}}\\

\caption{The pairwise comparison of the clustering results of six methods on four large-scale networks. The relative clustering performance are measured by ARI. Larger ARI, i.e., {larger circles} in the figure, indicates that the clustering results of the two methods are more close.}\label{comparari}
\end{figure*}

Figure \ref{comparari} shows the pairwise comparison of the clustering results of six methods on these four networks. The relative clustering performance are measured by ARI. It turns out that the random projection-based method and the random sampling-based method yield similar results with other compared methods, though the random sampling-based method seems to behave slightly different.

Overall, as indicated by our theory and experiments, the randomized methods could bring high efficiency while slightly sacrificing the accuracy. In real world applications, one should balance the accuracy-efficiency trade-off via selecting appropriate hyper parameters according to the real setting.

\begin{table*}[!htbp]
\centering
\footnotesize
\caption{ A summary of the four large-scale undirected networks.}\vspace{0.5cm}
\def\arraystretch{1.5}
\begin{tabular}{lccc}
{Networks}&{No. of nodes}&{No. of edges}&{Target rank}\\
\hline
DBLP collaboration network&317,080&1,049,866&3\\
Youtube social network& 1,134,890&2,987,624&7\\
Internet topology graph&1,696,415&11,095,298&4\\
LiveJournal social network&3,997,962&34,681,189&4\\

\hline
\end{tabular}
\label{table1}
\end{table*}

\begin{table*}[!htbp]
\centering
\footnotesize
\caption{ Median time (seconds) of each method for computing the (approximated) eigenvectors of four real network adjacency matrices over 20 replications, where for the random sampling, the time with the sampling time included and excluded (shown in the parentheses) are reported, respectively.
}
\vspace{0.5cm}
\def\arraystretch{1.5}
\begin{tabular}{p{1.6cm}p{1.6cm}p{3cm}p{1.6cm}p{1.6cm}p{1.6cm}p{1.6cm}}
\hline
{Networks}&Random projection&{Random sampling}&{\textsf{irlba}}&\textsf{svds}&\textsf{svdr}&\textsf{partial\_eigen}\\
\hline
DBLP &0.369&0.280(0.248)&0.341&0.411&6.132&0.346\\
Youtube & 2.037&2.302(2.204)&2.487&3.043&35.595&9.111  \\
Internet &2.773&2.072(1.774)&3.404&3.332&30.900&7.706\\
LiveJournal &13.213&7.207(6.216)&15.179&20.077&106.166&15.080\\
\hline
\end{tabular}
\label{table2}
\end{table*}

%\newpage
\section{Conclusion}
\label{sec:dissc}
In this paper, we used randomized sketching techniques to accelerate the spectral clustering when facing large-scale networks, say networks with millions of nodes, and studied how well the resulting algorithms perform under the SBMs and DC-SBMs. We studied two randomized spectral clustering algorithms. The first one is random projection-based, which reduces the computational cost by projecting the columns and rows of the adjacency matrix to a lower-dimensional space. The second one is random sampling-based, which samples the edges to obtain a sparsified adjacency matrix, and thus reducing the computational cost of partial eigen-decomposition. In the framework of SBMs, we studied these two randomized spectral clustering algorithms in terms of the approximation error that measures the deviation of the randomized adjacency matrix $\tilde{A}$ from the population matrix $P$, the misclassification error that measures the fraction of the number of mis-clustered nodes over the total number of nodes, and {the estimation error} for the link probability matrix $B$. In particular, we considered a {more generalized content} where ${\rm rank}(B)$ could be smaller than the cluster number $K$. Under mild conditions, the approximation error turns out to be statistically optimal, which shows that the randomized matrix behaves as if it was sampled from the SBM. We also extend theoretical results to DC-SBMs. Experimental results showed the merits of randomized spectral clustering on networks with up to millions of nodes. For practical convenience, we developed an R package \textsf{Rclust}.

There are many ways that the content in this paper can be extended. First, we studied the weak consistency of the pure spectral clustering without any regularization or refinement, and we mainly used the Davis-Kahan theorem to study the eigenvector perturbation. There exist several works on trimming or refining the pure spectral clustering to help the refined spectral clustering achieve the information-theoretic limit of the exact recovery (strong consistency) or minimax optimal rate of the partial recovery (weak consistency) under SBMs; see \citet{gao2017achieving,yun2016optimal}, among others. It would be interesting to study whether one could use similar treatments on the randomized spectral clustering in order to improve its theoretical performance while without increasing the time cost. On the other hand, a few works study the entry-wise perturbation of eigenvectors very recently; see \citet{cape2019signal,tang2021asymptotically,abbe2020entrywise,su2019strong}, among others. It would be important and insightful to study the entry-wise perturbation of eigenvectors after randomization, and also study the related strong consistency in SBMs. Second, although the approximation error is minimax optimal under SBMs, it would be important to develop advanced randomization techniques in order to weaken the condition on $\alpha_n$, $q$, $p_{ij}$'s. In addition, we mainly focused on the adjacency matrix sampled from the SBMs. It would be interesting to generalize the results to the Laplacian matrix and other network generating models, say--the latent space model, and the graphon models, among others. Finally, it would be important to study the estimation of the model parameters $K$ and ${\rm rank}(B)$ \citep{fishkind2013consistent,ma2021determining}.
\begin{center}
{\large ACKNOWLEDGMENT}
\end{center}
We are grateful to the editor, associate editor, and two reviewers for their consideration of our paper and for their helpful suggestions, which led to an improved version of this paper. We also thank Professor Yixuan Qiu for his great help in developing the R package.
Our research is partially supported by National Natural Science Foundation of China (No.U1811461), National Natural Science Foundation for Outstanding Young Scholars (No.72122018), and Natural Science Foundation of Shaanxi Province (No. 2021JQ-429 and No.2021JC-01).

\bibliographystyle{plainnat}
\bibliography{randspectral}

\pagestyle{empty}
%\clearpage
\newpage
\begin{center}
{\large\bf Supplemental Material}
\end{center}

\subsection*{Proof of Lemma \ref{lem:eigen}}
Denote ${\rm rank}(B)=K'$ and define $\Delta={\rm diag}(\sqrt{n_1},...,\sqrt{n_{K}})$. It is easy to see that $\Theta\Delta^{-1}$ has orthogonal columns. Write $P$ as
\begin{equation}
\phantomsection
\label{A.1}P=\Theta B\Theta^\intercal=\Theta\Delta^{-1}\Delta B\Delta\Delta^{-1}\Theta^\intercal=\Theta\Delta^{-1}LDL^\intercal\Delta^{-1}\Theta^\intercal,\tag{A.1}
\end{equation}
where we denote the eigenvalue decomposition of $\Delta B\Delta $ by $L_{K\times K'}D_{K'\times K'}L^\intercal_{K'\times K}$. Recall that the eigenvalue decomposition of $P$ is $U_{n\times K'}\Sigma_{K'\times K'} U^\intercal_{K'\times n}$ and note that $\Theta\Delta^{-1}L$ has orthogonal columns, so we obtain ${\Sigma}=D$ and
\begin{equation}
\phantomsection
\label{A.2}{U}=\Theta\Delta^{-1}L.\tag{A.2}
\end{equation}

Next we discuss the structure of $U$ respectively when $B$ is of full rank ($K'=K$) and rank deficient ($K'<K$).

(a) When $K'=K$, $\Delta^{-1}L$ is invertible. Thus in this case, $U_{i\ast}=U_{j\ast}$ if and only if $\Theta_{i\ast}=\Theta_{j\ast}$. Moreover, we can verify easily that $\Delta^{-1}L$ has perpendicular rows and the $k$th row has length $\sqrt{1/n_k}$, which indicate
  $\|{U}_{i\ast}-{U}_{j\ast}\|_2=\sqrt{(n_{g_i})^{-1}+(n_{g_j})^{-1}}.$

(b) When $K'<K$, $\Delta^{-1}L$ is not invertible. In this case, $\Theta_{i\ast}=\Theta_{j\ast}$ can imply $U_{i\ast}=U_{j\ast}$ by (\ref{A.2}).
On the other hand, by (\ref{A.2}), we have $\|{U}_{i\ast}-{U}_{j\ast}\|_2:=\|\frac{L_{g_{i}\ast}}{\sqrt{n_{g_i}}}-\frac{L_{g_{j}\ast}}{\sqrt{n_{g_j}}}\|_2.$ Hence if the rows of $\Delta^{-1}L$ are mutually distinct with their minimum Euclidean distance being larger than a deterministic sequence $\{\xi_n\}_{n\leq 1}$, then $\|{U}_{i\ast}-{U}_{j\ast}\|_2\geq \xi_n$ for any $\Theta_{i\ast}\neq\Theta_{j\ast}$.
\QEDA

\subsection*{Proof of Lemma \ref{lem:eigencondition}}
Suppose $g_i=k$ and $g_j=l$ ($l\neq k$), and recall that $B=\Delta^{-1}L_{K\times K'}D_{K'\times K'}L^\intercal_{K'\times K}\Delta^{-1}$, where $D=\Sigma$ and $P=U\Sigma U^\top$. Then we have
\begin{align}
\phantomsection
\label{A.3}
\iota_n\|{U}_{i\ast}-{U}_{j\ast}\|_2^2&=\sum_{k_1=1}^{K'} \iota_n (\frac{L_{kk_1}}{\sqrt{n_{k}}}-\frac{L_{lk_1}}{\sqrt{n_{l}}})^2\nonumber \\
&\geq \sum_{k_1=1}^{K'} D_{k_1k_1}(\frac{L_{kk_1}}{\sqrt{n_{k}}}-\frac{L_{lk_1}}{\sqrt{n_{l}}})^2\nonumber \\
&=\sum_{k_1=1}^{K'} D_{k_1k_1}(\frac{L_{kk_1}}{\sqrt{n_{k}}})^2  + \sum_{k_1=1}^{K'} D_{k_1k_1}(\frac{L_{lk_1}}{\sqrt{n_{l}}})^2 -2 \sum_{k_1=1}^{K'} D_{k_1k_1}\frac{L_{kk_1}L_{lk_1}}{\sqrt{n_{k}n_{l}}}\nonumber\\
&=B_{kk}+B_{ll}-2B_{kl}\nonumber\\
&\geq \eta_n,
\tag{A.3}
\end{align}
where the first and last inequalities are implied by our condition. As a result, for $\Theta_{i\ast}\neq \Theta_{j\ast}$, we obtain $\|{U}_{i\ast}-{U}_{j\ast}\|_2\geq \sqrt{\frac{\eta_n}{\iota_n}}$.\QEDA

\subsection*{Proof of Theorem \ref{rproappro}}
We use the concentration bound of the non-randomized $A$ around $P$ (\cite{lei2015consistency}; \cite{chin2015stochastic}; \cite{gao2017achieving}) and the argument about the low-rank randomized approximation in \cite{halko2011finding} to bound the derivation of $\tilde{A}^{\rm rp}$ from $P$.
We begin by noting that
\begin{align}
\phantomsection
\label{A.4}
\|\tilde{A}^{\rm rp}-P\|_2&=\|QQ^{\intercal}AQQ^{\intercal}-P\|_2\nonumber \\
&\leq \|A-P\|_2+\|QQ^{\intercal}AQQ^{\intercal}-A\|_2\nonumber \\
&=:\mathcal I_1+\mathcal I_2.
\tag{A.4}
\end{align}

For $\mathcal I_1$, \cite{lei2015consistency} use the delicate combinatorial argument to provide a sharp bound. That is, assume
\begin{equation}
{\rm max}_{kl}B_{kl}\leq \alpha_n \;{\rm for\; some}\; \alpha_n\geq c_0\,{\rm log}n/n,\nonumber
\end{equation}
then for any $s>0$, there exists a constant $c$ such that
\begin{align}
\phantomsection
\label{A.5}
\|A-P\|_2\leq c\sqrt{n\alpha_n}.
\tag{A.5}
\end{align}
with probability at least $1-n^{-s}$.

For $\mathcal I_2$, we first note that
\begin{align}
\phantomsection
\label{A.6}
\|A- QQ^{\intercal}AQQ^{\intercal}\|_2&=\|A-QQ^{\intercal}A+QQ^{\intercal}A-QQ^{\intercal}AQQ^{\intercal}\|_2\nonumber \\
&\leq \|A-QQ^{\intercal}A\|_2+\|QQ^{\intercal}(A-AQQ^{\intercal})\|_2 \nonumber \\
& \leq 2\|A-QQ^{\intercal}A\|_2.
\tag{A.6}
\end{align}
By the Corollary 10.9 and Theorem 9.2 of \cite{halko2011finding}, when $r\geq 4$, $r{\rm log r}\leq n$ and $q\geq1$, the following inequality holds with probability at least $1-6\cdot r^{-r}$,
\begin{equation}
\phantomsection
\label{A.7}\|A-QQ^{\intercal}A\|_2\leq \sigma_{K'+1} (A)(1+11\sqrt{K'+r}\cdot\sqrt{n})^{\frac{1}{2q+1}},\tag{A.7}
\end{equation}
where $\sigma_{K'+1} (\cdot)$ denotes the $K'+1$th largest eigenvalue of a symmetric matrix. Now we bound $\sigma_{K'+1} (A)$. Recall that $P=\Theta B \Theta ^{\intercal}$ is of rank $K'$, then we have
\begin{equation}
\phantomsection
\label{A.8}\sigma_{K'+1} (A)=\sigma_{K'+1} (A)-\sigma_{K'+1} (P)\leq \|A-P\|_2.\tag{A.8}
\end{equation}
As a result, with probability at least $1-6\cdot r^{-r}-n^{-s}$
\begin{equation}
\phantomsection
\label{A.9}\|A-QQ^{\intercal}A\|_2\leq c'\sqrt{n\alpha_n}(1+11\sqrt{K'+r}\cdot\sqrt{n})^{\frac{1}{2q+1}}\leq c''\sqrt{n\alpha_n}(\sqrt{K'+r}\cdot\sqrt{n})^{\frac{1}{2q+1}}.\tag{A.9}
\end{equation}
When $q=cn^{1/\tau}$ for any $\tau>0$, we can easily prove that $(\sqrt{K'+r}\cdot\sqrt{n})^{\frac{1}{2q+1}}=O(1)$ when $n$ goes to infinity. Therefore, we have
\begin{equation}
\phantomsection
\label{A.10}\mathcal I_2\leq c\sqrt{n\alpha_n}.\tag{A.10}
\end{equation}

Finally, combining (\ref{A.10}) with (\ref{A.5}), we arrive the results of Theorem \ref{rproappro}.\QEDA

\subsection*{Proof of Theorem \ref{rpromis}}
We make use of the framework in \cite{lei2015consistency} to bound the misclustered rate. To fix ideas, we recall some notation now. $U$ and ${U}^{\rm rp}$ denote the $K'$ leading eigenvectors of $P=\Theta B\Theta^{\intercal}$ and $\tilde{A}^{\rm rp}$ (the output of Algorithm \ref{randsp}), respectively. ${\tilde{U}}^{\rm rp }:={\tilde{\Theta}}^{\rm rp}{\tilde{X}}^{\rm rp}$ denotes output of the randomized spectral clustering (Algorithm \ref{randsp}). Recall that the heuristic of spectral clustering under the SBM lies in that two nodes are in the same community if and only if their corresponding rows of $U$ are the same (\cite{lei2015consistency}; \cite{rohe2011spectral}). Based on these facts, in what follows, we first bound the derivation of ${\tilde{U}}^{\rm rp }$ from $U$. Then, for those nodes within each true cluster that correspond to a large derivation of ${\tilde{U}}^{\rm rp}$ from $U$, we bound their size. At last, we show that for the remaining nodes, the estimated and true clusters coincide.

First, we bound the derivation of $\tilde{U}^{\rm rp}$ from $U$. Davis-Kahan $\rm sin \Theta$ theorem (Theorem VII.3.1 of \cite{bhatia1997graduate}) provides a useful tool for bounding the perturbation of eigenvectors from the perturbation of matrices. Specifically, by Proposition 2.2 of \cite{vu2013minimax}, there exists a $K'\times K'$ orthogonal matrix $O$ such that,
\begin{equation}
\phantomsection
\label{A.11}
\|{U}^{\rm rp}- UO\|_{\tiny \rm F}\leq \frac{2\sqrt{2K'}}{\gamma_n}\|\tilde{A}^{\rm rp}-P\|_2.\tag{A.11}
\end{equation}
Now we proceed to derive the Frobenius error of ${\tilde{U}}^{\rm rp}$. Note that
\begin{align}
\phantomsection
\label{A.12}
\|{\tilde{U}}^{\rm rp}- UO\|_{\tiny \rm F}^2&=\|{\tilde{U}}^{\rm rp}-{U}^{\rm rp}+{U}^{\rm rp}-UO\|_{\tiny \rm F}^2\nonumber \\
& \leq \|UO-{U}^{\rm rp}\|_{\tiny \rm F}^2+\|{U}^{\rm rp}-UO\|_{\tiny \rm F}^2 \nonumber \\
&=2\|{U}^{\rm rp}-UO\|_{\tiny \rm F}^2,\tag{A.12}
\end{align}
where the first inequality follows from our assumption that ${\tilde{U}}^{\rm rp}$ is the global solution minimum of the following $k$-means objective and $UO$ is a feasible solution,
\begin{equation}
({\tilde{\Theta}}^{\rm rp}, {\tilde{X}}^{\rm rp})=\underset{{\Theta\in  \mathbb M_{n,K},X\in \mathbb R^{K\times K'}}}{{\rm arg\;min}}\;\|\Theta X-{U}^{\rm rp}\|_{\rm \tiny F}^2.\nonumber
\end{equation}
Then combine (\ref{A.12}) with (\ref{A.11}) and the bound of $\|\tilde{A}^{\rm rp}-P\|_2$ in Theorem \ref{rproappro}, we have with probability larger than $1-6r^{-r}-n^{-s}$ that
\begin{align}
\phantomsection
\label{A.13}
\|{\tilde{U}}^{\rm rp}- UO\|_{\tiny \rm F}^2\leq \frac{cK'{n\alpha_n}}{\gamma_n^2}.
\tag{A.13}
\end{align}
For notational convenience, we denote the right hand side of (\ref{A.13}) as ${\rm err}(K',n,\alpha_n,\gamma_n)$ in what follows.

Then, we proceed to bound the fraction of misclustered nodes. Define
\begin{equation}
\phantomsection
\label{A.14}
S_k=\{i\in G_k(\Theta):\; \|({\tilde{U}})_{i\ast}^{\rm rp}- (UO)_{i\ast}\|_{\tiny \rm F}>\frac{\delta_n}{2}\},
\tag{A.14}
\end{equation}
where $\delta_n$ is defined in Theorem \ref{rpromis} and $S_k$ is essentially the number of misclustered nodes in the true cluster $k$ (after some permutation) as we will see soon. By the definition of $S_k$, it is
easy to see
\begin{equation}
\phantomsection
\label{A.15}
\sum_{k=1}^K|S_k|\delta_n^2/4\leq \|{\tilde{U}}^{\rm rp}- UO\|_{\tiny \rm F}^2={\rm err}(K',n,\alpha_n,\gamma_n).
\tag{A.15}
\end{equation}
Hence,
\begin{equation}
\phantomsection
\label{A.16}
\sum_{k=1}^K\frac{|S_k|}{n_k}\leq c\frac{{\rm err}(K',n,\alpha_n,\gamma_n)}{\delta_n^2{\rm min}\,n_k}.
\tag{A.16}
\end{equation}

Next, we show that the nodes outside $S_k$ are correctly clustered. Before that, we first prove $|S_k|<n_k$. We have by (\ref{A.16}) that
\begin{equation}
\phantomsection
\label{A.17}
\frac{|S_k|}{n_k}\leq c\frac{{\rm err}(K',n,\alpha_n,\gamma_n)}{\delta_n^2{\rm min}\,n_k}.
\tag{A.17}
\end{equation}
Thus it suffices to prove
\begin{equation}
\phantomsection
\label{A.18}
c\frac{{\rm err}(K',n,\alpha_n,\gamma_n)}{\delta_n^2{\rm min}\,n_k}<1.
\tag{A.18}
\end{equation}
which actually follows from the assumption (\ref{A4}). As a result, we have
$|S_k|<n_k$ for every $1\leq k\leq K$. Therefore, $T_k\equiv G_k\backslash S_k\neq \emptyset$, where we recall that $G_k$ denotes the nodes in the true cluster $k$. Let $T=\cup _{k=1}^KT_k$, we now show that the rows in $(UO)_{T\ast}$ has a one to one correspondence with those in ${\tilde{U}}_{T\ast}^{\rm rp}$. On the one hand, for $i\in T_k$ and $j\in T_l$ with $l\neq k$,
${\tilde{U}}_{i\ast}^{\rm rp}\neq {\tilde{U}}_{j\ast}^{\rm rp}$, otherwise we have the following contradiction
\begin{align}
\phantomsection
\label{A.19}
\delta_n&\leq \|(UO)_{i\ast}-(UO)_{j\ast}\|_2\nonumber\\
&\leq \|(UO)_{i\ast}-{\tilde{U}}_{i\ast}^{\rm rp}\|_2+\|(UO)_{j\ast}-{\tilde{U}}_{j\ast}^{\rm rp}\|_2 \nonumber\\
&<\frac{\delta_n}{2}+\frac{\delta_n}{2},
\tag{A.19}
\end{align}
where the first inequality follows from Lemma \ref{lem:eigen}. On the other hand, for $i,j\in T_k$,
${\tilde{U}}_{i\ast}^{\rm rp}= {\tilde{U}}_{j\ast}^{\rm rp}$, because otherwise $\tilde{U}_{T\ast}$ has more than $K$ distinct rows which contradicts the fact that the output cluster size is $K$.

Till now, we have proved the membership is correctly recovered outside of $\cup _{k=1}^K S_k$ and the rate of misclustered nodes in $S_k$ is bounded as in (\ref{A.16}).
Therefore we obtain the claim of Theorem \ref{rpromis}.\QEDA

\subsection*{Proof of Theorem \ref{rprolink}}
We first bound the the derivation of ${\tilde{B}}_{ql}^{\rm rp}$ from $B_{ql}$ for each pair of $1 \leq q,l\leq K$, then we use the union bound to obtain a bound of $\|{\tilde{B}}^{\rm rp}-B\|_\infty$. Denote $\mathcal E$ be the event that (\ref{4.5}) and (\ref{4.6}) in Theorem \ref{rpromis} hold, which holds with probability larger than $1-6r^{-r}-n^{-s}$ for any $s>0$. In what follows, we derive the bound under the event $\mathcal E$.

Note that for any $1\leq q,l\leq K$,
\begin{equation}
\phantomsection
\label{A.20} {B}_{ql}=\frac{\sum_{1\leq i,j\leq n}P_{ij}\Theta_{iq}\Theta_{jl}}{\sum_{1\leq i,j\leq n}\Theta_{iq}\Theta_{jl}}.\tag{A.20}
\end{equation}
Then we have the following observations,
\begin{align}
\phantomsection
\label{A.21}
&|{\tilde{B}}_{ql}^{\rm rp}-B_{ql}|\nonumber\\
=&\Big|\frac{\sum_{1\leq i,j\leq n}\tilde{A}^{\rm rp}_{ij}{\tilde{\Theta}}^{\rm rp}_{iq}{\tilde{\Theta}}^{\rm rp}_{jl}}{\sum_{1\leq i,j\leq n}{\tilde{\Theta}}^{\rm rp}_{iq}{\tilde{\Theta}}^{\rm rp}_{jl}}-\frac{\sum_{1\leq i,j\leq n}P_{ij}\Theta_{iq}\Theta_{jl}}{\sum_{1\leq i,j\leq n}\Theta_{iq}\Theta_{jl}}\Big|\nonumber\\
\leq& \Big|\frac{\sum_{1\leq i,j\leq n}\tilde{A}^{\rm rp}_{ij}{\tilde{\Theta}}^{\rm rp}_{iq}{\tilde{\Theta}}^{\rm rp}_{jl}}{\sum_{1\leq i,j\leq n}{\tilde{\Theta}}^{\rm rp}_{iq}{\tilde{\Theta}}^{\rm rp}_{jl}}-\frac{\sum_{1\leq i,j\leq n}\tilde{A}_{ij}^{\rm rp}\Theta_{iq}\Theta_{jl}}{\sum_{1\leq i,j\leq n}\Theta_{iq}\Theta_{jl}}\Big|+
\Big|\frac{\sum_{1\leq i,j\leq n}\tilde{A}_{ij}^{\rm rp}\Theta_{iq}\Theta_{jl}}{\sum_{1\leq i,j\leq n}\Theta_{iq}\Theta_{jl}}-\frac{\sum_{1\leq i,j\leq n}P_{ij}\Theta_{iq}\Theta_{jl}}{\sum_{1\leq i,j\leq n}\Theta_{iq}\Theta_{jl}}\Big| \nonumber\\
=:&\mathcal I_1+\mathcal I_2.
\tag{A.21}
\end{align}

First, for $\mathcal I_2$, we have
\begin{align}
\phantomsection
\label{A.22}
\mathcal I_2&\leq \frac{\|\tilde{A}^{\rm rp}-P\|_{\tiny \rm F}(\sum (\Theta_{iq}\Theta_{jl})^2)^{1/2}}{n_qn_l}=\frac{\|\tilde{A}^{\rm rp}-P\|_{\tiny \rm F}}{(n_qn_l)^{1/2}}\nonumber\\
&\leq \frac{\sqrt{K'+r}\|\tilde{A}^{\rm rp}-P\|_{2}}{(n_qn_l)^{1/2}}\nonumber\\
&\leq \frac{c\sqrt{K'+r}\sqrt{n\alpha_n}}{(n_qn_l)^{1/2}}\nonumber
\tag{A.22}
\end{align}
where the first inequality follows from the Cauchy-Schwarz's inequality and the fact that $\sum_{1\leq i,j\leq n}\Theta_{iq}\Theta_{jl}=n_qn_l$, the second inequality follows from $\|A\|_{\rm F}\leq \sqrt{{\rm rank} (A)}\|A\|_2$ for any matrix $A$ and the fact that $\tilde{A}^{\rm rp}-P$ has rank at most $K'+r$, and the last inequality is implied by the spectral bound of $\tilde{A}^{\rm rp}-P$ (see (\ref{4.1})).

Next, we bound $\mathcal I_1$. We have
\begin{align}
\phantomsection
\label{A.23}
\mathcal I_1
\leq&\Big|\frac{\sum_{1\leq i,j\leq n}\tilde{A}^{\rm rp}_{ij}{\tilde{\Theta}}^{\rm rp}_{iq}{\tilde{\Theta}}^{\rm rp}_{jl}}{\sum_{1\leq i,j\leq n}{\tilde{\Theta}}^{\rm rp}_{iq}{\tilde{\Theta}}^{\rm rp}_{jl}}-
\frac{\sum_{1\leq i,j\leq n}\tilde{A}^{\rm rp}_{ij}{{\Theta}}^{\rm rp}_{iq}{{\Theta}}^{\rm rp}_{jl}}{\sum_{1\leq i,j\leq n}{\tilde{\Theta}}^{\rm rp}_{iq}{\tilde{\Theta}}^{\rm rp}_{jl}}\Big|+
\Big|\frac{\sum_{1\leq i,j\leq n}\tilde{A}^{\rm rp}_{ij}{{\Theta}}^{\rm rp}_{iq}{{\Theta}}^{\rm rp}_{jl}}{\sum_{1\leq i,j\leq n}{\tilde{\Theta}}^{\rm rp}_{iq}{\tilde{\Theta}}^{\rm rp}_{jl}}-
\frac{\sum_{1\leq i,j\leq n}\tilde{A}^{\rm rp}_{ij}{{\Theta}}^{\rm rp}_{iq}{{\Theta}}^{\rm rp}_{jl}}{\sum_{1\leq i,j\leq n}{{\Theta}}^{\rm rp}_{iq}{{\Theta}}^{\rm rp}_{jl}}\Big|
\nonumber\\
=:&\mathcal I_{11}+\mathcal I_{12}.
\tag{A.23}
\end{align}
For $1\leq q\leq K$, denote the $\hat{n}_q$ be the number of nodes in the $q$th estimated cluster, that is, $\sum_{i}\tilde{{\Theta}}^{\rm rp}_{iq}=\hat{n}_q$. Then we have for $\mathcal I_{11}$ that,
\begin{align}
\phantomsection
\label{A.24}
\mathcal I_{11}&\leq \frac{1}{\hat{n}_q\hat{n}_l}\|\tilde{A}^{\rm rp}\|_{\rm F}(\sum_{i,j}({\Theta}^{\rm rp}_{iq}{{\Theta}}^{\rm rp}_{jl}+{\tilde{\Theta}}^{\rm rp}_{iq}{\tilde{\Theta}}^{\rm rp}_{jl})^2)^{1/2}\nonumber\\
&\leq \frac{1}{\hat{n}_q\hat{n}_l}\|\tilde{A}^{\rm rp}\|_{\rm F}((n_qn_l)^{1/2}+(\hat{n}_q\hat{n}_l)^{1/2}) \nonumber\\
&=\|\tilde{A}^{\rm rp}\|_{\rm F}(\frac{1}{(\hat{n}_q\hat{n}_l)^{1/2}}+\frac{({n}_q{n}_l)^{1/2}}{\hat{n}_q\hat{n}_l}), \nonumber
\tag{A.24}
\end{align}
where the first and second inequality follows from the Cauchy-Schwarz's inequality and the triangle inequality, respectively.
Using the Cauchy-Schwarz's inequality again, we have for $\mathcal I_{12}$ that,
\begin{align}
\phantomsection
\label{A.25}
\mathcal I_{12}\leq |\sum_{ij}\tilde{A}^{\rm rp}_{ij}{\Theta}^{\rm rp}_{iq}{{\Theta}}^{\rm rp}_{jl}||\frac{1}{n_qn_l}-\frac{1}{\hat{n}_q\hat{n}_l}|\leq \|\tilde{A}^{\rm rp}\|_{\rm F}(\frac{1}{({n}_q{n}_l)^{1/2}}+\frac{({n}_q{n}_l)^{1/2}}{\hat{n}_q\hat{n}_l}).\nonumber
\tag{A.25}
\end{align}
Putting (\ref{A.25}) and (\ref{A.24}) together, we have for $\mathcal I_1$ that,
\begin{align}
\phantomsection
\label{A.26}
\mathcal I_{1}&\leq \|\tilde{A}^{\rm rp}\|_{\rm \tiny F}(\frac{1}{({n}_q{n}_l)^{1/2}}+\frac{1}{(\hat{n}_q\hat{n}_l)^{1/2}}+2\frac{({n}_q{n}_l)^{1/2}}{\hat{n}_q\hat{n}_l})\nonumber\\
&\leq( \|\tilde{A}^{\rm rp}-P\|_{\rm \tiny F}+\|P\|_{\rm \tiny F})(\frac{1}{({n}_q{n}_l)^{1/2}}+\frac{1}{(\hat{n}_q\hat{n}_l)^{1/2}}+2\frac{({n}_q{n}_l)^{1/2}}{\hat{n}_q\hat{n}_l})\nonumber\\
&\leq (c \sqrt{K'+r}\sqrt{n\alpha_n}+\sqrt{K'}\sigma_n)(\frac{1}{({n}_q{n}_l)^{1/2}}+\frac{1}{(\hat{n}_q\hat{n}_l)^{1/2}}+2\frac{({n}_q{n}_l)^{1/2}}{\hat{n}_q\hat{n}_l}),\nonumber
\tag{A.26}
\end{align}
where the last inequality is implied by $\|A\|_{\rm F}\leq \sqrt{{\rm rank} (A)}\|A\|_2$ for any matrix $A$ and the following facts, $\tilde{A}^{\rm rp}-P$ has rank at most $K'+r$, $P$ has rank $K'$, the spectral bound of $\tilde{A}^{\rm rp}-P$ (see (\ref{4.1})), and the largest eigenvalue of $P$ is $\sigma_n$.
To further bound (\ref{A.26}), we now discuss the relationship between $n_{k}$ and $\hat{n}_k$. Recall (\ref{4.6}), we see that $S_k$ is the number of misclustered nodes in the $k$th true cluster. Hence we have
\begin{align}
\phantomsection
\label{A.27}
\hat{n}_k\geq n_k-S_k \geq n_k-n_kc_3^{-1}\frac{K'n\alpha_n}{\gamma_n^2}=n_k(1-c_3^{-1}\frac{K'n\alpha_n}{\gamma_n^2\delta_n^2{\rm min}\, n_k}),\nonumber
\tag{A.27}
\end{align}
where the second inequality follows from (\ref{4.6}), namely, the error bound for the sum of misclassification error over all $K$ clusters. Combining (\ref{A.27}) with (\ref{A.26}), we have the bound for $\mathcal I_1$,
\begin{align}
\phantomsection
\label{A.28}
\mathcal I_{1}&\leq c( \frac{\sqrt{K'+r}\sqrt{n\alpha_n}}{{\rm min}\, n_k}+\frac{\sqrt{K'}\sigma_n}{{\rm min}\, n_k})\left(1+(1-\Phi_n)^{-1}+\frac{2{\rm max}\, n_k}{{\rm min}\, n_k}(1-\Phi_n)^{-2}\right),\nonumber
\tag{A.28}
\end{align}
where $\Phi_n:=c_3^{-1}\frac{K'n\alpha_n}{\gamma_n^2\delta_n^2{\rm min}\, n_k}$
Consequently, combining (\ref{A.28}) with (\ref{A.23}), we obtain the bound for $|{\tilde{B}}_{ql}^{\rm rp}-B_{ql}|$.
Finally, considering the event $\mathcal E$ and using the union bound, we obtain the desired bound for $\|{\tilde{B}}^{\rm rp}-B\|_\infty$.\QEDA

\subsection*{Proof of Theorem \ref{rsamappro}}
Before deriving the spectral error bound of $\tilde{A}^{\rm rs}$ from $P$, we first give some notation. Recall that $\tilde{A}^{\rm rs}$ is obtained by two steps:
\emph{(a)} Randomly select pair $(i,j)$ of the adjacency matrix $A$ independently with probability $p_{ij}$ regardless of the value of $A_{ij}$, and for each pair $(i,j)(i<j)$, the symmetric sparsified matrix $\tilde{A}^{\rm s}$ is defined as $\tilde{A}_{ij}^{\rm s}=\frac{A_{ij}}{p_{ij}}$ if $(i,j)$ is selected, and $\tilde{A}_{ij}^{\rm s}=0$ otherwise, \emph{(b)} Apply an {iterative algorithm} to find the nearly-optimal rank $K'$ approximation $\tilde{A}^{\rm rs}$ of $A^{\rm s}$. Let ${G}$ be the adjacency matrix of an Erod\"{o}s-Renyi graph with edge $(i,j)$ being with probability $0<p_{ij}<1$. Define $\bar{P}=\left(\frac{1}{p_{ij}}\right)\in\mathbb{R}^{n^2}$. Then it is obvious that $\tilde{A}_s$ in \emph{(a)} can be written as ${\tilde{A}_s=\bar{P}\circ G\circ A}$, where $\circ$ denotes the element-wise multiplication. To simplify the proof, we assume that \emph{(b)} finds the exactly optimal rank $K'$ approximation $\tilde{A}^{\rm rs}$ of $\tilde{A}^{\rm s}$, i.e.,
\begin{align}
\phantomsection
\label{A.29}
\tilde{A}^{\rm rs}={\rm arg\,min}\;_{{\rm rank}(M)\leq K'}\|\tilde{A}_s-M\|_2={\rm arg\,min}\;_{{\rm rank}(M)\leq K'}\|\bar{P}\circ G\circ A-M\|_2.
\tag{A.29}
\end{align}
Now we proceed to derive the error bound of $\tilde{A}^{\rm rs}$ from $P$. Note that
\begin{align}
\phantomsection
\label{A.30}
\|\tilde{A}^{\rm rs}-P\|_2&\leq\|\tilde{A}^{\rm rs}-\bar{P}\circ G\circ A\|_2+\|\bar{P}\circ G\circ A-P\|_2\nonumber\\
&\leq 2\|\bar{P}\circ G\circ A-P\|_2 =2\|\bar{P}\circ G\circ  (A-P)+\bar{P}\circ G\circ P-P\|_2 \nonumber\\
&\leq 2\|\bar{P}\circ G\circ (A-P)\|_2+ 2 \|\bar{P}\circ G\circ P-P\|_2,\nonumber\\
&=\mathcal I_1+\mathcal I_2,\nonumber\tag{A.30}
\end{align}
where the second inequality follows from (\ref{A.29}) and the fact that ${\rm rank}(P)=K'$.

To bound $\mathcal I_1$ and $\mathcal I_2$, we need the following results on the spectral-norm bound of a random matrix with symmetric independent and bounded entries (see Proposition 1 of \citet{chen2015fast}; Corollaries 3.6 and 3.12 in \citet{bandeira2016sharp}).
\begin{proposition}\label{propsition1}
Let $X$ be an $n\times n$ symmetric random matrix whose entries $X_{ij}$'s are independent symmetric random variables and bounded such that ${\rm max}_{ij}|X_{ij}|\leq \sigma_1$. Define \begin{align}
\phantomsection
\label{A.31}
\sigma_2={\rm max}_i \sqrt{\mathbb E\sum_jX_{ij}^2}.\nonumber
\tag{A.31}
\end{align}
Then there exists universal constants $c$ and $c'$ such that,
\begin{align}
\phantomsection
\label{A.32}
&\quad\quad\quad\quad\mathbb  E\|X\|_2\leq 3\sigma_2+c\sigma_1\sqrt{{\rm log} n}, \nonumber \\
&\mathbb P(\|X\|_2\geq 3\sigma_2+t)\leq n\cdot {\rm exp}(-\frac{t^2}{c'\sigma_1^2})\; {\mbox{ for any}}\;\, t\geq 0.
\tag{A.32}
\end{align}
\end{proposition}

We first bound $\mathcal I_1=2\|\bar{P}\circ G\circ (A-P)\|_2$ by conditioning on $A-P\equiv W$, where $\bar{P}\circ G\circ W$ is the $X$ in Proposition \ref{propsition1}.  We have $(\bar{P}\circ G\circ W)_{ij}=b_{ij}\frac{W_{ij}}{p_{ij}}$, where $b_{ij}\sim {\rm Bernoulli}(p_{ij})$. It is easy to see that ${\rm max}_{ij}|b_{ij}\frac{W_{ij}}{p_{ij}}|\leq \frac{2}{p_{\rm min}}$, so we can set $\sigma_1= \frac{2}{p_{\rm min}}$. And we also have,
\begin{align}
\phantomsection
\label{A.33}
\sigma_2&={\rm max}_i \sqrt{\mathbb E(\sum_jb_{ij}^2\frac{W_{ij}^2}{p_{ij}^2}| W)}={\rm max}_i \sqrt{\sum_j\frac{W_{ij}^2}{p_{ij}^2} \mathbb E( b_{ij}^2| W)}\nonumber\\
&\leq{\rm max}_i\;p_{\rm min}^{-1/2}\sqrt{\|W_{i\ast}\|_2^2}= p_{\rm min}^{-1/2}\sqrt{\|W\|_{2,\infty}^2}\leq  p_{\rm min}^{-1/2}\|W\|_2,\nonumber
\tag{A.33}
\end{align}
where for any matrix $B$, $\|B\|_{2,\infty}:= {\rm max}_i(\sum_jB_{ij}^2)^{1/2}$ and the last inequality follows from the fact that $\|W\|_{2,\infty} \leq \|W\|_2$. Alternatively, we can obtain $\sigma_2\leq c\sqrt{{\rm max}_i\sum_j \frac{1}{p_{ij}}}$ if we repeat the steps in (\ref{A.33}) and take summation over $\mathbb E(b_{ij}^2|W)/p_{ij}^2$. Hence, $\sigma_2\leq c {\rm min}\,\{ p_{\rm min}^{-1/2}\|W\|_2, \sqrt{{\rm max}_i\sum_j \frac{1}{p_{ij}}}\}$.
Choosing $t=c\sigma_1\sqrt{{\rm log} n}$ in Proposition \ref{propsition1} for large enough constant $c>0$, then there exists constant $\nu_1>0$ such that with probability larger than $1-n^{\nu_1}$,
\begin{align}
\phantomsection
\label{A.34}
\mathcal I_1=2\|\bar{P}\circ G\circ (A-P)\|_2\leq c\,{\rm max}\, \Big\{{\rm min}\,\big\{ p_{\rm min}^{-1/2}\|W\|_2, \sqrt{{\rm max}_i\sum_j {p_{ij}^{-1}}}\big\},\;\frac{\sqrt{{\rm log} n}}{p_{\rm min}}\Big\}.
\tag{A.34}
\end{align}
To further bound $\mathcal I_1$, we use the following spectral norm error bound of $A-P$ proved in \cite{lei2015consistency}. That is, under assumption (\ref{A2}),
\begin{align}
\phantomsection
\label{A.35}
\|W\|_2=\|A-P\|_2\leq c \sqrt{n\alpha_n},
\tag{A.35}
\end{align}
with probability larger than $1-n^{-\nu_1}$, where we set $\nu_1$ to be identical to that corresponds to (\ref{A.34}).
As a result, we have with probability larger than $1-2n^{-\nu_1}$ that,
\begin{align}
\phantomsection
\label{A.36}
\mathcal I_1\leq c\,{\rm max}\, \Big\{{\rm min}\,\big\{ p_{\rm min}^{-1/2}\sqrt{n\alpha_n}, \sqrt{{\rm max}_i\sum_j {p_{ij}^{-1}}}\big\},\;\frac{\sqrt{{\rm log} n}}{p_{\rm min}}\Big\}.
\tag{A.36}
\end{align}

Next, we use Proposition \ref{propsition1} again to bound $\mathcal I_2=2 \|\bar{P}\circ G\circ P-P\|_2$. We first have
\begin{align}
\phantomsection
\label{A.37}
{\rm max}_{ij}|\frac{1}{p_{ij}}G_{ij}P_{ij}-P_{ij}|\leq {\rm max} \{1,\frac{1}{p_{\rm min}}-1\}\cdot\alpha_n.
\tag{A.37}
\end{align}
Hence we can set $\sigma_1= {\rm max} \{1,\frac{1}{p_{\rm min}}-1\}\cdot\alpha_n$. For $\sigma_2$, we have
\begin{align}
\phantomsection
\label{A.38}
\sigma_2&={\rm max}_i \sqrt{\mathbb E(\sum_j[(\frac{1}{p_{ij}}G_{ij}-1)^2P_{ij}^2)]}={\rm max}_i \sqrt{\sum_j\alpha_{n}^2\mathbb E(\frac{1}{p_{ij}}G_{ij}-1)^2}\leq  \sqrt{\alpha_n^2{\rm max}_i\sum_j(\frac{1}{p_{ij}}-1)}\nonumber.
\tag{A.38}
\end{align}
Selecting $t=c\sigma_1\sqrt{{\rm log} n}$ in Proposition \ref{propsition1} for large enough constant $c>0$, then there exists constant $\nu_2>0$ such that with probability larger than $1-n^{-\nu_2}$,
\begin{align}
\phantomsection
\label{A.39}
\mathcal I_2\leq c\,{\rm max} \left( \sqrt{\alpha_n^2{\rm max}_i\sum_j(\frac{1}{p_{ij}}-1)},\;{\rm max} \{1,\frac{1}{p_{\rm min}}-1\}\cdot\alpha_n\sqrt{{\rm log}n}\right).
\tag{A.39}
\end{align}

Consequently, combining (\ref{A.36}) with (\ref{A.39}), we have with probability larger than $1-c_6n^{-\nu}$ that
\begin{align}
\phantomsection
\label{A.40}
\|\tilde{A}^{\rm rs}-P\|_2&=\mathcal I_1+\mathcal I_2\nonumber\\
&\leq c\,  {\rm max}\Big\{I_1,\;\frac{\sqrt{{\rm log} n}}{p_{\rm min}},\, \sqrt{n\alpha_n^2(\frac{1}{p_{\rm min}}-1)},\;\sqrt{\alpha_n^2{\rm log}n{\rm max} \{1,\frac{1}{p_{\rm min}}-1\}^2}\Big\} ,\nonumber
\tag{A.40}
\end{align}
where $\nu={\rm min}\{\nu_1,\nu_2\}$ and $I_1:={\rm min}\Big\{\sqrt{\frac{n\alpha_n}{p_{\rm min}}},\sqrt{{\rm max}_i\sum_j\frac{1}{p_{ij}}}\Big\}$. The conclusion in Theorem \ref{rsamappro} is arrived. \QEDA

\subsection*{Proof of Lemma \ref{lem:eigen2}}
Let $\tilde{\vartheta}$ be an $n\times 1$ vector such that the $i$th element is $\vartheta_i/\|\phi_{g_i}\|_2$, where recall that $\phi_k$ is an $n\times1$ vector that consistent with $\vartheta$ on $G_k$ and zero otherwise. Let $\bar{\Theta}$ be the normalized membership matrix such that $\bar{\Theta}(i,k)=\bar{\vartheta}_i$ if $i\in G_k$ and $\bar{\Theta}(i,k)=0$. And it can be verified that $\bar{\Theta}^\intercal \bar{\Theta}=I.$ Recall $\Omega={\rm diag} (\|\phi_1\|_2,...,\|\phi_{K}\|_2)$. Then we have
\begin{equation}
\phantomsection
\label{A.41}
{\rm diag}(\vartheta)\Theta=\bar{\Theta}\Omega.
\tag{A.41}
\end{equation}
As a result,
\begin{equation}
\phantomsection
\label{A.42}
P={\rm diag}(\vartheta)\Theta B\Theta^\intercal{\rm diag}(\vartheta)=\bar{\Theta}\Omega B\Omega\bar{\Theta}^\intercal=\bar{\Theta}HDH^{\intercal}\bar{\Theta}^\intercal,
\tag{A.42}
\end{equation}
where we denote the eigenvalue decomposition of $\Omega B\Omega$ by
\begin{equation}
\phantomsection
\label{A.43}
\Omega B\Omega=H_{K\times K'}D_{K'\times K'}H_{K'\times K}^\intercal,
\tag{A.43}
\end{equation}
where $H$ has orthogonal columns. Recall that we also have $P=U\Sigma U^\intercal$, hence by the orthonormality of $\bar{\Theta}$ and $H$, we have
\begin{equation}
\phantomsection
\label{A.44}
{U}=\bar{\Theta}H,  \quad {\Sigma}=D.
\tag{A.44}
\end{equation}

Specifically,
${U}_{i\ast}=\tilde{\vartheta}_iH_{k\ast}$ for $i\in G_k$.

Now we discuss the structure of $U$ respectively when $B$ is of full rank ($K'=K$) and
rank deficient ($K'<K$).

(a) When $K'=K$, $H$ is a square matrix with orthogonal columns, implying that the rows of $H$ are perpendicular with each other. As a result, ${\rm cos}({U}_{i\ast},{U}_{j\ast})=0$ if $g_i\neq g_j$ and ${\rm cos}({U}_{i\ast},{U}_{j\ast})=1$ if $g_i= g_j$.

(b) When $K'<K$, it is straightforward that ${\rm cos}({U}_{i\ast},{U}_{j\ast})=1$ if $g_i= g_j$. For $g_i\neq g_j$, without loss of generality we assume $g_i=k,g_j=l(l\neq k)$. In this case, we observe that ${\rm cos}({U}_{i\ast},{U}_{j\ast})={\rm cos}({H}_{k\ast},{H}_{l\ast})$. Therefore, by the condition that $\max_{k,l}{\rm cos}({H}_{k\ast},{H}_{l\ast})<\xi'_n<1$, we arrive the conclusion of Lemma \ref{lem:eigen2}.\QEDA

\subsection*{Proof of Lemma \ref{lem:eigen2condition}}
First, we show that for any $1\leq k\leq K$, $\|{H}_{k\ast}\|_2\neq 0$, which excludes the trivial case that ${\rm cos}({H}_{k\ast},{H}_{l\ast})=1$ for $k\neq l$. To see this, by (\ref{A.43}), (\ref{A.44}) and the condition that for any $1\leq i\leq K'$, $0<\Sigma_{ii}<\overline{\iota}_n$, we have
$$\overline{\iota}_n\|{H}_{k\ast}\|_2^2\geq \sum_{k_1=1}^{K'}D_{k_1k_1}H_{kk_1}^2=\bar {B}_{kk}>0,$$
where the last inequality follows from the assumption that $B_{kk}>0$ for any $1\leq k\leq K$ and the definition of $\bar{B}$.

Second, we show that for any $\lambda$ and any $1\leq k<l\leq K$, $\|{H}_{k\ast}-\lambda{H}_{l\ast}\|_2^2>0$ under our condition, which indicates that ${\rm cos}({H}_{k\ast},{H}_{l\ast})<1$. In fact, we can observe that
\begin{align}
\phantomsection
\label{A.45}
\overline{\iota}_n\|{H}_{k\ast}-\lambda {H}_{l\ast}\|_2^2\geq \sum_{k_1=1}^{K'}D_{k_1k_1}(H_{kk_1}-\lambda H_{lk_1})^2=\lambda^2 \bar{B}_{ll} -2\lambda \bar{B}_{kl}+\bar{B}_{kk}.
\tag{A.45}
\end{align}
Note that the RHS of (\ref{A.45}) is a parabola of form $a\lambda^2+b\lambda+c$ with $a:=\bar{B}_{ll}$, $b:=-2\bar{B}_{kl}$, and $c:=\bar{B}_{kk}$. Also note that $\bar{B}_{ll}>0$, hence the RHS of (\ref{A.45}) is always larger than 0 if the discriminant $b^2-4ac:=4 \bar{B}_{kl}^2-4\bar{B}_{kk}\bar{B}_{ll}<0$, which is actually our condition.

Third, we give an explicit bound for ${\rm cos}({H}_{k\ast},{H}_{l\ast})$. Choosing $\lambda=-\frac{2a}{b}$, we obtain that
$$\lambda^2 \bar{B}_{ll} -2\lambda \bar{B}_{kl}+\bar{B}_{kk}\geq \frac{-b^2+4ac}{4a}:=\frac{-\bar{B}_{kl}^2+\bar{B}_{kk}\bar{B}_{ll}}{\bar{B}_{ll}}\geq \frac{\eta'_n}{\beta_n}>0,$$
where the last inequality follows from our conditions. Hence by (\ref{A.45}), for any $\lambda$, we have
\begin{align}
\phantomsection
\label{A.46}
\|{H}_{k\ast}-\lambda {H}_{l\ast}\|_2^2\geq \frac{\eta'_n}{\overline{\iota}_n \beta_n}.
\tag{A.46}
\end{align}
Note that the LHS of (\ref{A.46}) is also a parabola. Choosing $\lambda=\frac{{H}_{k\ast}{H}_{l\ast}^\intercal}{\|{H}_{l\ast}\|_2^2}$, we thus have
$$\frac{-({H}_{k\ast}{H}_{l\ast}^\intercal)^2+\|{H}_{l\ast}\|_2^2\|{H}_{k\ast}\|_2^2}{\|{H}_{l\ast}\|_2^2} \geq \frac{\eta'_n}{\overline{\iota}_n \beta_n},$$
which implies
$${\rm cos}({H}_{k\ast},{H}_{l\ast})\leq \sqrt{1-\frac{\eta'_n}{\overline{\iota}_n\beta_n\|{H}_{k\ast}\|_2^2}}\leq \sqrt{1-\frac{\eta'_n}{\overline{\iota}_n \beta_n^2/\underline{\iota}_n}},$$
where in the last inequality, we used the fact that $$\underline{\iota}_n\|{H}_{k\ast}\|_2^2\leq \sum_{k_1=1}^{K'}D_{k_1k_1}H_{kk_1}^2=\bar {B}_{kk}<\beta_n.$$
The proof is completed. \QEDA

\subsection*{Proof of Theorem \ref{rpromis2}}
The proof is similar to that of Theorem \ref{rpromis} except that we need to handel the normalized eigenvectors and the eigenvectors with 0 norm. To fix ideas, we recall and introduce some notation now. With slight abuse of notation, $U$ and $\hat{U}$ denote the $K'$ leading eigenvectors of $P=\Theta B\Theta^{\intercal}$ and $\tilde{A}$, respectively. ${\tilde{U}}$ denotes the output of the randomized spherical spectral clustering. ${{U}}'$ denotes the normalized ${U}$, namely, ${U}'_{i\ast}=U_{i\ast}/\|U_{i\ast}\|_2$. ${\hat{U}}'$ denotes the normalized ${\hat{U}}$, namely, ${\hat{U}}'_{i\ast}={\hat{U}}_{i\ast}/\|{\hat{U}}_{i\ast}\|_2$ and ${\hat{U}}'_{i\ast}:=0$ if $\|{\hat{U}}_{i\ast}\|_2=0$.

First, following the same proof strategy with that in Theorem \ref{rpromis}, there exists a $K'\times K'$ orthogonal matrix $O$ such that,
\begin{align}
\phantomsection
\label{A.47}
\|{\hat{U}}- UO\|_{\tiny \rm F}^2\leq \frac{cK'{\|A-P\|_2^2}}{\gamma_n^2}.
\tag{A.47}
\end{align}
For notational simplicity, in what follows we assume the orthogonal matrix $O$ is the identity matrix.
Note that $U$'s rows are all non-zero. To see this, recall in (\ref{A.43}), we have shown that ${U}_{i\ast}=\tilde{\vartheta}_iH_{k\ast}$ for $i\in G_k$, where $H_{K\times K'}$ has orthogonal columns. When $K'=K$, $H$ is an orthonormal matrix, hence we have
\begin{align}
\phantomsection
\label{A.48}
\|U_{i\ast}\|_2^2=\tilde{\vartheta}_i^2.
\tag{A.48}
\end{align}
When $K'<K$, by our condition that $0<\Sigma_{ii}<\overline{\iota}_n$ and ${B}_{kk}>0$ we have
$$\overline{\iota}_n\|{H}_{k\ast}\|_2^2\geq \sum_{k_1=1}^{K'}D_{k_1k_1}H_{kk_1}^2=\bar {B}_{kk}>0,$$
which implies that
\begin{align}
\phantomsection
\label{A.49}
\|U_{i\ast}\|_2^2=\tilde{\vartheta}_i^2\|H_{k\ast}\|_2^2\geq \tilde{\vartheta}_i^2\bar {B}_{kk}/\overline{\iota}_n.
\tag{A.49}
\end{align}
For any vectors $a$ and $b$, the fact that $$\|\frac{a}{\|a\|_2}-\frac{b}{\|b\|_2}\|_2\leq 2\frac{\|a-b\|_2}{{\rm max}(\|a\|_2,\|b\|_2)}$$
holds, and for any $a=0$, $\|0-\frac{b}{\|b\|_2}\|_2\leq 2\frac{\|0-b\|_2}{\|b\|_2}$ holds trivially. We thus have
\begin{align}
\phantomsection
\label{A.50}
\|{\hat{U}}'- U'\|_{\tiny \rm F}^2\leq c\sum_{i=1}^n\frac{\|{\hat{U}}_{i\ast}-{U}_{i\ast}\|_2^2}{\|{U}_{i\ast}\|_2^2}
\leq \frac{\|{\hat{U}}- U\|_{\tiny \rm F}^2}{\min_i \|{U}_{i\ast}\|_2^2}\leq c\frac{K'{\|A-P\|_2^2}}{\gamma_n^2\min_i \|{U}_{i\ast}\|_2^2},
\tag{A.50}
\end{align}
where the last inequality follows from (\ref{A.47}) and $\min_i \|{U}_{i\ast}\|_2^2$ can be lower bounded differently depending on $K'=K$ (see (\ref{A.48})) or $K'<K$ (see (\ref{A.49})). Further, by the fact that ${\tilde{U}}$ is the $k$-means solution of ${\hat{U}}'$, we have
\begin{align}
\phantomsection
\|{\tilde{U}}-U'\|_{\tiny \rm F}^2\leq \|{\tilde{U}}-{\hat{U}}'\|_{\tiny \rm F}^2+\|{\hat{U}}'- U'\|_{\tiny \rm F}^2\leq 2\|{\hat{U}}'- U'\|_{\tiny \rm F}^2\leq  c\frac{K'{\|A-P\|_2^2}}{\gamma_n^2\min_i \|{U}_{i\ast}\|_2^2},\nonumber
\end{align}

Next, we proceed to bound the fraction of misclustered nodes. Define
\begin{equation}
\phantomsection
\label{A.51}
S_k=\{i\in G_k(\Theta):\; \|{\tilde{U}}_{i\ast}-{U}'_{i\ast}\|_2>\frac{\mu_n}{2}\},
\tag{A.51}
\end{equation}
where $\mu_n={\sqrt{2}}$ if $K'=K$, and $ \mu_n={\sqrt{2(1-\xi'_n)}}$ if $K'<K$, where $\xi'_n$ is defined in (\ref{A7}). By the definition of $S_k$, it is
easy to see
\begin{equation}
\phantomsection
\label{A.52}
\sum_{k=1}^K|S_k|\mu_n^2/4\leq \|{\tilde{U}}-{U}'\|_{\tiny \rm F}^2\leq c\frac{K'{\|A-P\|_2^2}}{\gamma_n^2\min_i \|{U}_{i\ast}\|_2^2},
\tag{A.52}
\end{equation}
where the last inequality follows from (\ref{A.50}). And thus
\begin{equation}
\phantomsection
\label{A.53}
\sum_{k=1}^K\frac{|S_k|}{n_k}\leq c\frac{K'{\|A-P\|_2^2}}{\gamma_n^2\min_i \|{U}_{i\ast}\|_2^2\cdot \mu_n^2{\rm min}\,n_k}.
\tag{A.53}
\end{equation}

Now, we show that the nodes outside $S_k$ are correctly clustered. We first note that the nodes corresponding to zero rows of ${\tilde{U}}$ are in $S_k$. Therefore those outside $S_k$ correspond to nonzero ${\tilde{U}}_{i\ast}$'s. By (\ref{A.53}) and our condition (\ref{A8}) and (\ref{A9}), we first have $|S_k|<n_k$. Hence, $T_k\equiv G_k\backslash S_k\neq \emptyset$, where we recall that $G_k$ denotes the nodes in the true cluster $k$. Let $T=\cup _{k=1}^KT_k$, we now show that the rows in $U'_{T\ast}$ has a one to one correspondence with those in ${\tilde{U}}'_{T\ast}$. On the one hand, for $i\in T_k$ and $j\in T_l$ with $l\neq k$,
${\tilde{U}}'_{i\ast}\neq {\tilde{U}}'_{j\ast}$, otherwise we have the following contradiction
\begin{align}
\phantomsection
\label{A.54}
\mu_n\leq \sqrt{2-2{\rm cos}(U'_{i\ast},U'_{j\ast})}&= \|U'_{i\ast}-U'_{j\ast}\|_2\nonumber\\
&\leq \|U'_{i\ast}-{\tilde{U}}_{i\ast}\|_2+\|{\tilde{U}}_{j\ast}-U'_{j\ast}\|_2 \nonumber\\
&<\frac{\mu_n}{2}+\frac{\mu_n}{2},
\tag{A.54}
\end{align}
where the first inequality follows from Lemma \ref{lem:eigen2}. On the other hand, for $i,j\in T_k$,
${\tilde{U}}'_{i\ast}= {\tilde{U}}'_{j\ast}$, because otherwise $\tilde{U}'_{T\ast}$ has more than $K$ distinct rows which contradicts the fact that the output cluster size is $K$.

Consequently, we obtain the claim of Theorem \ref{rpromis2}.\QEDA

\end{document}